\documentclass{aa}

\usepackage{graphicx}
\usepackage{txfonts}
\usepackage{xcolor}
\usepackage{ulem}

\usepackage[hidelinks]{hyperref}
\usepackage{orcidlink}

\usepackage{siunitx}
\DeclareSIUnit\parsec{pc}
\DeclareSIUnit\gauss{G}
\DeclareSIUnit\erg{erg}
\DeclareSIUnit\year{yr}
\DeclareSIUnit\arcmin{arcmin}
\DeclareSIUnit\arcsec{arcsec}
\DeclareSIUnit\counts{cts}
\DeclareSIUnit\jansky{Jy}
\DeclareSIUnit\steradian{sr}

\usepackage{ctable}

\begin{document}

    \title{\textit{eROSITA} X-ray analysis of the PeVatron candidate Westerlund 1}
    %\subtitle{}

    %List of authors
    \author{
            Konstantin Haubner\orcidlink{0009-0007-7808-4653}\inst{\ref{inst1},\ref{inst2},\ref{inst3}}
        \and
            Manami Sasaki\orcidlink{0000-0001-5302-1866}\inst{\ref{inst1}}
        \and
            Alison Mitchell\orcidlink{0000-0003-3631-5648}\inst{\ref{inst4}}
        \and
            Gabriele Ponti\orcidlink{0000-0003-0293-3608}\inst{\ref{inst5},\ref{inst6}}
        \and
            Gavin Rowell\orcidlink{0000-0002-9516-1581}\inst{\ref{inst7}}
        \and
            Sabrina Einecke\orcidlink{0000-0001-9687-8237}\inst{\ref{inst7}}
        \and
            Miroslav Filipovi\'c\orcidlink{0000-0002-4990-9288}\inst{\ref{inst8}}
        \and
            Sanja Lazarevi\'c\orcidlink{0000-0001-6109-8548}\inst{\ref{inst8},\ref{inst9},\ref{inst10}}
        \and
            Gerd P\"uhlhofer\orcidlink{0000-0003-4632-4644}\inst{\ref{inst11}}
        \and
            Andrew Strong\inst{\ref{inst6}}}

    %Institutes of the authors
    \institute{
            Dr. Karl Remeis Observatory, Erlangen Centre for Astroparticle Physics, Friedrich-Alexander-Universität Erlangen-Nürnberg, Sternwartstraße 7, 96049 Bamberg, Germany\label{inst1}
        \and
            INAF -- Arcetri Astrophysical Observatory, Largo Enrico Fermi 5, 50125 Firenze, Italy\\
            \email{konstantin.haubner@inaf.it}\label{inst2}
        \and
            Dipartimento di Fisica e Astronomia, Università degli Studi di Firenze, via G. Sansone 1, 50019 Sesto Fiorentino, Firenze, Italy\label{inst3}
        \and 
            Erlangen Centre for Astroparticle Physics, Friedrich-Alexander-Universit\"at Erlangen-N\"urnberg, Nikolaus-Fiebiger-Str. 2, 91058 Erlangen, Germany\label{inst4}
        \and
            INAF -- Osservatorio Astronomico di Brera, Via E. Bianchi 46, 23807 Merate, Italy\label{inst5}  
        \and
            Max-Planck-Institut f{\"u}r extraterrestrische Physik, Gießenbachstraße 1, 85748 Garching, Germany\label{inst6}
        \and
            School of Physics, Chemistry and Earth Sciences,  The University of Adelaide, Adelaide 5005, Australia\label{inst7}
        \and
            Western Sydney University, Locked Bag 1797, Penrith South DC, NSW 2751, Australia\label{inst8} 
        \and
            CSIRO, Space and Astronomy, PO Box 76, Epping, NSW 1710, Australia\label{inst9}
        \and
            Astronomical Observatory, Volgina 7, 11060 Belgrade, Serbia\label{inst10}
        \and
            Institut f{\"u}r Astronomie und Astrophysik, Universit{\"a}t T{\"u}bingen, Sand 1, 72076 T{\"u}bingen, Germany\label{inst11}
}

    %Receive and acceptance date of the paper; added by the editor
    \date{Received 23 August 2024; accepted 19 January 2025}

    %Abstract of the paper
    \abstract
    %Context heading (optional)
    {It is still unclear which fraction of cosmic rays above an energy of \SI{1}{\peta\electronvolt} is accelerated by the observed Galactic PeVatron population. These sources of unknown physical origin are detected through their $\gamma$-ray emission, which also identifies them as particle accelerators. However, their $\gamma$-ray data are typically degenerate between hadronic and leptonic emission scenarios, which hinders their straightforward association with the mainly hadronic cosmic ray population.}
    %Aims heading (mandatory)
    {In this study, we aimed to distinguish between leptonic and hadronic particle acceleration scenarios for the PeVatron candidate HESS~J1646$-$458, which is associated with the star cluster Westerlund~1 (Wd~1). To this end, we first studied the diffuse X-ray emission from Wd~1 to better understand if its origin is of thermal or nonthermal nature. In addition, we searched for X-ray synchrotron emission from the associated PeVatron candidate HESS~J1646$-$458 to put new constraints on the magnetic field strength and the leptonic particle population of this source.}
    %Methods heading (mandatory)
    {We used data from the all-sky surveys 1 to 4 of the \textit{extended Roentgen Survey with an Imaging Telescope Array} (\textit{eROSITA}) on board the Spectrum-Roentgen-Gamma orbital platform to spectrally analyze the diffuse emission from Wd~1 and HESS~J1646$-$458. For Wd~1, we fitted and compared a purely thermal model and a model with a thermal and a nonthermal component. Next, we analyzed the spectra of four annuli around Wd~1 that coincide with HESS~J1646$-$458 to search for synchrotron radiation.}
    %Results heading (mandatory)
    {We find that \textit{eROSITA} data cannot distinguish between thermal and nonthermal source scenarios for the diffuse emission from Wd~1 itself. For a thermal source scenario, the observed X-ray flux can be explained in large part by unresolved pre-main sequence stars or by thermalized stellar wind shocks. In the case of the PeVatron candidate HESS~J1646$-$458, we find no evidence of synchrotron emission. We estimated an upper confidence bound of the synchrotron flux up to \SI{40}{\arcmin} around Wd~1 of \SI{1.9e-3}{\per\kilo\electronvolt\per\square\centi\meter\per\second}. We used this result to study the spectral energy distribution of the source. From that, we obtained an upper $1\sigma$ confidence bound on the magnetic field strength of HESS~J1646$-$458 of $\SI{7}{\micro\gauss}$.}
    %Conclusions heading (optional), leave it empty if necessary
    {Our upper bound on the magnetic field strength in HESS~J1646$-$458 is compatible with a previous estimate in the literature for a fully leptonic source scenario. Therefore, a purely leptonic emission scenario is compatible with our results. The same is the case for hadronic and hybrid scenarios, for which even less synchrotron flux is expected compared to the leptonic scenario.}

    %Keywords to categorize the paper
    \keywords{acceleration of particles --
            cosmic rays --
            shock waves --
            stars: winds, outflows --
            open clusters and associations: individual: Westerlund 1 --
            X-rays: stars
            }

    \maketitle

\nolinenumbers

%%% Start of the main text of the paper %%%
\section{Introduction}

    {\setlength{\parskip}{.04cm plus0mm minus1mm}Cosmic rays (CRs) are one of the main drivers of the physics of the interstellar medium (ISM). They change metal abundances in the ISM and influence star formation therein \citep{Padovani2020}, they likely accelerate high-velocity clouds in the Galactic halo \citep{Wiener2019}, and they might drive galactic outflows \citep{Ipavich1975, 2022Armillotta, Peschken2023}. Consequently, CRs could also help to regulate the baryon content of galaxies and to alleviate the missing baryons problem in cosmology \citep{Dashyan2020, Ruszkowski2023}.

    Therefore, a better understanding of the production of CRs is important for the study of galaxy evolution. In general, there is good evidence that the bulk of CRs is produced via diffusive shock acceleration in strong shocks, which typically arise in supernova remnants (SNRs; \citealt{2012Helder}, and references therein). However, the standard paradigm of CR acceleration, which posits that the bulk of CRs up to energies of a few petaelectronvolt are accelerated in isolated SNRs, is increasingly being challenged. On the theoretical side, it is still unclear whether SNRs can accelerate particles to the required energies above \SI{1}{\peta\electronvolt} \citep{Lagage1983, Hillas2005}. On the empirical side, the standard paradigm does not account for differences in the spectra of individual CR particle species, for example between hydrogen and helium \citep{AMS2015}. More examples of problems of the standard paradigm can be found in \citet{Gabici2019}.}

    The most important challenge in the context of this study is the discovery of a population of Galactic PeVatron sources, that is, sources that accelerate particles to energies above \SI{1}{\peta\electronvolt} \citep{Abeysekara2020, Cao2021}. Since such particles are charged and therefore deflected in the Galactic magnetic field, they cannot be traced back to their individual sources. Instead, PeVatrons are identified via the detection of secondary $\gamma$-ray spectra that extend well above \SI{100}{\tera\electronvolt}. Crucially, the physical nature of these sources detected in such a way is still unclear. However, since there are very often no promising SNRs near them, their detection has further increased the tension between observations and the standard SNR paradigm and fueled the interest in alternative sources of Galactic CRs.
    
    Among the alternative sources of Galactic CRs that have been put forward are pulsar wind nebulae \citep{Ohira2018, Albert2021} and young star clusters. Of the latter, several have been observationally established as sites of particle acceleration, including Westerlund~1 (Wd~1; \citeauthor{Abramowski2012} \citeyear{Abramowski2012}), Westerlund~2 \citep{Yang2018}, and Cygnus~OB2 \citep{Ackermann2011}. From a theoretical perspective, these objects could accelerate particles either at colliding shock waves inside the cluster itself \citep{Vieu2020} or, alternatively, at the cluster wind termination shock \citep{Morlino2021}.

    Still, a major problem with connecting PeVatron candidates to the production of CRs is the frequent degeneracy of their $\gamma$-ray spectra with respect to two emission scenarios: First, in the leptonic scenario, $\gamma$-rays are produced from accelerated electrons mainly via inverse Compton (IC) scattering, and second, in the hadronic scenario, accelerated hadrons, mainly protons, produce pions in interactions with ISM particles. Of these pions, the neutral ones subsequently decay into $\gamma$-rays. Consequently, disentangling between these two emission scenarios is a main goal of the study of PeVatrons.

    Here, the focus lies on the PeVatron candidate Wd~1. With a mass around $\SI{4e4}{}M_\sun$ \citep{Clark2005, Brandner2008}, Wd~1 is one of the most massive young star clusters in the Milky Way. Discovered in 1961 by Bengt Westerlund \citep{Westerlund1961}, the cluster is located at a right ascension of $\alpha_\mathrm{J2000} = 16^\mathrm{h}47^\mathrm{m}02.4^\mathrm{s}$ and a declination of $\delta_\mathrm{J2000} = -45\degr{}51'07"$ \citep{Tarricq2021} and therefore on the Galactic plane ($l = \SI{339.55}{\deg}$, $b = \SI{-0.40}{\deg}$). The distance to Wd~1 has been estimated using different methods, such as the luminosities of spectroscopically identified stars \citep{Westerlund1987}, the radial velocity of neutral hydrogen gas \citep{Kothes2007}, and Gaia parallax measurements \citep{Negueruela2022, Navarete2022}. The estimates mostly lie between $3$ and \SI{5}{\kilo\parsec}. Here, we assume a distance of \SI{3.9}{\kilo\parsec} to be consistent with previous $\gamma$-ray studies of Wd~1 \citep{Aharonian2022, Haerer2023}.

    Westerlund~1 is widely studied for its rich population of massive stars. According to \citet{Clark2020}, it contains more than 100 OB giants and supergiants, four red supergiants, one luminous blue variable, and possibly six yellow hypergiants (see \citet{Beasor2023} for a disputing argument on the hypergiants). Particularly interesting for this study, Wd~1 also hosts at least 24 Wolf-Rayet (WR) stars \citep{Clark2020}, whose stellar winds might act as sites of particle acceleration. Furthermore, Wd~1 harbors the magnetar CXOU~164710.2$-$455216 \citep{Borghese2019} in its southeast, which is also the only direct evidence of past supernova activity in the cluster.

    Coinciding with Wd~1, \citet{Abramowski2012} discovered the very-high-energy (VHE; $0.1 - \SI{100}{\tera\electronvolt}$) $\gamma$-ray source HESS~J1646$-$458. It has a complex shell-like but energy-independent morphology with a radius of ${\sim}\SI{30}{\arcmin}$, and Wd~1 lies roughly at its center.  With newer observations, the spectrum of the source is well described by a power law with an exponential cutoff at $44^{+17}_{-11}\,\SI{}{\tera\electronvolt}$ \citep{Aharonian2022}. The only weakly constrained position of this cutoff leaves open the possibility of HESS~J1646$-$458 being a PeVatron.  

    Furthermore, in the high-energy (HE; $0.1 - \SI{100}{\giga\electronvolt}$) $\gamma$-ray range, \citet{Ohm2013} detected extended emission ${\sim}\SI{1}{\deg}$ to the south of Wd~1 using the \textit{Fermi} Large Area Telescope (\textit{Fermi}-LAT). On the other hand, for the \SI{1.1}{\deg}-radius circle around Wd~1 they were only able to derive an upper flux limit. According to the incremental \textit{Fermi}-LAT fourth source catalog \citep{Abdollahi2022}, this region contains six HE $\gamma$-ray sources, two of which are the energetic pulsars PSR~J1648$-$4611 and PSR~J1650$-$4601 \citep{Smith2023}. Still, the pulsed nature of these two sources and the size of HESS~J1646$-$458, as well as its lack of an energy-dependent morphology, expected due to the rapid cooling of $\gamma$-ray emitting electrons, render pulsars an unlikely explanations for the VHE $\gamma$-ray emission. In general, based on energetic considerations, \citet{Abramowski2012} and \citet{Aharonian2022} find that Wd~1 is the most likely physical power source for the VHE-emitting relativistic particle distribution.

    Turning to the VHE $\gamma$-ray emission scenario, the spectrum of HESS~J1646$-$458 can be explained either by an underlying leptonic or hadronic particle distribution with an exponential cutoff power law shape \citep{Aharonian2022}. Due to the source's shell-like morphology and the small diffusion length of electrons, \citet{Aharonian2022} exclude a leptonic scenario if electrons are accelerated inside Wd~1 itself, but leave it open if electrons are accelerated at the cluster wind termination shock. Indeed, in a theoretical investigation of emission scenarios at the cluster wind termination shock, \citet{Haerer2023} find a preference for the leptonic scenario. This is based on the energy requirement for proton acceleration and on these particles' large diffusion length, which hinders an easy explanation of the shell-like morphology. \citet{Haerer2023} also derived lower and upper bounds on the magnetic field strength at the cluster wind termination shock of $\SI{0.7}{\micro\gauss} < B < \SI{4.5}{\micro\gauss}$ in the leptonic scenario.

    Consequently, the viable scenarios are hadron acceleration inside Wd~1 itself and lepton acceleration at the cluster wind termination shock. Disentangling these models requires further insights beyond the available $\gamma$-ray observations. In particular, X-ray observations are helpful due to their potential to detect synchrotron radiation. Since the synchrotron power from protons is suppressed by 13 orders of magnitude compared to electrons, its detection would indicate the presence of an accelerated electron population. As an example, the detection of X-ray synchrotron radiation has allowed \citet{Kavanagh2019} to confirm the star cluster Cygnus~OB2 as a leptonic $\gamma$-ray source.
    
    The diffuse X-ray emission from Wd~1 was previously studied by \citet{Muno2006} using the \textit{Chandra} X-ray observatory and by \citet{Kavanagh2011} using the \textit{X-ray Multi-mirror} (\textit{XMM}) \textit{Newton} telescope. They both found that the X-ray spectrum of Wd~1 is well fit with a model with two emission components, the softer one of which is thermal in nature. Specifically, \citet{Muno2006} discussed the possibility of a nonthermal hard emission component based on the nondetection of emission lines expected from a thermal hard component. This would indicate the presence of accelerated particles inside the star cluster.
    
    However, using the higher spectral resolution of \textit{XMM-Newton}, \citet{Kavanagh2011} discovered a significant emission line around \SI{6.7}{\kilo\electronvolt}. This is most likely associated with a K-shell transition of helium-like Fe~XXV and is expected from a plasma with a temperature above \SI{2}{\kilo\electronvolt}. The hard thermal component associated with this line leaves little room for additional nonthermal emission. Based on this result, a purely thermal explanation of the diffuse X-ray emission from Wd~1 seems preferable.

    Still, these studies only analyzed a few square arc minutes around Wd~1, since they were limited by the field of view sizes of \textit{Chandra} and \textit{XMM-Newton}. In particular, they covered the star cluster itself, which has an angular diameter of ${\sim} \SI{6}{\arcmin}$, but not HESS~J1646$-$458 in its entirety, which spans ${\sim} 1\deg$. This is even more important since HESS~J1646$-$458 has a shell-like morphology where the flux maximum is located at a distance of ${\sim}\SI{30}{\arcmin}$ from Wd~1.
    
    With the \textit{extended Roentgen Survey with an Imaging Telescope Array} (\textit{eROSITA}; \citeauthor{Predehl2021} \citeyear{Predehl2021}), it is now possible to overcome this field of view limitation. Its all-sky survey nature guarantees full coverage of Wd~1 and its surroundings and allows us to search for synchrotron radiation from the entirety of HESS~J1646$-$458. As a relatively young mission, \textit{eROSITA} was launched on board the Spectrum-Roentgen-Gamma orbital platform \citep{Sunyaev2021} in 2019 to perform eight all-sky surveys in the $0.2-\SI{10}{\kilo\electronvolt}$ energy band. The data of the first of these surveys have been made public recently, and the accompanying paper \citep{eROSITA2024} provides descriptions of the \textit{eROSITA} processing pipelines, the available types of data products, and the available catalogs.

    The rest of the paper is structured as follows: In Sect.~\ref{SecObservations}, we present the \textit{eROSITA} observations that we used. In Sect.~\ref{SecAnalysis}, we explain our analysis steps and their results. It is split into the following subsections: Subsection~\ref{SecImages} gives the images of Wd~1 and its surroundings, while Subsect.~\ref{SecContamination} covers the contamination from a bright point source, which was the main challenge to our analysis. Subsections~\ref{SecWd1Background} and \ref{SecWd1Spectrum} present our spectral analysis of the background region and of Wd~1, respectively. Our main analysis, the search for synchrotron radiation around Wd~1, is described in Subsect.~\ref{SecSynchrotron}. This is followed by the derivation of upper confidence bounds on the synchrotron radiation in Subsect.~\ref{SecUpperLimits} and by the spectral energy distribution fit to the data from \citet{Aharonian2022} and to our data in Subsect.~\ref{SecSED}. Section~\ref{SecDiscussion} then gives the discussion of our analysis and results. We wrap up with a summary of our conclusions in Sect.~\ref{SecConclusions}.
    
\section{Observations} \label{SecObservations}

    To date, four of the planned eight \textit{eROSITA} all-sky surveys (eRASS) have been completed. We used the combined data of these four surveys, commonly labeled eRASS:4, which covered Wd~1 for the first time on March 13, 2020 and for the last time on September 19, 2021. Importantly, \textit{eROSITA} contains seven individual telescope modules and seven corresponding individual CCDs. Of these, the detectors associated with the modules 5 and 7 suffer from optical light leak, most likely stemming from the Sun \citep{Predehl2021}. For this reason, we excluded these two telescope modules from our analyses. For better statistics, we then combined the data of the remaining five modules. The only exception to this are the background fits to the HESS~J1646$-$458 regions described in Sect.~\ref{SecUpperLimits}.

    We used data processed with the standard \textit{eROSITA} processing pipeline, version 020. We reduced this data with the \textit{eROSITA} Science Analysis Software System (\texttt{eSASS}, \citeauthor{Brunner2022} \citeyear{Brunner2022}) version \texttt{eSASSusers\_211214}. We merged the event files of eRASS 1 to eRASS 4 with the \texttt{evtool} task. During this, we also filtered out potential bad events by applying the 0xc00f7f30 flag. Next, we performed good time interval and flare filtering using the \texttt{FLAREGTI} flag with standard settings. The flare filtering reduced the exposure times of the investigated spectra by up to ${\sim}7\%$, for example, from ${\sim}\SI{1120}{\second}$ to ${\sim}\SI{1040}{\second}$ for the Wd~1 source region. We checked that this does not change the results of our analyses in a significant way by also performing the main steps without prior flare filtering.

    The masks which we applied to the event files are different for the study of Wd~1 itself and for the study of HESS~J1646$-$458. For the analysis of the diffuse emission from Wd~1, we only masked the magnetar CXOU~164710.2$-$455216. This is because the magnetar is the only resolved point source inside the star cluster and because the automated point source masking would also remove large parts of the diffuse emission from Wd~1. The magnetar mask is described in Subsect.~\ref{SecWd1Background}.

    For the search for synchrotron radiation, the faintness of the expected source emission made more extensive masking of point sources necessary. Therefore, we masked circles with radius \SI{50}{\arcsec} around all point sources listed in the eRASS:4 internal point source catalog. In addition, we manually applied individual larger masks to bright sources that were not properly removed by the initial masking procedure. The only such sources overlapping with our analysis regions are the low-mass X-ray binary (LMXB) GX~340$+$0 \citep{Friedman1967} and the high-mass X-ray binaries 2MASS~J16415078$-$4532253 and 2MASS~J16480656$-$4512068 \citep{Cutri2003}. In particular, GX~340$+$0 has a strong halo that contaminates large parts of our source regions and makes masking alone insufficient to deal with its effect. Its treatment is described in detail in Sect. \ref{SecContamination}. On the other hand, the two high-mass X-ray binaries have no such prominent halos. They were masked using circular regions with radii of \SI{162}{\arcsec} and \SI{90}{\arcsec} and center positions of $\alpha = 16^\mathrm{h}41^\mathrm{m}50.9^\mathrm{s}$, $\delta = -45\degr{}32'24.0"$ and $\alpha = 16^\mathrm{h}48^\mathrm{m}06.5^\mathrm{s}$, $\delta = -45\degr{}12'11.7"$, respectively.

\section{Analysis and results} \label{SecAnalysis}

\subsection{Images} \label{SecImages}

    Using the \texttt{eSASS} task \texttt{evtool}, we created images of the surroundings of Wd~1. We set the \texttt{rebin} parameter to $100$, which results in an image pixel size of \SI{5}{\arcsec}, well below the field of view-averaged resolution of \textit{eROSITA} of ${\sim}\SI{26}{\arcsec}$. Furthermore, we did not perform exposure correction since the exposure times in the region of interest are close to homogeneity and the region is far from the image edges, rendering vignetting effects negligible.

\begin{figure}
        \centering
        \includegraphics[width = \linewidth]{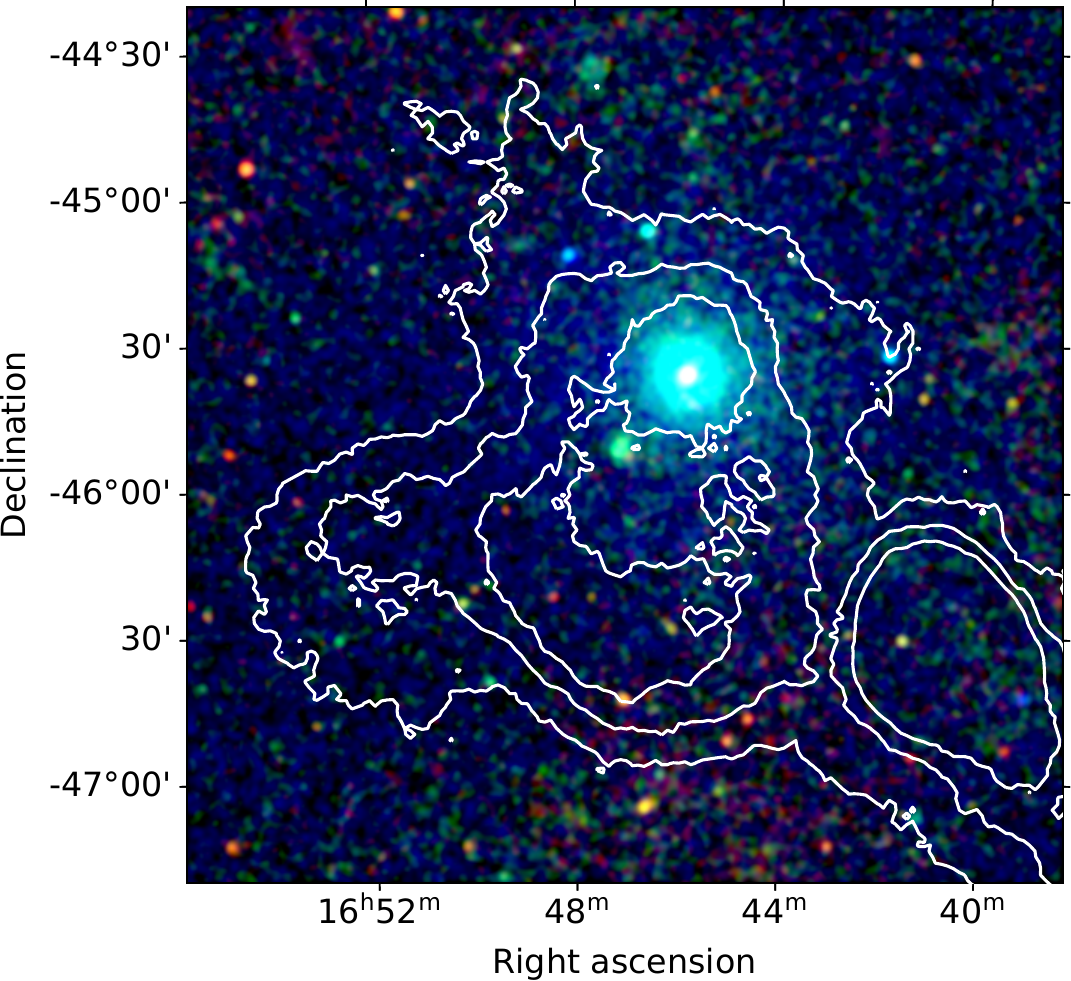}
        \caption{\textit{eROSITA} RGB image of the surroundings of Wd~1 with H.E.S.S. 4, 8, and $12\sigma$ contours from \citet{Aharonian2022} in white. Red, green, and blue colors correspond to $0.7$ to \SI{1.1}{\kilo\electronvolt}, $1.1$ to \SI{2.3}{\kilo\electronvolt}, and $2.3$ to \SI{10}{\kilo\electronvolt}, respectively. The image was smoothed using a Gaussian kernel with standard deviation \SI{40}{\arcsec}. Wd~1 itself is the diffuse green source near the center of the image, while the bright blue source to its northwest is the LMXB GX~340$+$0.}
        \label{FigTeVContours}
    \end{figure}

    Figure~\ref{FigTeVContours} shows an RGB \textit{eROSITA} image of the surroundings of Wd~1 with H.E.S.S. VHE $\gamma$-ray contours overlaid in white. These correspond to significances of 4, 8, and $12\sigma$, as in Fig.~1~(b) of \citet{Aharonian2022}. Red, green, and blue colors correspond to the X-ray energy bands between $0.7$ and \SI{1.1}{\kilo\electronvolt}, $1.1$ and \SI{2.3}{\kilo\electronvolt}, and $2.3$ and \SI{10}{\kilo\electronvolt}, respectively. The X-ray image was smoothed with a Gaussian kernel with standard deviation \SI{40}{\arcsec}.

    Wd~1 is visible as a diffuse green source near the center of Fig.~\ref{FigTeVContours}. However, the most striking source in the image is the LMXB GX~340$+$0 to the northwest of Wd~1. As a bright point source, it produces a hard and extended spherical halo that contaminates the region of the $\gamma$-ray source HESS~J1646$-$458. We do not note any indication of diffuse X-ray emission correlated with the morphology of HESS~J1646$-$458, which would be a sign of the presence of synchrotron radiation. However, the position of GX~340$+$0 coincides with the northern emission peak of the $\gamma$-ray source. This is also noted by \citet{Aharonian2022}, who ascribe it to a chance alignment, since pulsars are unlikely to contribute to HESS~J1646$-$458's emission.

\begin{figure}
        \centering
        \includegraphics[width = \linewidth]{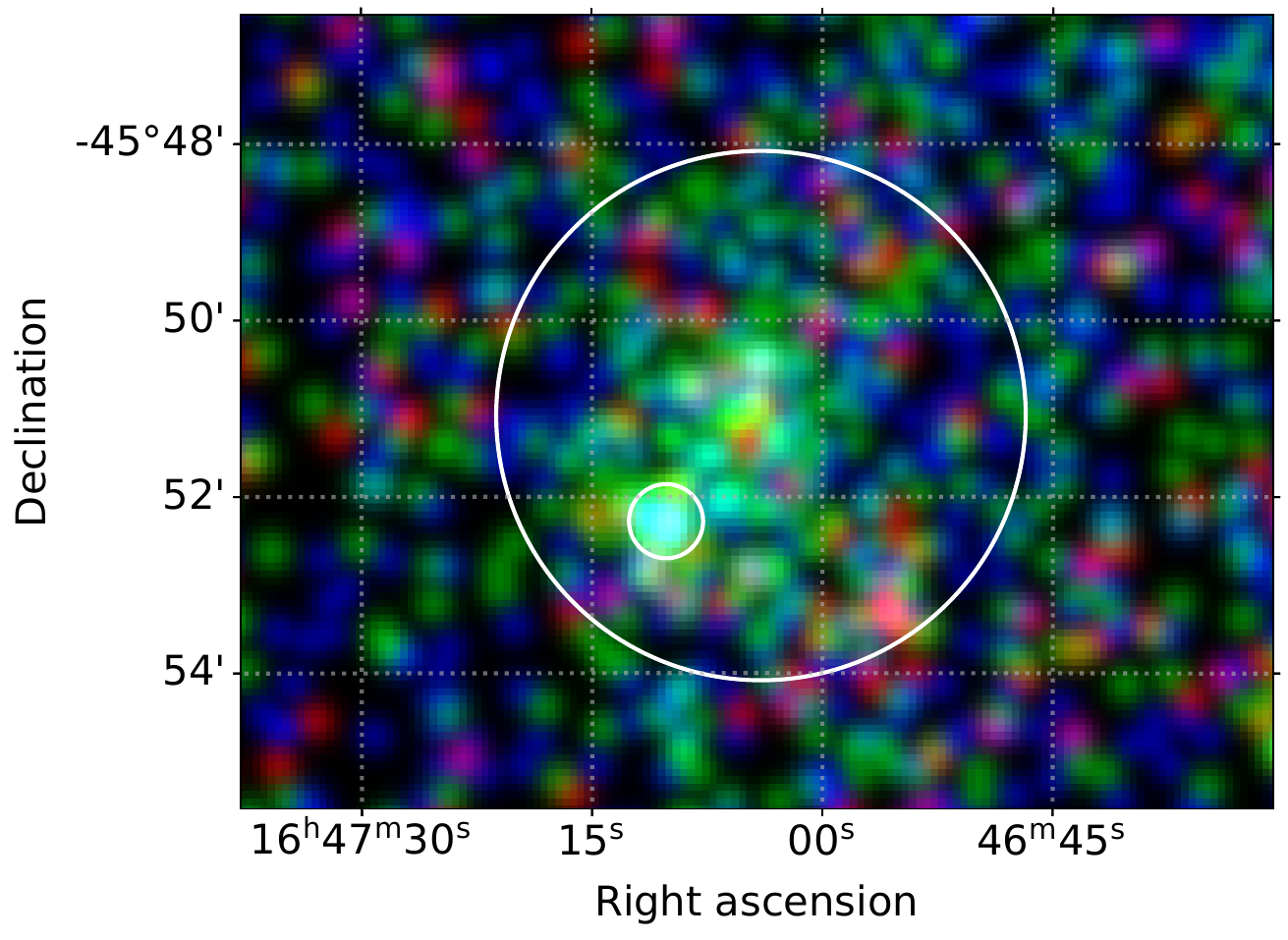}
        \caption{\textit{eROSITA} RGB image of Wd~1. Red, green, and blue colors correspond to $0.7$ to \SI{1.1}{\kilo\electronvolt}, $1.1$ to \SI{2.3}{\kilo\electronvolt}, and $2.3$ to \SI{10}{\kilo\electronvolt}, respectively. The image was smoothed using a Gaussian kernel with standard deviation \SI{10}{\arcsec}. The large circle indicates the source area defined for Wd~1, from which the region given by the small circle around the magnetar CXOU~164710.2$-$455216 was excluded.}
        \label{FigWd1Image}
    \end{figure}

    Next, Fig.~\ref{FigWd1Image} shows an RGB zoom-in on Wd~1, with the same bands as in Fig.~\ref{FigTeVContours}. Here, we smoothed the image with a Gaussian kernel with standard deviation \SI{10}{\arcsec}. The region highlighted with the large circle in the figure was used for the spectral analysis of Wd~1 as described in Subsect. \ref{SecWd1Spectrum}. The diffuse emission from Wd~1 can be seen to extend from southeast to northwest. The only clearly discernible point source associated with the star cluster is the magnetar CXOU~164710.2$-$455216 in its southeastern part. For the spectral analysis of Wd~1, we masked it with the indicated circular region.

\subsection{Contamination from GX 340+0}    \label{SecContamination}

    In X-ray astronomy, halos such as the one around GX~340$+$0 in Fig.~\ref{FigTeVContours} typically arise due to dust scattering in the Galactic plane \citep{Corrales2017, Lamer2021} and due to single-reflection stray light in the X-ray mirrors. These effects have different spectral and spatial properties, with dust scattering typically being important closer to the source, and stray light being important farther away. In particular, for \textit{eROSITA}, stray light can appear up to an angular distance of ${\sim}3\deg$ from a source \citep{Freyberg2020}.

    We spectrally inspected the surroundings of GX~340$+$0 in Fig.~\ref{FigTeVContours} and found Wd~1 and most of the region corresponding to HESS~J1646$-$458 to be affected by this contamination. Consequently, it is crucial for our spectral analysis to take this into account. We resorted to a phenomenological approach, which has the added advantage that we did not have to determine the exact contributions of dust scattering and stray light.

    First, we chose background regions for our spectral analyses that, under the assumption of a spherical LMXB halo, should contain the same relative amount of contamination as the source regions. Second, whenever spectral modeling was necessary, we modeled the contamination with an absorbed accretion disk model, that is, \texttt{tbabs}$\cdot$\texttt{diskbb} in \texttt{Xspec}. We found that this model provides a reasonable fit to the immediate surroundings of GX~340$+$0, which are most strongly affected by the contamination.

    The resulting fit parameters of this accretion disk component are phenomenological in nature. We used them only to properly model the background spectra for our studies of source components. In particular, we did not compare the fit parameters to literature studies of GX~340$+$0. We stress that no physical properties of the X-ray binary should be inferred from our accretion disk fit parameters.

\subsection{Fit methodology and background of Westerlund 1} \label{SecWd1Background}

    The first part of our study is the analysis of the diffuse X-ray emission from Wd~1 itself. We always extracted background and source spectra with the \texttt{eSASS} task \texttt{srctool}, using the region extent type \texttt{tophat}. For the Wd~1 source region, we chose a circular region of radius \SI{3}{\arcmin} centered on $\alpha = 16^\mathrm{h}47^\mathrm{m}04.8^\mathrm{s}$, $\delta = -45\degr{}51'05.1"$. This can be seen as the large white circle in Fig.~\ref{FigWd1Image} and corresponds to the optical extent of Wd~1. We also tested a smaller region with radius \SI{2.4}{\arcmin}, corresponding to the X-ray extent of Wd~1, and found it to lead to fit results consistent within 90\%. On the other hand, a much larger source region than the one we used would only increase the background component in the source spectrum, as can be seen in Fig.~\ref{FigWd1Image}. For our analysis, we only masked the magnetar CXOU~J164710.2$-$455216. The circular mask for the magnetar has a radius of \SI{0.42}{\arcmin} and is centered on $\alpha = 16^\mathrm{h}47^\mathrm{m}09.6^\mathrm{s}$, $\delta = -45\degr{}52'12.0"$. It is given as the small white circle in Fig.~\ref{FigWd1Image}. 

    Due to the contamination from GX~340$+$0, we paid particular attention to the choice of background. To keep the relative contamination contribution as large as in the source area, we chose a region with the same distance to the LMXB as Wd~1. Furthermore, to maximize the statistics of the background region, we chose an annulus centered on the position of GX~340$+$0. The inner and outer radii of this annulus correspond to the closest and farthest point of the source region from GX~340$+$0, respectively. Therefore, the center of the background region is at $\alpha = 16^\mathrm{h}45^\mathrm{m}48.0^\mathrm{s}$, $\delta = -45\degr{}36'36.0"$ and the two radii are \SI{16.62}{\arcmin} and \SI{22.62}{\arcmin}. Since the background region should not contain emission from the source, we excluded a \SI{6}{\arcmin}-radius circle concentric with the source region from the background region. The final background region can be seen in Fig.~\ref{FigWd1Bkg} in the appendix. We also tested two other background regions, both of ${\sim}\SI{1200}{\square\arcmin}$ size and located ${\sim}\SI{0.8}{\deg}$ to the northeast and southwest of GX~340$+$0, and found them to lead to similar fit results, but with larger uncertainties for some fit parameters.

    For our spectral analyses, we used \texttt{Xspec}\footnote{https://heasarc.gsfc.nasa.gov/xanadu/xspec/} version 12.13.1 with the Cash fit statistic \citep{Cash1979}. We performed all spectral fits in the energy range from $0.2$ to \SI{9.5}{\kilo\electronvolt}. After running initial fits, we used the \texttt{error} command to determine the confidence intervals of the fit parameters. Whenever this resulted in an improved fit, we accepted the new fit and repeated the procedure. For evaluating the goodness of our fits, we used the Pearson-$\chi^2$-test as well as \texttt{Xspec}'s \texttt{goodness} command. This command fits spectral models to simulated data that are based on the best-fit parameters. The fit statistic distribution of these fits to simulated data is then compared to the fit statistic of the best fit to the real data. When the real physical spectrum of the source appears similar to the best-fit model spectrum, ${\sim}50\%$ of the simulated data fits should have a smaller fit statistic than the real data fit. Furthermore, due to the low count number in the source spectrum of Wd~1, we combined the data from the five \textit{eROSITA} telescope modules into one spectrum. We did the same for the background spectrum. Next, we fitted the background spectrum, so that we could later fix the parameters of the background components during the source fits.
    
    In general, the \textit{eROSITA} background consists of two distinct physical components: First, the particle background is caused by charged particles hitting the detectors. Second, the astrophysical X-ray background stems from Galactic and extragalactic X-ray sources and is mostly diffuse in nature. In our case, a third background component comes from the halo of GX~340$+$0.

    The particle background is well described by the so-called filter wheel closed models, which are available to the \textit{eROSITA} consortium. These models were constructed based on observations with closed filter wheels, that is, no astrophysical X-rays were reaching the detectors. The filter wheel closed models are described in more detail in \citet{Yeung2023}. Specifically, we used the model for the combination of all telescope modules except numbers 5 and 7. The normalizations of these models have to be set to the \texttt{BACKSCAL} parameters of the respective fits files. This parameter gives the average source area in square degrees overlapping with the detector during the good time intervals. Here, we list it in the respective tables together with our best-fit parameters.

    After fixing the particle background in such a way, we fitted the combined X-ray background and contamination components. We modeled the X-ray background after \citet{Ponti2023}, who studied the diffuse \textit{eROSITA} background in the \SI{142}{\square\deg} \textit{eROSITA} final equatorial depth survey (eFEDS) area, which was observed during the telescope's calibration and performance verification phase. Notably, this background model also includes a Galactic corona component, studied in more detail by \citet{Locatelli2024}. Following \citet{Ponti2023}, our background model has the following components:
    \begin{enumerate}
        \item The local hot bubble (LHB), described by a thermal \texttt{apec} component with temperature $k_\mathrm{B}T_\mathrm{LHB} = \SI{0.11}{\kilo\electronvolt}$ and metallicity $Z_\mathrm{LHB} = 1Z_\sun$, where $Z_\sun$ is the Solar metal abundance.
        \item The circumgalactic medium (CGM), described by an absorbed thermal component, that is, \texttt{tbabs}$\cdot$\texttt{apec}. Temperature and metallicity were set to $k_\mathrm{B}T_\mathrm{CMB} = \SI{0.16}{\kilo\electronvolt}$ and $Z_\mathrm{CMB} = 0.08Z_\sun$.
        \item The Galactic corona, described by a \texttt{tbabs}$\cdot$\texttt{apec} component. Here, temperature and metallicity were set to $k_\mathrm{B}T_\mathrm{cor} = \SI{0.65}{\kilo\electronvolt}$ and $Z_\mathrm{cor} = 1Z_\sun$.
        \item The cosmic X-ray background (CXB), which consists mainly of unresolved active galactic nuclei. It is well described by an absorbed power law component. We used \texttt{tbabs}$\cdot$\texttt{powerlaw} with photon index $\Gamma_\mathrm{CXB} = 1.46$ and normalization $\eta_\mathrm{CXB} = \SI{8.88e-7}{\per\kilo\electronvolt\per\square\centi\meter\per\second\per\square\arcmin}$.
    \end{enumerate}
    Furthermore, we modeled the contamination based on our fits to the emission around GX~340$+$0:
    \begin{enumerate}
        \item[5.] The GX~340$+$0 halo was fitted with a phenomenological \texttt{tbabs}$\cdot$\texttt{diskbb} model. The parameters of this model are the hydrogen column density $N_\mathrm{H}^\mathrm{LMXB}$, the temperature $k_\mathrm{B}T_\mathrm{in}$, and the normalization $\eta_\mathrm{LMXB}$. These parameters were left free during the fit to the Wd~1 background region. The resulting best-fit values except for the normalization were then used as fixed parameters for the following search for synchrotron radiation from HESS~J1646$-$458.
    \end{enumerate}
    We tried adding a solar wind charge exchange (SWCX) component described by \citet{Ponti2023} via the AtomDB charge exchange model\footnote{http://www.atomdb.org/CX/}. SWCX results from the interaction of highly ionized atoms in the Solar wind with neutral atoms in the near-Earth environment. However, this additional SWCX component was consistently set to a normalization of zero. Therefore, we did not include it in our final fits.
    
    For constraining the neutral hydrogen column densities of the \texttt{tbabs} components, we used the \texttt{HEASARC} tool \texttt{nH}\footnote{https://heasarc.gsfc.nasa.gov/cgi-bin/Tools/w3nh/w3nh.pl} with data from the HI 4 Pi survey \citep{Dickey1990}. For the source region, we used the \texttt{nH} value at its center, namely \SI{2.20e22}{\per\square\centi\meter}. For the larger background region, we calculated the arithmetic mean and standard deviation of eleven equally distributed \texttt{nH} values along the annulus, which resulted in \SI{2.11\pm0.10e22}{\per\square\centi\meter}. For the GX~340$+$0 halo component, we left $N_\mathrm{H}^\mathrm{LMXB}$ unconstrained.

    Initially, we fixed the column densities of the CGM, the Galactic corona, and the CXB to the values determined via \texttt{nH}. However, this led to poor fits and in particular to CGM and corona normalizations several orders of magnitude larger than for \citet{Ponti2023}. Therefore, we allowed the CGM and corona column densities $N_\mathrm{H}^\mathrm{CGM}$ and $N_\mathrm{H}^\mathrm{cor}$ to vary up to maximum values given by the \texttt{nH} values. This is justified by Wd~1 lying on the Galactic plane, due to which some of the emissions attributed to the CGM and corona components might stem from sources inside the Galactic disk. This is different from the situation of \citet{Ponti2023}, who studied the eFEDS region, which is located farther away from the Galactic disk.

    The best-fit parameters for the background region of Wd~1 can be found in the appendix in Table~\ref{TabWd1Bkg}. The reduced Pearson-$\chi^2$ for the background fit is $\chi^2_\mathrm{P}/\nu = 918.41/897 = 1.024$, where $\nu$ is the number of degrees of freedom. The output of the \texttt{goodness} command is $67\%$, indicating a good fit.

\subsection{Spectrum of Westerlund 1} \label{SecWd1Spectrum}

    The spectral models for the Wd~1 source region consist of the particle background model, the X-ray background model fixed to the best-fit parameters from Table~\ref{TabWd1Bkg}, as well as the actual source models. The X-ray background component was scaled according to the different areas of the source and background regions that were taken from the \texttt{REGAREA} keywords in the respective headers. The values can be found in Tables~\ref{TabWd1Bkg} and \ref{Tab2apec} for the background and the source region, respectively. The only background parameters that were left free during the source fits are the column densities $N_\mathrm{H}^\mathrm{CGM}$ and $N_\mathrm{H}^\mathrm{cor}$, since these are expected to vary across the sky.

    We find that the source spectrum can be well fitted with an absorbed model that contains two additive components. For consistency with the studies of \citet{Muno2006} and \citet{Kavanagh2011}, and to investigate the possible presence of nonthermal emission inside Wd~1, we focused on two models: First, a model consisting of two \texttt{apec} components absorbed by a common \texttt{tbabs} component, which we call \texttt{2apec}, and second, a model consisting of an \texttt{apec} and a nonthermal \texttt{powerlaw} component with a common \texttt{tbabs} absorption, which we call \texttt{apec+pl}.

\begin{figure}
        \centering
        \includegraphics[width = \linewidth]{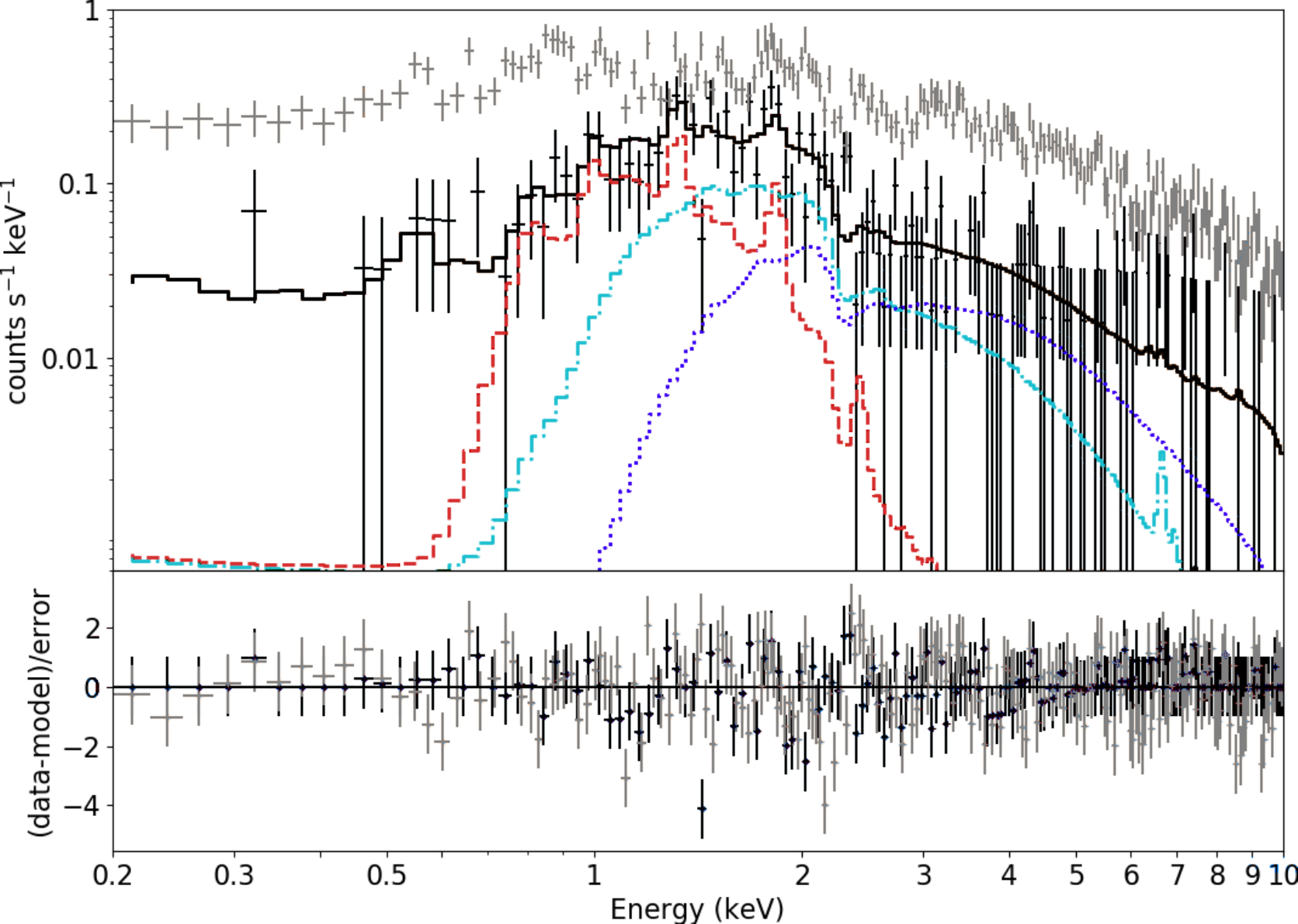}
        \caption{Results of the \texttt{2apec} fit to the spectrum of Wd~1. \textit{Upper panel:} Black data points are the source spectrum, gray data points are the background as measured in the background region. Both are binned to a minimum significance of $5\sigma$. The source model is given in black, while the soft \texttt{apec} component is red and dashed, the harder \texttt{apec} component is light blue and dashed-dotted, and the GX~340$+$0 halo component is dark blue and dotted. \textit{Lower panel:} Normalized residuals of the source spectrum (black) and of the background spectrum (gray).}
        \label{Fig2apec}%
    \end{figure}

\begin{figure}
        \centering
        \includegraphics[width = \linewidth]{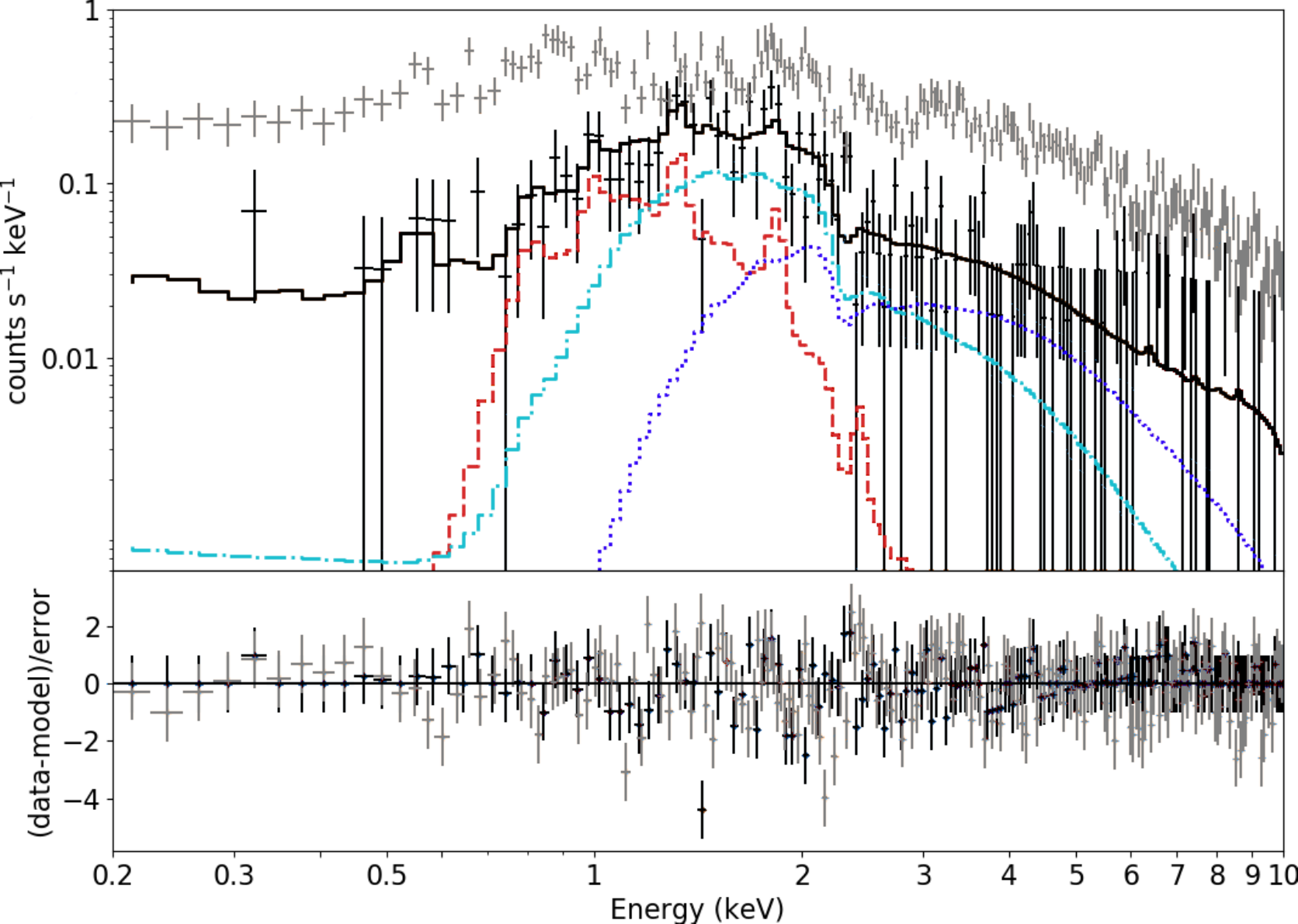}
        \caption{Results of the \texttt{apec+pl} fit to the spectrum of Wd~1. \textit{Upper panel:} Black data points are the source spectrum, gray data points are the background as measured in the background region. Both are binned to a minimum significance of $5\sigma$. The overall model is given in black, while the \texttt{apec} component is red and dashed, the \texttt{powerlaw} component is light blue and dashed-dotted, and the GX~340$+$0 halo component is dark blue and dotted. \textit{Lower panel:} Normalized residuals of the source spectrum (black) and of the background spectrum (gray).}
        \label{FigAPECPL}%
    \end{figure}

    Visually, both models provide equally good descriptions of the data. The \texttt{2apec} fit can be seen in Fig.~\ref{Fig2apec}, while the \texttt{apec+pl} fit is given in Fig.~\ref{FigAPECPL}. The main difference between the two models lies in the \SI{6.7}{\kilo\electronvolt} peak of the \texttt{2apec} model that stems from He-like Fe~XXV in the hotter thermal component. Since the role of this component is taken by the power law in the \texttt{apec+pl} model, it lacks this emission line. Importantly, the detection of this peak allowed \citet{Kavanagh2011} to identify the bulk of the hard emission from Wd~1 as thermal. However, in our case, there are no noteworthy residuals around \SI{6.7}{\kilo\electronvolt} in either of the two models.

    The \texttt{2apec} fit has a reduced Pearson-$\chi^2$ of $\chi^2_\mathrm{P}/\nu = 1743.02/1803 = 0.967$, while the goodness command resulted in 32\% of the simulations having a better fit statistic than the best fit to the data. The best-fit parameters with $1\sigma$ uncertainties can be found in the second block of Table~\ref{Tab2apec} together with the corresponding parameters from \citet{Muno2006} and \citet{Kavanagh2011}. The metal abundances were poorly constrained during the fit, so we fixed them to $Z_1 = 2Z_\sun$ and $Z_2 = 0.62Z_\sun$ as determined by \citet{Muno2006} and \citet{Kavanagh2011}. We also determined the flux in the fitted source components using \texttt{Xspec}'s \texttt{cflux} component. We did this for the energy band from $2$ to \SI{8}{\kilo\electronvolt} to be consistent with \citet{Kavanagh2011}. The total flux per source area in this band in the \texttt{2apec} model is $10_{-5}^{+5}\cdot\SI{e-14}{\erg\per\square\centi\meter\per\second\per\square\arcmin}$.

    \citet{Muno2006} fitted three circular regions of different radii around Wd~1. To allow a comparison to our study, we summed their fit values for normalizations in the three innermost regions, that is, up to a distance of \SI{3.5}{\arcmin} from Wd~1. For temperatures and column densities, we calculated area weighted arithmetic means for these regions. Uncertainties were calculated via Gaussian error propagation. Similarly, \citet{Kavanagh2011} used an analysis region with a radius of \SI{2}{\arcmin}. In Table~\ref{Tab2apec}, this is taken into account by giving all normalizations and fluxes per \SI{1}{\square\arcmin}. For the \texttt{2apec} model, our fit parameters are consistent with the literature values within the $1\sigma$ intervals. The only exceptions are the temperature of the softer component $k_\mathrm{B}T_1$, which is hotter in the case of \citet{Muno2006}, and similarly its normalization $\eta_1$, which is smaller in their study. Also note that \citet{Kavanagh2011} do not state confidence intervals for the integrated flux $F_\mathrm{2-\SI{8}{\kilo\electronvolt}}$, which hinders a proper comparison.
    
    Similarly, the \texttt{apec+pl} model has a reduced Pearson-$\chi^2$ of $\chi^2_\mathrm{P}/\nu = 1754.42/1803 = 0.973$ and a goodness of 29\%. The fit parameters are given in the third block of Table~\ref{Tab2apec}. We set the metal abundance of the thermal component to $Z_\texttt{apec} = 2Z_\sun$ to be consistent with  \citet{Muno2006} and \citet{Kavanagh2011}. The total flux per source area between $2$ and \SI{8}{\kilo\electronvolt} in the \texttt{apec+pl} model is $9_{-2.7}^{+5}\cdot\SI{e-14}{\erg\per\square\centi\meter\per\second\per\square\arcmin}$. Compared to the literature studies, our $k_\mathrm{B}T_\texttt{apec}$ value is relatively small. Otherwise, all parameters of the \texttt{apec+pl} model are in $1\sigma$ agreement. Finally, we note that both column densities $N_\mathrm{H}^\texttt{2apec}$ and $N_\mathrm{H}^\texttt{apec+pl}$ ran into their hard upper limit in our fits. However, we kept them this way to reduce parameter degeneracies in our models.

    \begin{table*}[h]
        \caption{Best-fit parameters of the \texttt{2apec} model (\textit{second block}) and the \texttt{apec+pl} model (\textit{third block}), with corresponding parameters from \citet{Muno2006} and \citet{Kavanagh2011}.}
        \label{Tab2apec}
        \centering
        \begin{tabular}{llll}
            \hline
            \hline
            Parameter & This study & \citet{Muno2006} & \citet{Kavanagh2011}\\
            \hline
            \texttt{BACKSCAL} [\SI{}{\square\arcmin}] & 23.11\\
            \texttt{REGAREA} [\SI{}{\square\arcmin}] & 28.57\\
            $N_\mathrm{H}$ [\SI{e22}{\per\square\centi\meter}] & $2.20$\\
            \specialrule{.125em}{.1em}{.1em}   
            $N_\mathrm{H}^\mathrm{CGM}$ [\SI{e22}{\per\square\centi\meter}] & $0.4_{-0.1}^{+0.1}$\\
            $N_\mathrm{H}^\mathrm{cor}$ [\SI{e22}{\per\square\centi\meter}] & $1.7_{-0.5}^{2.2}$\\
            \hline
            $N_\mathrm{H}^\texttt{2apec}$ [\SI{e22}{\per\square\centi\meter}] & $2.2_{-0.22}^{2.2}$ & $2.2_{-0.2}^{+0.2}$ & $2.0_{-0.2}^{+0.2}$\\
            $k_\mathrm{B}T_1$ [\SI{}{\kilo\electronvolt}] & $0.46_{-0.13}^{+0.11}$ & $0.90_{-0.14}^{+0.14}$ & $0.68_{-0.13}^{+0.12}$\\
            $\eta_1$ [\SI{e-14}{\centi\meter\tothe{-5}\per\square\arcmin}] & $1.4_{-0.6}^{+1.2}\cdot10^{-4}$ & $5.4_{-0.6}^{+1.5}\cdot10^{-5}$\\
            $k_\mathrm{B}T_2$ [\SI{}{\kilo\electronvolt}] & $5.7_{-2.5}^{+10}$ & $9.1_{-1.4}^{+2.8}$ & $3.1_{-0.4}^{+0.6}$\\
            $\eta_2$ [\SI{e-14}{\centi\meter\tothe{-5}\per\square\arcmin}] & $1.1_{-0.3}^{+0.4}\cdot10^{-4}$ & $1.5_{-0.1}^{+0.1}\cdot10^{-4}$\\
            $F_\mathrm{2-\SI{8}{\kilo\electronvolt}}$ [\SI{}{\erg\per\square\centi\meter\per\second\per\square\arcmin}] & $10_{-5}^{+5}\cdot10^{-14}$ & & $14\cdot10^{-14}$\\ 
            \specialrule{.125em}{.1em}{.1em} 
            $N_\mathrm{H}^\mathrm{CGM}$ [\SI{e22}{\per\square\centi\meter}] & $0.43_{-0.09}^{+0.13}$\\
            $N_\mathrm{H}^\mathrm{cor}$ [\SI{e22}{\per\square\centi\meter}] & $1.4_{-0.4}^{2.2}$\\
            \hline
            $N_\mathrm{H}^\texttt{apec+pl}$ [\SI{e22}{\per\square\centi\meter}] & $2.2_{-0.22}^{2.2}$ & $2.4_{-0.2}^{+0.2}$ & $2.1_{-0.3}^{+0.3}$\\
            $k_\mathrm{B}T_\texttt{apec}$ [\SI{}{\kilo\electronvolt}] & $0.43_{-0.12}^{+0.13}$ & $0.96_{-0.08}^{+0.14}$ & $0.81_{-0.09}^{+0.16}$\\
            $\eta_\texttt{apec}$ [\SI{e-14}{\centi\meter\tothe{-5}\per\square\arcmin}] & $1.2_{-0.6}^{+1.4}\cdot10^{-4}$ & $6.2_{-1.2}^{+1.1}\cdot10^{-5}$\\
            $\Gamma$ & $2.1_{-0.6}^{+0.5}$ & $2.1_{-0.1}^{+0.1}$ & $2.4_{-0.2}^{+0.2}$\\
            $\eta_\mathrm{PL}$ [\SI{}{\per\kilo\electronvolt\per\square\centi\meter\per\second\per\square\arcmin}] & $4.8_{-2.0}^{+1.5}\cdot10^{-5}$ & $6.5_{-0.9}^{+1.0}\cdot10^{-4}$\\
            $F_\mathrm{2-\SI{8}{\kilo\electronvolt}}$ [\SI{}{\erg\per\square\centi\meter\per\second\per\square\arcmin}] & $9_{-2.7}^{+5}\cdot10^{-14}$ & & $12\cdot10^{-14}$\\
            \hline
        \end{tabular}
        \tablefoot{The \texttt{BACKSCAL} and \texttt{REGAREA} parameters were taken from the header of the Wd~1 source spectrum fits file, while the $N_\mathrm{H}$ parameter was determined using \texttt{nH}. These three parameters are identical for both models. $F_\mathrm{2-\SI{8}{\kilo\electronvolt}}$ was determined using the \texttt{cflux} component from \texttt{Xspec}. All other parameters are \texttt{Xspec} best-fit parameters with 1$\sigma$ confidence intervals determined via \texttt{Xspec}'s \texttt{error} command. All normalizations and fluxes were divided by the source areas in square arcminutes. Confidence intervals given without a plus or minus sign indicate that a parameter ran into a hard limit before it could sample its full $1\sigma$ region. The values from \citet{Muno2006} and \citet{Kavanagh2011} were determined as described in the main text.}
    \end{table*}
    
\subsection{Search for synchrotron radiation from HESS J1646--458} \label{SecSynchrotron}

\begin{figure}
        \centering
        \includegraphics[width = \linewidth]{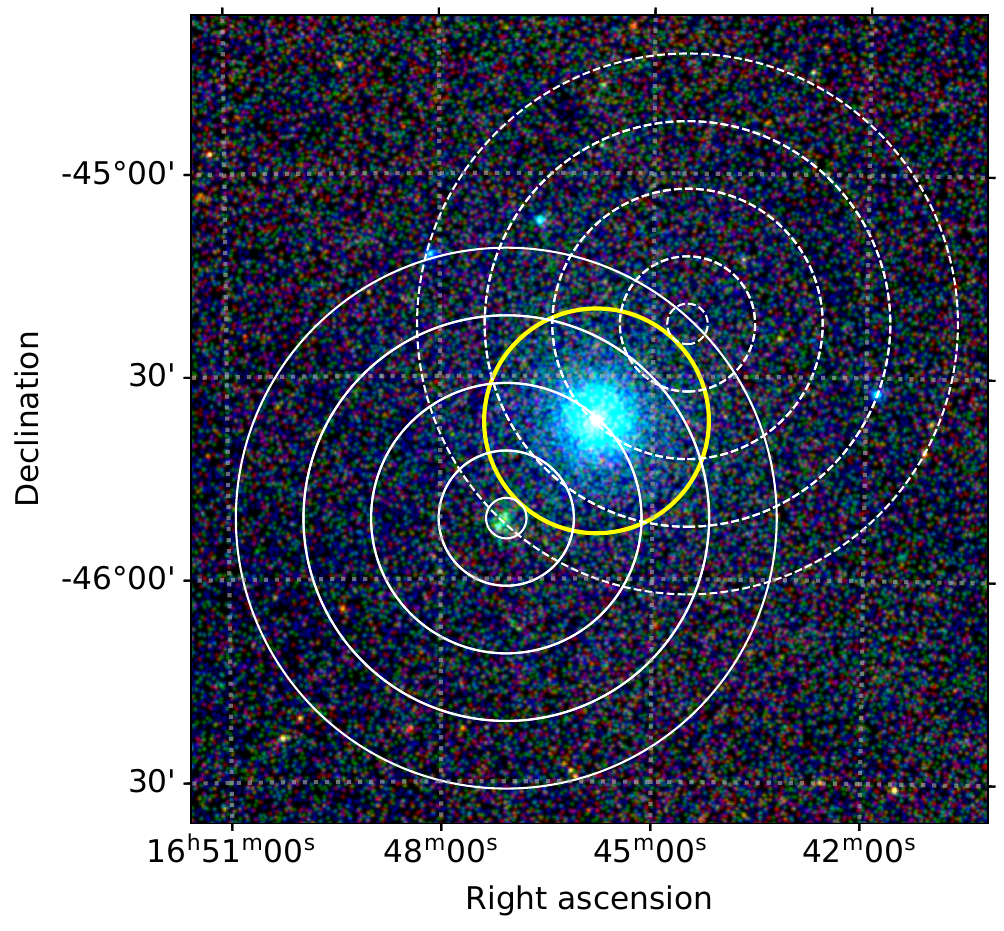}
        \caption{RGB image as in Fig.~\ref{FigWd1Image}. The solid white annuli are the four source regions around Wd~1 used for the study of HESS~J1646$-$458, while the dashed white regions are the corresponding background regions. The yellow circle around GX 340$+$0 was masked for the analysis of the three innermost annulus pairs. For the analysis of the outermost annulus pair, the mask was larger and had a radius of \SI{30}{\arcmin}.}
        \label{FigAnalysisRegions}%
    \end{figure}

    The morphology of HESS~J1646$-$458 roughly describes a circle of radius ${\sim}\SI{30}{\arcmin}$ around Wd~1, as shown in Fig.~\ref{FigTeVContours}. To search for synchrotron radiation, we defined four annulus-shaped source regions around Wd~1 (center $\alpha = 16^\mathrm{h}47^\mathrm{m}04.8^\mathrm{s}$, $\delta = -45\degr{}51'05.1"$). As can be seen in Fig.~\ref{FigAnalysisRegions}, these annuli have radii from 3 to \SI{10}{\arcmin}, 10 to \SI{20}{\arcmin}, 20 to \SI{30}{\arcmin}, and 30 to \SI{40}{\arcmin}. They are labeled as region 1, 2, 3, and 4, respectively.

    In contrast to the analysis of Wd~1, we used the event files with masked point sources for the search for synchrotron radiation. Furthermore, to reduce the contamination in the source regions, we applied a manual mask to GX~340$+$0. This mask is centered on $\alpha = 16^\mathrm{h}45^\mathrm{m}48.0^\mathrm{s}$, $\delta = -45\degr{}36'36.0"$. For regions 1, 2, and 3, its radius is \SI{16.6}{\arcmin}, while for region 4, it is \SI{30.0}{\arcmin}. The mask for the first three regions can be seen in Fig.~\ref{FigAnalysisRegions} in yellow. Since \textit{eROSITA} stray light halos can extend to radii of $3\deg$, we nevertheless expect contamination to be present in the source regions. Therefore, we compared the source region spectra to suitable background regions to check if a significant source component is actually present. We constructed the background regions via mirroring the four source regions on the position of GX~340$+$0. The four resulting background regions are centered on $\alpha = 16^\mathrm{h}44^\mathrm{m}31.2^\mathrm{s}$, $\delta = -45\degr{}22'12.0"$ and are otherwise identical to the source regions. Together with the solid source regions, they can be seen as dashed circles in Fig.~\ref{FigAnalysisRegions}.

\begin{figure}
        \centering
        \includegraphics[width = \linewidth]{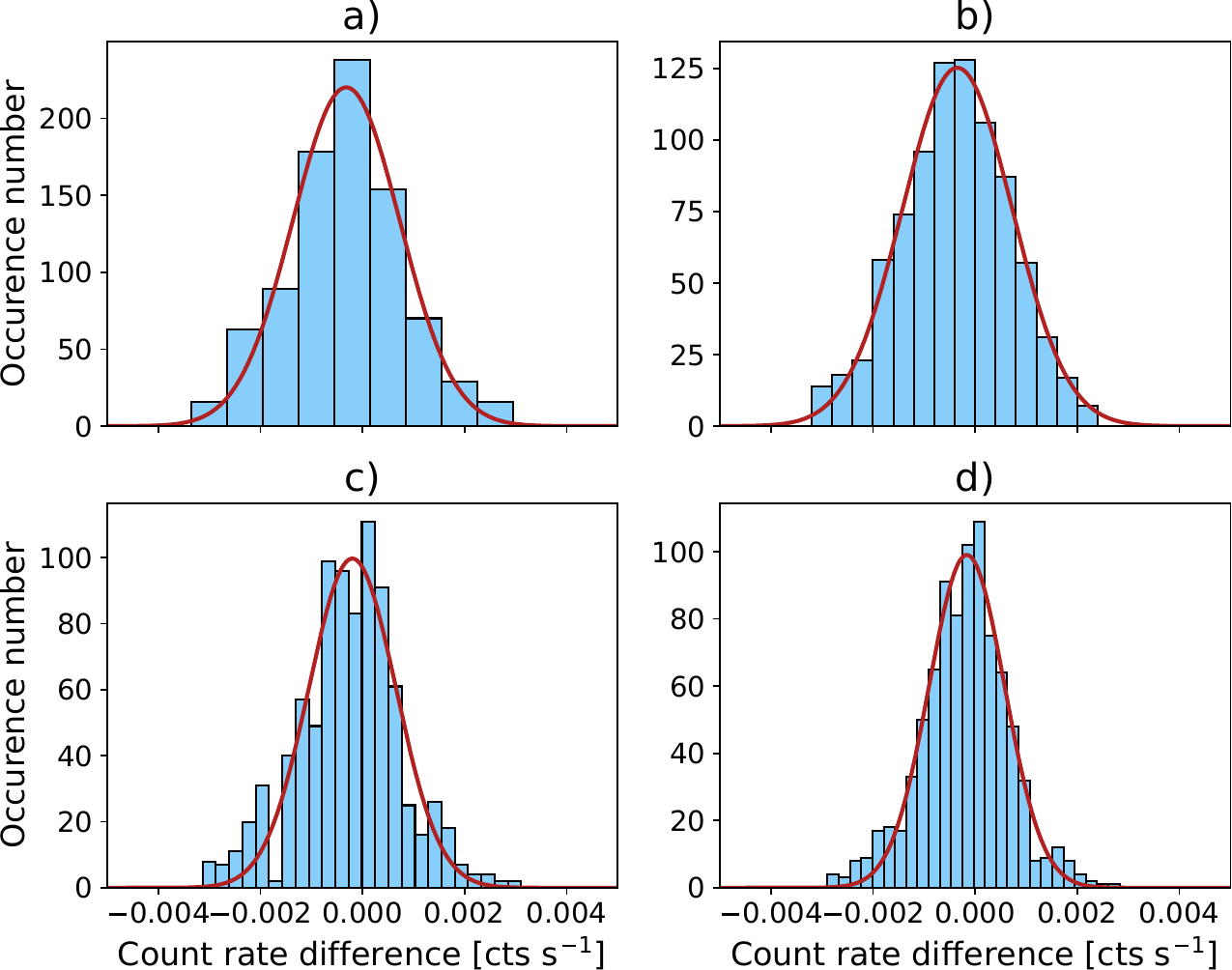}
        \caption{Histograms of the residuals obtained via subtracting background from source spectra in blue with corresponding Gaussian fits in red. Region pairs 1, 2, 3, and 4 correspond to panels a), b), c), and d), respectively. For each region pair, we employed the minimum bin size that resulted in no empty bins}
        \label{FigGaussFits}%
    \end{figure}

    We extracted the spectra of the combined telescope modules for all eight regions and compared the source and background regions pairwise. To this end, we subtracted each background spectrum from the corresponding source spectrum and plotted the residual histograms of the four different region pairs in Fig.~\ref{FigGaussFits}. The count rate differences between source and background regions were binned with the minimum bin size which did not result in any empty bins.

    With mean counts larger than 100, we treated the spectral residuals as Gaussian distributed and fitted them accordingly. The resulting fit parameters for the four region pairs are given in Table~\ref{TabGauss}. In particular, one can see that the mean values $\mu$ of the fitted Gaussian distributions are negative in all cases. This indicates larger measured count rates from the background regions than from the corresponding source regions, which is the opposite of what would be expected from an additional synchrotron source component in the source regions.

    \begin{table*}[h]
        \caption{Gaussian fit parameters and t-test results for the four source region and background region pairs.}
        \label{TabGauss}
        \centering
        \begin{tabular}{clll|lll}
            \hline
            \hline
            Region & $A$ & $\mu$ [\SI{}{\counts\per\second}] & $\sigma$ [\SI{}{\counts\per\second}] & $t$ & $\nu$ & $p$\\
            \hline
            1 & $0.58 \pm 0.04$ & $(-3.2 \pm 0.8)\cdot10^{-4}$ & $(10.5 \pm 0.8)\cdot10^{-4}$ & $-3.41$ & $874$ & $3.40\cdot10^{-4}$ \\
            2 & $0.34 \pm 0.01$ & $(-3.5 \pm 0.3)\cdot10^{-4}$ & $(10.7 \pm 0.3)\cdot10^{-4}$ & $-5.06$ & $874$ & $2.61\cdot10^{-7}$ \\
            3 & $0.21 \pm 0.02$ & $(-2.0 \pm 0.7)\cdot10^{-4}$ & $(9.0 \pm 0.7)\cdot10^{-4}$ & $-3.95$ & $874$ & $4.19\cdot10^{-5}$ \\
            4 & $0.18 \pm 0.01$ & $(-1.6 \pm 0.3)\cdot10^{-4}$ & $(7.4 \pm 0.3)\cdot10^{-4}$ & $-2.34$ & $874$ & $9.65\cdot10^{-3}$\\
            \hline
        \end{tabular}
        \tablefoot{Columns to the left of the vertical line give the fit parameters of the Gaussian fits to the residual histograms of the four source and background region pairs. The parameters are the normalization of the distribution $A$, its mean value $\mu$, and its standard deviation $\sigma$. Columns to the right of the vertical line give the t-statistics $t$, the numbers of degrees of freedom $\nu$, and the p-values $p$ for the t-tests that tested whether the mean value of the underlying difference distributions is at least $0$.}
    \end{table*}

    Following the results of the Gaussian fits, we also performed one-sample t-tests to check for the presence of a source component in the source regions. Such test can be used to determine whether the mean value $\mu$ of a sample significantly deviates from some assumed underlying mean value $\mu_0$. In particular, we chose a one-sided t-test with the null hypothesis $\mu_0 \geq 0$ and the alternative hypothesis $\mu_0 < 0$. Therefore, we tested whether the count rates from the background regions were significantly larger than the count rates from the source regions.

    The resulting t-statistics $t$ can be found in Table~\ref{TabGauss}. It also gives the numbers of degrees of freedom $n -1 $, where $n$ is the number of channels of the analyzed spectra in the employed range between $0.2$ and \SI{9.5}{\kilo\electronvolt}, as well as the resulting p-values $p$. For region pairs 1 and 3, we find $p \lesssim 3\cdot10^{-3}$, that is, a significance of more than $3\sigma$, while for region pair 2, $p \sim 3\cdot10^{-7}$, which corresponds to $5\sigma$. Only for region pair 4 lies the significance between $2$ and $3\sigma$ with $p \sim 10^{-2}$. These low p-values indicate that it would be unlikely to obtain the results measured by us if the flux from any source region was truly at least as large as the flux from the corresponding background region. Based on the negative means of the Gaussian fits, and the results of the t-tests, we conclude that no evidence for an additional X-ray source component from HESS~J1646$-$458 is present in the eRASS:4 data.

    \begin{figure*}
        \centering
        \includegraphics[width = \linewidth]{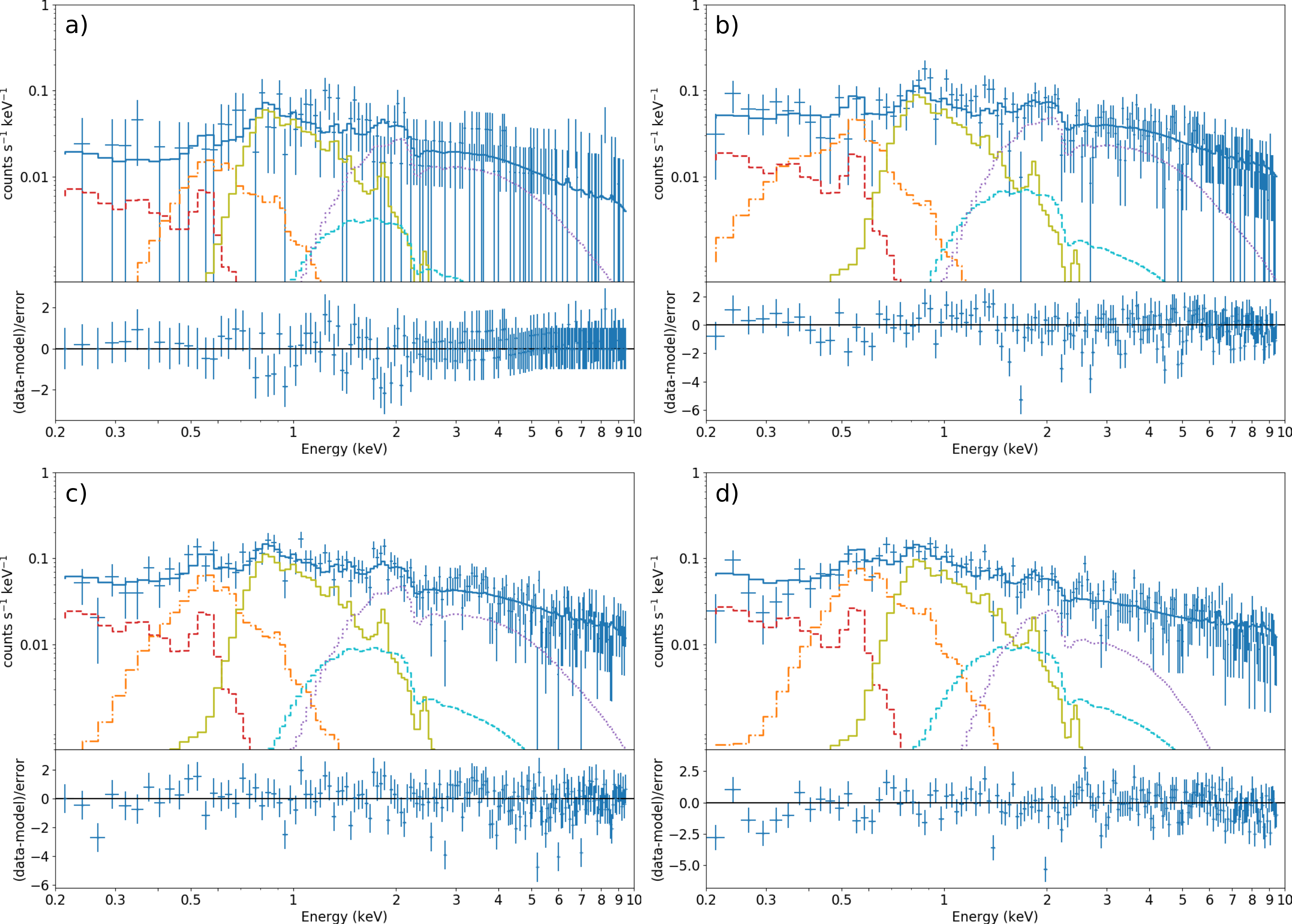}
        \caption{Background fits to the four source regions 1, 2, 3, and 4 in panels a), b), c), and d), respectively. The data are given in blue. \textit{Upper panels:} The main model components are shown using different colors and line styles: The LHB is red and dashed, the CGM is orange and dashed-dotted, the corona is yellow and solid, the CXB is light blue and dashed, and the GX~340$+$0 halo components are purple and dotted. \textit{Lower panels:} Normalized residuals of the fits.}
        \label{FigBackgroundFits}%
    \end{figure*}

\subsection{Synchrotron upper bounds} \label{SecUpperLimits}

    Based on our nondetection of an X-ray source component from HESS~J1646$-$458, we derived upper bounds on the flux of a potential additional synchrotron component. To this end, we fitted the spectra of the four source regions around HESS~J1646$-$458 with pure background models. Following this, we then tested how large a potential additional power law component can get without affecting the fit quality too much. The background spectra for doing this were of the same type as the background for Wd~1 described in Sec. \ref{SecWd1Background}. The only difference is that we fixed $N_\mathrm{H}^\mathrm{LMXB}$ and $k_\mathrm{B}T_\mathrm{in}$ in advance to their best-fit values from Table~\ref{TabWd1Bkg}.

    Since the count numbers in the four HESS~J1646$-$458 source regions are substantially higher than in the Wd~1 source region, we decided to not use the combined data from the different telescope modules, but to fit the spectra of the five modules separately. This allows us to separately model the instrumental effects of the different modules. In particular, they were fitted simultaneously whereby the fit parameters of the five modules were restricted to be identical. The only differences between the five models for the five telescope modules were their overall normalizations: The background model of each telescope module was scaled with a constant factor $w$. These factors were empirically determined by the \textit{eROSITA} consortium and are listed in Table~\ref{TabJ1646Scale} in the appendix. Since we fitted the individual telescope module spectra, we also used the appropriate FWC models for the individual modules. The corresponding \texttt{BACKSCAL} values for this can also be found in Table~\ref{TabJ1646Scale}.

    Again, the $N_\mathrm{H}$ values of the four regions were determined using the \texttt{nH} tool. This was done by determining the $N_\mathrm{H}$ values of five equal-distance positions along the annulus of source region 1, and for seven such positions for each of the other three source regions. We then calculated the arithmetic means and standard deviations of these values to characterize the absorption of the four source regions. As before, fixing the $N_\mathrm{H}$ parameters for the CGM and the Galactic corona resulted in large residuals in the final fit and in strong disagreement with the fit parameters of \citet{Ponti2023}. Therefore, we used the \texttt{nH} values only as upper limits in the cases of $N_\mathrm{H}^\mathrm{CGM}$ and $N_\mathrm{H}^\mathrm{cor}$.

    The background fits to the four source regions are shown in Fig.~\ref{FigBackgroundFits}. The final fit parameters and 90\% confidence intervals for each of the four regions are given in Table~\ref{TabJ1646Bkg}. For region 2, the fit was unstable and tended to very low values for $N_\mathrm{H}^\mathrm{CGM}$. The consequence of this was a strong CGM component that caused a negligibly small normalization of the LHB. Therefore, we fixed $N_\mathrm{H}^\mathrm{CGM}$ of region 2 to \SI{0.4e22}{\per\square\centi\meter}, the best-fit value of region 4, and repeated the fit. Table~\ref{TabJ1646Bkg} also states the reduced Pearson-$\chi^2$-values and the outputs of the goodness command $P$ for the different regions. In general, $P$ lies between $0.49$ and $0.90$, which indicates satisfactory fits. The best-fit parameters of the four regions agree within their 90\% confidence intervals, which is in line with the idea that they contain no additional source component corresponding to the shell-like HESS~J1646$-$458. The only exception from this is $\eta_\mathrm{LMXB}$, which has particularly small relative uncertainties for all regions.

    \begin{table*}[h]
        \caption{Background fit parameters, reduced Pearson-$\chi^2$-values, and goodness of fit values $P$ for the four source regions around HESS~J1646$-$458.}
        \label{TabJ1646Bkg}
        \centering
        \begin{tabular}{cllll}
            \hline
            \hline
            Parameter & Region 1 & Region 2 & Region 3 & Region 4\\
            \hline
            \texttt{REGAREA} $[\SI{}{\square\arcmin}]$ & $203.4$ & $657.2$ & $1217$ & $1488$ \\
            $N_\mathrm{H}$ [\SI{e22}{\per\square\centi\meter}] & $2.0 \pm 0.1$ & $2.0 \pm 0.2$ & $1.9 \pm 0.3$ & $1.8 \pm 0.3$ \\
            $N_\mathrm{H}^\mathrm{Bkg}$ [\SI{e22}{\per\square\centi\meter}] & $1.9 \pm 0.1$ & $1.9 \pm 0.2$ & $1.9 \pm 0.3$ & $1.7 \pm 0.4$ \\
            \hline
            $\eta_\mathrm{LHB}$ [\SI{e-14}{\centi\meter\tothe{-5}\per\square\arcmin}] & $6_{6}^{+5}\cdot10^{-7}$ & $1.0_{-0.3}^{+0.3}\cdot10^{-6}$ & $6.0_{-2.5}^{+2.5}\cdot10^{-7}$ & $8.5_{-2.7}^{+2.2}\cdot10^{-7}$ \\        
            $N_\mathrm{H}^\mathrm{CGM}$ [\SI{e22}{\per\square\centi\meter}] & $0.9_{-0.9}^{+0.4}$ & $0.416$ & $0.3_{-0.2}^{+0.2}$ & $0.4_{-0.3}^{+0.2}$\\
            $\eta_\mathrm{CGM}$ [\SI{e-14}{\centi\meter\tothe{-5}\per\square\arcmin}] & $7_{-7}^{+70}\cdot10^{-4}$ & $5_{-2}^{+2}\cdot10^{-5}$ & $3_{-2}^{+6}\cdot10^{-5}$ & $1.0_{-0.7}^{+2}\cdot10^{-4}$\\
            $N_\mathrm{H}^\mathrm{cor}$ [\SI{e22}{\per\square\centi\meter}] & $1.3_{-0.4}^{2.02}$ & $0.9_{-0.2}^{+0.2}$ & $1.1_{-0.2}^{+0.2}$ & $1.0_{-0.2}^{+0.2}$\\
            $\eta_\mathrm{cor}$ [\SI{e-14}{\centi\meter\tothe{-5}\per\square\arcmin}] & $1.3_{-0.7}^{+1.0}\cdot10^{-5}$ & $6.1_{-1.5}^{+1.8}\cdot10^{-6}$ & $8.2_{-1.4}^{+1.6}\cdot10^{-6}$ & $7.1_{-1.2}^{+1.4}\cdot10^{-6}$\\
            $\eta_\mathrm{LMXB}$ [\SI{}{\per\kilo\electronvolt\per\square\centi\meter\per\second}] & $8.4_{-0.9}^{+1.0}\cdot10^{-4}$ & $2.4_{-0.2}^{+0.2}\cdot10^{-3}$ & $3.4_{-0.2}^{+0.2}\cdot10^{-3}$ & $2.4_{-0.2}^{+0.2}\cdot10^{-3}$\\
            \hline
            $\chi^2_\mathrm{P}/\nu$ & $4485/4349 = 1.03$ & $4369/4350 = 1.00$ & $4467/4349 = 1.03$ & $4456/4349 = 1.02$\\
            $P$ & $0.90$ & $0.49$ & $0.90$ & $0.85$\\
            \hline
        \end{tabular}
        \tablefoot{The \texttt{REGAREA} parameters were taken from the headers of the respective source spectrum fits files. The $N_\mathrm{H}$ parameters are the arithmetic means and standard deviations of several equal-distance column densities determined along the four source regions via \texttt{nH}. For region 1, we used 5 such points, and for the other regions 7. $N_\mathrm{H}^\mathrm{Bkg}$ are the same values for the four corresponding background regions. All other parameters are \texttt{Xspec} best-fit parameters with their respective 90\% confidence intervals. All normalizations except for $\eta_\mathrm{LMXB}$ are scaled with the background area in square arcminutes. Confidence intervals given without a plus or minus indicate a hard limit allowed for the fit into which the parameter ran. For region~2, $N_\mathrm{H}^\mathrm{CGM}$ was fixed as described in the text.}
    \end{table*}

    This demonstrates that the four source regions can be well described by pure background models without the need for an additional source component, for example from synchrotron radiation. This matches the results from the Gaussian fits in the previous section. These background fits form the basis for determining upper confidence bounds on the synchrotron flux from the four source regions. Specifically, we fixed the background models to the parameters from Table~\ref{TabJ1646Bkg}. We then added a \texttt{powerlaw} component to the spectral model of each region and determined the upper 90\% confidence bounds on its normalization. We tried out three different values for the power law index $\Gamma$, namely $1.5$, $2.0$, and $2.5$. 
    
    To be conservative, we also added absorbing \texttt{tbabs} components to the power law components, which resulted in higher upper bounds than without absorption. Since we considered a synchrotron component from the vicinity of Wd~1, we used the same $N_\mathrm{H}$ value as for the star cluster itself. In particular, if the synchrotron radiation came from the stellar wind termination shock or from inside the star cluster, there should be negligible absorbing material between Wd~1 and the synchrotron radiation compared to the kiloparsec-scale column between Earth and Wd~1. Therefore, we adopted $N_\mathrm{H} = \SI{2.2e22}{\per\square\centi\meter}$, consistent with our fits to the Wd~1 source region.

    The resulting upper bounds are given in Table~\ref{TabUpperLimits}. They refer to the flux of the power law components at \SI{1}{\kilo\electronvolt}. For all four regions, the upper bounds monotonically get larger for larger power law indices. The likely reason is that the main constraint on the synchrotron components comes from energies above \SI{1}{\kilo\electronvolt}, because the lower energy parts of the synchrotron components are absorbed by the \texttt{tbabs} component. Indeed, we find that without absorption the situation is reversed and the upper bounds get smaller for larger power law indices. For our further analysis, we then used the upper bounds for $\Gamma = 2.5$, which gives the highest possible flux values. Summing up the limits for the four regions, we obtain \SI{1.9e-3}{\per\kilo\electronvolt\per\square\centi\meter\per\second} as the upper confidence bound on the X-ray synchrotron flux from the combined source regions around Wd~1, corresponding to an area of \SI{3566}{\square\arcmin}.

    \begin{table}[h]
        \caption{Upper 90\% confidence bounds on the synchrotron flux in the four HESS~J1646$-$458 source regions for three choices of the power law index $\Gamma$.}
        \label{TabUpperLimits}
        \centering
        \begin{tabular}{cllll}
            \hline
            \hline
            $\Gamma$ & Region 1 & Region 2 & Region 3 & Region 4\\
            \hline
            1.5 & \SI{2.6e-4}{} & \SI{1.6e-4}{} & \SI{2.9e-4}{} & \SI{3.7e-4}{} \\
            2.0 & \SI{3.6e-4}{} & \SI{2.0e-4}{} & \SI{3.7e-4}{} & \SI{4.9e-4}{} \\
            2.5 & \SI{4.7e-4}{} & \SI{2.6e-4}{} & \SI{4.9e-4}{} & \SI{6.5e-4}{} \\
            \hline
        \end{tabular}
        \tablefoot{The values refer to the flux of the power law components at \SI{1}{\kilo\electronvolt} in \SI{}{\per\kilo\electronvolt\per\square\centi\meter\per\second}.}
    \end{table}
    
\subsection{Spectral energy distribution} \label{SecSED}

    To obtain an upper bound on the magnetic field strength at the cluster wind termination shock of Wd~1, we fitted the spectral energy distribution (SED) of HESS~J1646$-$458. For this, we used the VHE $\gamma$-ray data from \citet{Aharonian2022}, displayed in their Fig.~7, as well as the X-ray upper bound obtained by us. We also added a radio upper limit from \textit{Planck} described in \citet{Aharonian2022} and the HE $\gamma$-ray upper limit from \textit{Fermi}-LAT derived by \citet{Ohm2013}. The radio upper limit takes the value \SI{0.55}{\mega\jansky\per\steradian} at \SI{30}{\giga\hertz} within a radius of \SI{1}{\deg} around Wd~1, while the HE $\gamma$-ray upper limit is \SI{1.4e-11}{\erg\per\square\centi\meter\per\second} at \SI{30}{\giga\electronvolt} within a radius of \SI{1.1}{\deg} around Wd~1. \citet{Ohm2013} did not state this upper limit explicitly, but showed it in their Fig.~2, from which we extracted it. For comparing to the H.E.S.S. data, we rescaled all three values to the source area of \citet{Aharonian2022}, that is, \SI{11664}{\square\arcmin}. Therefore, the adopted values for the X-ray, radio, and HE $\gamma$-ray data are, in energy-flux units, \SI{9.8e-12}{\erg\per\square\centi\meter\per\second}, \SI{1.6e-10}{\erg\per\square\centi\meter\per\second}, and \SI{1.2e-11}{\erg\per\square\centi\meter\per\second}, respectively.

    We fitted an SED to the data using the fitting software \texttt{naima}\footnote{https://naima.readthedocs.io/} \citep{Zabalza2015}. It is based on the \texttt{emcee} package \citep{ForemanMackey2013} and enables Markov chain Monte Carlo fits of underlying electron and proton particle distributions that produce different types of nonthermal radiation. These include IC scattering, synchrotron radiation, and pion decay. Since the main new result of our study is the X-ray synchrotron upper confidence bound and since the production of synchrotron radiation from hadrons is highly suppressed, we performed a leptonic SED fit.

    For the underlying electron distribution, we used a standard exponential cutoff power law,
    \begin{equation}
        \frac{\mathrm{d}N}{\mathrm{d}E_\mathrm{e}} = \Phi_0^\mathrm{e}\left(\frac{E_\mathrm{e}}{\SI{1}{\tera\electronvolt}}\right)^{-\Gamma}\exp{\left(-\frac{E_\mathrm{e}}{E_\mathrm{c}}\right)},
    \end{equation}
    with the normalization $\Phi_0^\mathrm{e}$, the power law index $\Gamma$, and the cutoff energy $E_\mathrm{c}$. Together with the magnetic field strength $B$, these are the fit parameters of our SED fit. Furthermore, at the energies around the X-ray upper bound, the SED fit is influenced by the minimum energy of accelerated electrons $E_\mathrm{min}$. Since we are not aware of tight constraints on this value for Wd~1, we tested the effect of different values between \SI{10}{\mega\electronvolt} and \SI{1}{\giga\electronvolt}.

    On the radiation side, we included synchrotron radiation \citep{Aharonian2010} and IC scattering \citep{Khangulyan2014}  into our fit. For the latter, we used the following target photon fields: the cosmic microwave background, the diffuse Galactic infrared field with temperature \SI{26}{\kelvin} and energy density \SI{0.74}{\electronvolt\per\cubic\centi\meter}, the diffuse optical field with \SI{2400}{\kelvin} and \SI{1.4}{\electronvolt\per\cubic\centi\meter}, and the radiation field from Wd~1 with \SI{40000}{\kelvin} and \SI{30}{\electronvolt\per\cubic\centi\meter} \citep{Aharonian2022}.
  
    We imposed extremely loose priors on the fit parameters, namely $\Phi_0^\mathrm{e} \geq 0$, $-1 \leq \Gamma \leq 5$, $\SI{1}{\tera\electronvolt} \leq E_\mathrm{c} \leq \SI{e5}{\tera\electronvolt}$, and $0 \leq B \leq \SI{100}{\micro\gauss}$. $\Phi_0^\mathrm{e}$ and $E_\mathrm{c}$ were implemented as log-uniform priors. These priors allow for the sampling of a wide range of plausible parameters around the best-fit values of \citet{Aharonian2022}. We used 50 walkers for the MCMC chains and indicated the \citet{Aharonian2022} best-fit parameters as start values, that is, $\Phi_0^\mathrm{e} = \SI{4.7e35}{\per\electronvolt}$, $\Gamma = 2.97$, $E_\mathrm{c} = \SI{180}{\tera\electronvolt}$. For the magnetic field, we used a typical Galactic value of $B = \SI{2}{\micro\gauss}$. We started with a burn in period of 500 iterations, reset the sampler, and ran it for another 1000 steps. Furthermore, we demanded a minimum number of steps at least 10 times as large as the autocorrelation times of the chains. We checked the integrated autocorrelation times for the four parameters with the built in \texttt{emcee} command \texttt{autocorr.integrated\_time}. For our final fits, this resulted in values between 52 and 72 iterations. Finally, we checked the acceptance fractions of the steps and found them to be close to 0.4 for all fits. We conclude that our fits converged.

    In Table~\ref{TabSED}, we show the best-fit parameters for the $E_\mathrm{min}$ values \SI{100}{\mega\electronvolt}, \SI{500}{\mega\electronvolt}, and \SI{1}{\giga\electronvolt}. The first three parameters, that is, $\Phi_0^\mathrm{e}$, $\Gamma$, and $E_\mathrm{c}$, are in good agreement with each other and with the values of \citet{Aharonian2022} for different choices of $E_\mathrm{min}$. This does not come as a surprise, since these parameters are mainly constrained by the VHE $\gamma$-ray data of these authors. Also, pay attention to the strong uncertainty on $E_\mathrm{c}$, which is in line with the work of these authors. This indicates that the presence of a $\gamma$-ray energy cutoff in HESS~J1646$-$458's spectrum cannot be firmly established yet.

    \begin{table}[h]
        \caption{Spectral energy distribution fit results for different assumed minimum electron energies.}
        \label{TabSED}
        \centering
        \begin{tabular*}{\linewidth}{@{\extracolsep{\fill}} cllll }
            \hline
            \hline
            $E_\mathrm{min}$ [\SI{}{\mega\electronvolt}] & $\Phi_0^\mathrm{e}$ [\SI{e35}{\per\electronvolt}] & $\Gamma$ & $E_\mathrm{c}$ [\SI{}{\tera\electronvolt}] & $B$ [\SI{}{\micro\gauss}]\\
            \hline
            100 & $4.5_{-0.6}^{+0.7}$ & $2.9_{-0.1}^{+0.1}$ & $160_{-60}^{+400}$ & $6_{-5}^{+13}$ \\
            500 & $4.8_{-0.6}^{+0.6}$ & $3.0_{-0.1}^{+0.1}$ & $240_{-110}^{+700}$ & $2.6_{-1.8}^{+4}$ \\
            1000 & $4.8_{-0.6}^{+0.6}$ & $3.0_{-0.1}^{+0.1}$ & $240_{-110}^{+600}$ & $2.9_{-2.0}^{+4}$\\
            \hline
        \end{tabular*}
    \end{table}
    
    The main result of our study is the inclusion of X-ray data, which allow us to constrain the magnetic field strength around Wd~1. From the posterior distributions of $B$, we find that its value is consistent with 0 in all cases. However, the value of the upper confidence bound on this parameter is affected by the value of $E_\mathrm{min}$. For $E_\mathrm{min} = \SI{500}{\mega\electronvolt}$, we have a relatively tight $1\sigma$ constraint of $B \lesssim \SI{6.6}{\micro\gauss}$, as can be seen in Table~\ref{TabSED}. We show the SED fit for this value of $E_\mathrm{min}$ in Fig.~\ref{FigSED}, while the fits for the two other values are given in Fig.~\ref{FigSEDDouble} in the appendix. For larger values of $E_\mathrm{min}$ (e.g., \SI{1}{\giga\electronvolt}) the upper bound on $B$ remains basically the same. For lower values of $E_\mathrm{min}$ however, the constraint on $B$ gets looser. For example, for $E_\mathrm{min} = \SI{100}{\mega\electronvolt}$, we have only $B \leq \SI{19}{\micro\gauss}$ at a $1\sigma$ confidence.
    
    \begin{figure}
        \centering
        \includegraphics[width = \linewidth]{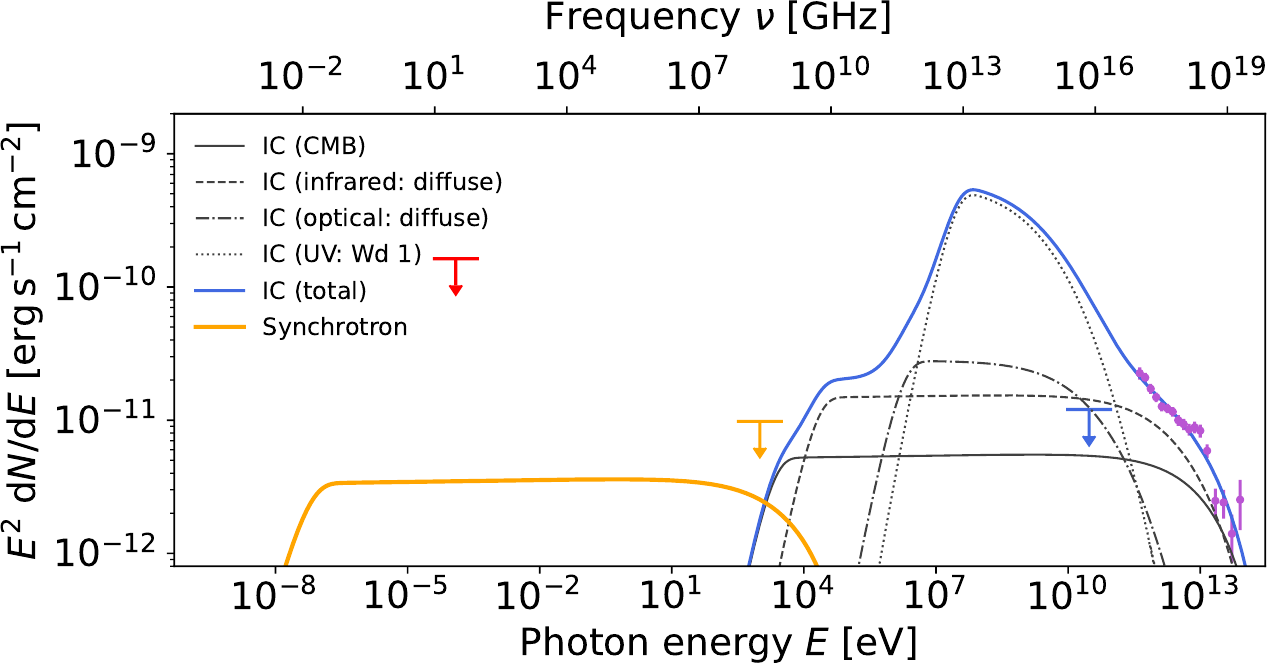}
        \caption{Spectral energy distribution fit to HESS~J1646$-$458, assuming $E_\mathrm{min} = \SI{500}{\mega\electronvolt}$. The data are, from left to right: the \textit{Planck} radio upper limit in red, the \textit{eROSITA} X-ray upper confidence bound in orange, the \textit{Fermi}-LAT HE $\gamma$-ray upper limit in blue, and the H.E.S.S. VHE $\gamma$-ray data from \citet{Aharonian2022} in purple. The solid orange line is synchrotron radiation and the solid blue line is IC scattering. The different target photon contributions to the IC scattering are given by the thin black lines. The CMB is solid, diffuse Galactic infrared radiation dashed, diffuse Galactic optical light dashed-dotted, and Wd~1's photon field dotted.}
        \label{FigSED}%
    \end{figure}

\section{Discussion} \label{SecDiscussion}

\subsection{Background fits}

    Even though our study is not aimed at the analysis of the \textit{eROSITA} background, it is still worthwhile to briefly discuss our background fits. First, our background fits tended to negligible SWCX contributions. This is in contrast to \citet{Ponti2023}, who find negligible SWCX for eRASS1 and 2, but somewhat increased values for eRASS 3 and 4. However, \citet{Yeung2023} also find the SWCX intensity to increase with ecliptic latitude. The lower ecliptic latitude of Wd~1 of $\sim\SI{-23}{\deg}$ compared to the eFEDS's $\sim\SI{-15}{\deg}$ might then potentially explain the subdominant SWCX component in our analysis. For a more detailed discussion of SWCX in \textit{eROSITA}, we refer to Dennerl et al. (in prep.).

    In general, our background fit parameters are consistent within their 90\% confidence intervals between the different regions. This is the case for the four regions around Wd~1 given in Table~\ref{TabJ1646Bkg} and for the background for Wd~1 given in Table~\ref{TabWd1Bkg}. The only exceptions are the GX~340$+$0 halo normalizations $\eta_\mathrm{LMXB}$. This is to be expected, since the contaminating halo's flux should decline with the distance from Wd~1. Indeed, while the absolute values of $\eta_\mathrm{LMXB}$ increase for regions further outward, this trend is reversed when normalizing to the same source area.

    However, despite the self consistency of our fit parameters, they are different from the ones in \citet{Ponti2023}. These authors find that the Galactic corona component has the smallest normalization, the LHB is roughly an order of magnitude stronger, and the CGM is yet another order of magnitude stronger. In our fits, we find that the LHB is smallest, the corona is roughly an order of magnitude stronger, and the CGM is roughly another order of magnitude stronger. Again, this difference might be attributable to the difference in source regions between the two studies: The eFEDS region lies away from the Galactic plane, while Wd~1 lies directly in the Galactic plane. Therefore, we might expect more Galactic emission along the line of sight in our fits. This could either stem from diffuse gas or unresolved point sources. In particular, M dwarfs have been suggested as an additional contribution to the Galactic X-ray background with a strong dependence on Galactic latitude \citep{Masui2009, Wulf2019}.

    In line with this explanation, we find that acceptable fits require $N_\mathrm{H}^\mathrm{CGM}$ and $N_\mathrm{H}^\mathrm{cor}$ values below the full line of sight values obtained via the \texttt{nH} tool. Due to the larger extent of the Galactic disk along our line of sight, we expect Galactic neutral hydrogen and dust to also extend over a longer line of sight fraction. In particular, this might cause some of the emission regions contributing to our spectra to lie in front of some of the absorbing material, therefore resulting in lower $N_\mathrm{H}$ values.

\subsection{Diffuse emission from Westerlund 1}

    Regarding the fits to Wd~1 itself, the goodness of the \texttt{2apec} model is 32\% and the goodness of the \texttt{apec+pl} model is 29\%. This indicates that the fits are similar in quality. Looking at the plots in Fig.~\ref{Fig2apec} and \ref{FigAPECPL}, we note that the two fits are very similar visually, too. Most importantly, there are no notable residuals left around \SI{6.7}{\kilo\electronvolt} in either of the fits. This is the energy of the Fe~XXV emission line expected for a hot plasma, as can be seen in the \texttt{2apec} model in Fig.~\ref{Fig2apec}.

    The detection of this emission line allowed \citet{Kavanagh2011} the identification of the bulk of the emission from the inner \SI{2}{\arcmin} of Wd~1 as thermal hot plasma emission. Since we do not see any significant residuals at the respective energy, we conclude that the eRASS:4 data are insufficient to confirm or refute the presence of the Fe~XXV emission line within \SI{3}{\arcmin} around Wd~1. This is mainly driven by the poor statistics of the survey-type \textit{eROSITA} observation of Wd~1 and by the lower spectral resolution compared to \textit{XMM-Newton}. An independent confirmation of the thermal nature of the emission from Wd~1 would likely require pointed observations with higher spectral and spatial resolution, for example with the \textit{Athena} X-ray Integral Field Unit (X-IFU, \citeauthor{Barret2023} \citeyear{Barret2023}). In addition, the \textit{NuSTAR} telescope might help to confirm or refute the presence of nonthermal radiation due to its large sensitivity at high X-ray energies.
    For the time being, we base our following discussion of potential source components on the result by \citet{Kavanagh2011} that the bulk of the emission from Wd~1 is thermal in nature. From the fitted value for $F_\mathrm{2-\SI{8}{\kilo\electronvolt}}$ from the \texttt{2apec} model and the assumed distance to Wd~1 of \SI{3.9}{\kilo\parsec}, we calculate a luminosity of Wd~1 of $L_\mathrm{2-\SI{8}{\kilo\electronvolt}} = 5.2_{-2.6}^{+2.6}\cdot10^{33}\,\SI{}{\erg\per\second}$.

    In the case of \citet{Muno2006}, the luminosity is $3_{-1}^{+1}\cdot10^{34}\,\SI{}{\erg\per\second}$. However, they employed a larger source radius of \SI{5}{\arcmin} and a larger distance of \SI{5}{\kilo\parsec}. On the other hand, \citet{Kavanagh2011} find $1.7\cdot10^{33}\SI{}{\erg\per\second}$ within a radius of \SI{2}{\arcmin}, assuming a distance of \SI{3.5}{\kilo\parsec}. Finally, it is important to keep in mind that the authors of these studies masked many point sources inside of Wd~1. In our study, this was only done for the magnetar CXOU~J164710.2-455216, which is the only point source inside Wd~1 resolved by \textit{eROSITA}.

\subsubsection{Pre-main sequence stars}

    \citet{Clark2008} detected 45 candidate pre-main sequence (PMS) stars within \SI{5}{\arcmin} of the cluster core of Wd~1. These stars are X-ray sources since they produce magnetic reconnection events that heat the surrounding plasma. Based on comparisons to the Orion Nebular Cluster's PMS population, \citet{Clark2008} estimated that the PMS population of Wd~1 could be larger than 36,000 objects. In the following, we base our own estimation of the unresolved PMS flux from Wd~1 on the analysis by \citet{Kavanagh2011}.

    These authors used data from the Chandra Orion Ultradeep Project \citep{Getman2005}, which identified point sources in the Orion Nebular Cluster. Assuming that the initial mass function of this cluster is comparable to the initial mass function of Wd~1, \citet{Kavanagh2011} then used the age, distance, and absorbing column density of Wd~1 to rescale the PMS flux distribution of the Orion cluster. This way, the arrived at an unresolved PMS luminosity of ${\sim}\SI{1.3e33}{\erg\per\second}$ between $2$ and \SI{8}{\kilo\electronvolt}.

    For their analysis, \citet{Kavanagh2011} masked resolved PMS stars inside Wd~1. We re-added the luminosities of these PMS stars, which are given in their Table~1, to their diffuse PMS luminosity estimate. This way, we arrive at an estimated PMS luminosity of $\SI{1.7e33}{\erg\per\second}$. Next, we corrected this luminosity for the differing adopted distances to Wd~1 in their study and ours: These authors used \SI{3.55}{\kilo\parsec}, while we used \SI{3.9}{\kilo\parsec}. This way, we arrive at an estimated contribution of ${\sim}\SI{2e33}{\erg\per\second}$ from unresolved PMS stars to the diffuse Wd~1 X-ray luminosity between $2$ and \SI{8}{\kilo\electronvolt}. Based on this, unresolved PMS stars might be able to explain around 40\% of the diffuse X-ray luminosity of Wd~1.

    This is more than two times larger than the fraction of 15\% of Wd~1's surface brightness obtained by \citet{Kavanagh2011}. We find that this difference can be explained by two factors: First, there are the excluded resolved PMS stars in the study of \citet{Kavanagh2011} that we re-added to our estimate. Second, as can be seen in Table~\ref{Tab2apec}, the surface brightness of Wd~1 is larger by a factor of around $1.4$ in the study of \citet{Kavanagh2011} compared to our study. Most probably, this comes from the fact that these authors only studied the innermost \SI{2}{\arcmin} of the star cluster, where the surface brightness is higher. However, \citet{Kavanagh2011} still estimated the PMS flux for a radius of \SI{3}{\arcmin} around Wd~1. Therefore, they rescaled their PMS estimate to the smaller source radius of \SI{2}{\arcmin}, which results in a lower PMS-to-measured luminosity fraction. No such rescaling was performed in our case, since the two areas already coincided for our study.

\subsubsection{Stellar winds}

    Wd~1 is often studied for its rich stellar population, which in particular harbors at least 24 WR stars. These stars are characterized by strong stellar winds that are expected to collide and heat up the medium inside the star cluster, therefore resulting in the production of an X-ray emitting gas. These stellar winds are particularly interesting since they are also candidate sites for the acceleration of CRs.

    As in the cases of \citet{Muno2006} and \citet{Kavanagh2011}, we estimate the stellar wind properties inside Wd~1 based on the analytic cluster wind model of \citet{Canto2000}. We assume monoatomic hydrogen gas and consequentially use the solution for a supersonic gas with an adiabatic index of $5/3$. Following this, the density and the temperature of the thermalized stellar wind shocks can be estimated as
    \begin{equation}
        n = 0.19N\left(\frac{\dot{M}}{10^{-5}M_\odot\SI{}{\per\year}}\right)\left(\frac{v_\mathrm{w}}{\SI{1000}{\kilo\meter\per\second}}\right)^{-1}\left(\frac{R_\mathrm{c}}{\SI{1}{\parsec}}\right)^{-2}\SI{}{\per\cubic\centi\meter}
    \end{equation}
    and
    \begin{equation}
        k_\mathrm{B}T = 1.3\left(\frac{v_\mathrm{w}}{\SI{1000}{\kilo\meter\per\second}}\right)^{2}\SI{}{\kilo\electronvolt},
    \end{equation}
    with $N$ the number of WR stars, $\dot{M}$ their mean mass loss rate, $v_\mathrm{w}$ their mean wind speed, and $R_\mathrm{c}$ the spherical cluster radius.
        
    We assume 24 WR stars \citep{Clark2020} with mean mass loss rates of $1.4\cdot10^{-5}M_\sun\SI{}{\per\year}$ and wind speeds of \SI{1320}{\kilo\meter\per\second}, as estimated by \citet{Kavanagh2011} based on the spectral types of WR stars in Wd~1's core \citep{Crowther2006}. Furthermore, at a distance of \SI{3.9}{\kilo\parsec}, \SI{3}{\arcmin} correspond to a cluster radius of $R_\mathrm{c} = \SI{3.4}{\parsec}$. This results in a wind density of \SI{0.4}{\per\cubic\centi\meter} and a temperature of \SI{2.3}{\kilo\electronvolt}. This estimated temperature lies within the lower $2\sigma$ bound of the hotter temperature of our \texttt{2apec} fit, $5.7_{-2.5}^{+10}\,\SI{}{\kilo\electronvolt}$. Given the roughness of the estimate, the agreement seems reasonable.

    Next, we predict an X-ray flux from the calculated wind density $n$. To do this, we first calculate the emission measure $K_\mathrm{EM} = (4/3)\pi R_\mathrm{c}^3 n^2 = \SI{7.8e56}{\per\cubic\centi\meter}$. With the distance $d = \SI{3.9}{\kilo\parsec}$, this results in the \texttt{apec} normalization $\eta_\mathrm{wind} = 10^{-14}K_\mathrm{EM}/(4\pi d^2) = \SI{0.004}{\per\centi\meter\tothe{5}}$. Based on this normalization and the wind temperature, we defined an \texttt{apec} model in \texttt{Xspec} and calculated its flux with \texttt{Xspec}'s \texttt{flux} command. This way, we obtain \SI{2.0e-12}{\erg\per\square\centi\meter\per\second} between $2$ and \SI{8}{\kilo\electronvolt}, corresponding to a luminosity of \SI{3.7e33}{\erg\per\second}. This amounts to ${\sim}70\%$ of the measured X-ray luminosity from Wd~1. However, we note the large uncertainties on the luminosity, which means that the true value could well be anywhere between 50\% and 100\%. Nevertheless, we deem it plausible that stellar winds strongly contribute to Wd~1's X-ray flux and that they might even be responsible for the bulk of the X-ray emission.

    However, the estimate gets even more uncertain when one considers the dependency of the wind luminosity prediction on the initial parameters. For example, \citet{Muno2006} used a WR mean mass loss rate of $6\cdot10^{-5}M_\sun\SI{}{\per\year}$ and a mean wind speed of \SI{1700}{\kilo\meter\per\second}, which gives a wind density of $n = \SI{1.4}{\per\cubic\centi\meter}$. The \texttt{apec} normalization depends quadratically on this quantity, and consequentially, these authors predict a wind luminosity 15 times larger than our estimate.

\subsubsection{Supernova remnants}

    Apart from the presence of the magnetar CXOU~164710.2$-$455216, there is no direct evidence of past supernovae inside Wd~1. Still, given the high mass stellar content of the cluster and its age, several supernovae should have already exploded inside it. For example, \citet{Muno2006} estimated an average supernova rate inside Wd~1 of one per $7000-\SI{13000}{\year}$. With a standard supernova energy of \SI{e51}{\erg}, this results in a power output between $2.5$ and \SI{4.9e39}{\erg\per\second}. On energetic grounds, SNRs could therefore account for the observed diffuse X-ray luminosity of Wd~1.

    However, this estimated power is only a time averaged quantity. In particular, SNR ejecta propagating with a typical speed of \SI{1000}{\kilo\meter\per\second} should leave Wd~1 within ${\sim}\SI{1000}{\year}$ \citep{Vink2012, Pedlar1999}. Based on this statistical argument, Wd~1 should be free of SNRs most of the time. In addition, the absence of direct evidence for SNRs associated with Wd~1 renders a supernova explanation of the diffuse X-ray luminosity implausible. Quite possibly, the stellar winds from the WR stars inside the cluster have cleared the nearby ISM to such a degree that potential SNRs cannot produce strong X-ray emission anymore.

\subsection{Magnetic field strength and very-high-energy emission scenario}

    The main result of this study is the nondetection of nonthermal X-ray emission from the PeVatron candidate HESS~J1646$-$458. This is based on the negative mean values of the Gaussian fits to the source area residuals and on the t-test results, as given in Table~\ref{TabGauss}. In particular, the t-tests rejected the hypothesis of a count rate from the source regions at least as large as the one from the background regions.
    
    The fact that the count rates from the background regions are larger than those from the source regions might be explained by the following two facts: First, the absorbing column densities in the background regions, as estimated with the \texttt{nH} tool, are systematically smaller than in the source regions at the 1\% level. This can be seen in Table~\ref{TabJ1646Bkg}. Second, from the headers of the respective spectral fits files, we found that the exposure times in the background regions are systematically larger than in the source regions at the $0.1$ to $1\%$ level. 
    
    Based on background fits to the four source regions, we then derived an upper confidence bound on the X-ray synchrotron flux from HESS~J1646$-$458. From this, we obtained am upper $1\sigma$ confidence bound on the average magnetic field strength from the four source regions, which depends on the assumed minimum energy $E_\mathrm{min}$ of accelerated electrons: For \SI{500}{\mega\electronvolt} and larger values, we have ${\sim}\SI{6.6}{\micro\gauss}$, while for \SI{1}{\giga\electronvolt}, we have only \SI{19}{\micro\gauss}, as shown in Table~\ref{TabSED}. The likely reason for this is the position of the IC lower energy cutoff determined by $E_\mathrm{min}$. For energies above \SI{500}{\mega\electronvolt}, the IC component is cut off at large enough energies to not have an effect on the X-ray upper bound, as can be seen in Fig.~\ref{FigSED} and Fig.~\ref{FigSEDDouble}~b). For lower values of $E_\mathrm{min}$ on the other hand, the IC component starts to contribute to the X-ray upper bound until it dominates it at $E_\mathrm{min} = \SI{100}{\mega\electronvolt}$, as can be seen in Fig.~\ref{FigSEDDouble}~a). For even lower values of $E_\mathrm{min}$, the IC component violates this upper bound.
    
    As a consequence, the likelihood of the \texttt{naima} fit gets penalized already for the IC component, which reduces the importance of the additional penalty on the synchrotron component for violating the X-ray upper bound. Therefore, the upper bound of $B$ gets looser. In fact, the upper $1\sigma$ confidence bound of \SI{19}{\micro\gauss} for the $E_\mathrm{min} = \SI{100}{\mega\electronvolt}$ fit matches well with the radio upper limit, which seems to be the new main constraint on $B$ in this case. Since this effect is owed to \texttt{naima}'s implementation of upper limits, and since an IC contribution to the X-ray upper bound could in reality only reduce the room for additional synchrotron radiation, we argue that our upper confidence bound of $B \leq \SI{6.6}{\micro\gauss}$ is valid irrespective of the value of $E_\mathrm{min}$. Importantly, this estimate applies to a purely leptonic $\gamma$-ray emission scenario. For a larger hadronic contribution, the upper bound becomes even looser. The most likely emission site in a leptonic scenario is the cluster wind termination shock, based on the morphology of the VHE $\gamma$-ray source region \citep{Aharonian2022}. 

    In principle, it would be interesting to trace a potential variability in the $B$ field upper bound with the radial distance from Wd~1. However, the SED fits necessary for this would require VHE $\gamma$-ray spectra for different annuli around Wd~1, which are not available. Still, it is possible to estimate a rough radial trend from the power law fit results for $\Gamma = 2.5$ in Table~\ref{TabUpperLimits}. When we normalize the derived fluxes with the source region areas, we get ${\sim}\SI{0.02}{\per\kilo\electronvolt\per\square\centi\meter\per\second\per\square\arcmin}$ for region~1 and ${\sim}\SI{0.004}{\per\kilo\electronvolt\per\square\centi\meter\per\second\per\square\arcmin}$ for the other three regions. This speaks against a pronounced radial dependence of the $B$-field apart from the already established shell-like morphology of the HESS~J1646$-$458 VHE flux. The higher flux bound in region~1 might stem from the stronger CGM and corona background components in this region, which leave more room for additional synchrotron radiation. These two components, however, have large uncertainties, as can be seen in Table~\ref{TabJ1646Bkg}.
    
    Returning to the average magnetic field strength at the Wd~1 cluster wind termination shock, this is estimated to be between \SI{0.7}{\micro\gauss} and \SI{4.5}{\micro\gauss} in a purely leptonic emission scenario \citep{Haerer2023}. The lower bound is obtained from the Hillas criterion \citep{Hillas2005} and the upper bound stems from the requirement of a shock with an Alfvénic Mach number larger than $1$. Our upper bound of $B \lesssim \SI{6.6}{\micro\gauss}$ is in line with this estimate. In particular, the upper bound on the magnetic field by \citet{Haerer2023} is compatible with our nondetection of synchrotron radiation from HESS~J1646$-$458. However, we note that while the upper bound of these authors stems from a theoretical argument, the limit derived by us is more empirical in nature and fundamentally based on the \textit{eROSITA} nondetection of a synchrotron source component.

    Furthermore, an older estimate of $B$ comes from the \textit{Planck} radio upper limit employed by \citet{Aharonian2022}. Based on an average flux within \SI{1}{\deg} around Wd~1 of \SI{0.55}{\mega\jansky\per\steradian}, these authors find an upper limit of \SI{10}{\micro\gauss}, that is, somewhat looser than our own estimate. However, for this estimate, \citet{Aharonian2022} assume that only ${\sim}\SI{0.3}{\mega\jansky\per\steradian}$, that is, half of the measured flux, actually stems from HESS~J1646$-$458. As a test of consistency, we checked and found that we can reproduce this flux with our best-fit SED parameters and $B = \SI{10}{\micro\gauss}$.

    Around \SI{30}{\giga\hertz}, a strong contribution from synchrotron radiation produced by electrons that diffuse through the Galaxy is expected \citep{Martire2022}. A continuous spectrum that extends from the \textit{Planck} upper limit to our X-ray upper bound then requires a relatively steep decrease in the Galactic synchrotron power between \SI{30}{\giga\hertz} and \SI{1}{\kilo\electronvolt}.  We find that a power law index of the underlying electron distribution at least as steep as ${\sim} 3.3$ is necessary to accomplish this. Indeed, this is in good agreement with measurements of the CR electron power law index between $10$ and \SI{100}{\giga\electronvolt}, which is found to be ${\sim}3.2$ by the Alpha Magnetic Spectrometer on the International Space Station \citep{Aguilar2019}. In particular, \SI{10}{\giga\electronvolt} electrons should be responsible for synchrotron radiation around \SI{30}{\giga\hertz}.

    Finally, regarding the likely $\gamma$-ray emission scenario in HESS~J1646$-$458, we note the following: In a hadronic scenario, accommodating the upper bound on the magnetic field is trivial since the synchrotron flux is suppressed by 13 orders of magnitude based on the proton-to-electron mass ratio. For energetic and morphological reasons, \citep{Haerer2023} find that such a hadronic scenario is unrealistic if particles are accelerated at the cluster wind termination shock. A leptonic scenario at the cluster wind termination shock is also considered in length by these authors and is found to provide a plausible explanation of the spectrum and the morphology of the source. Since our estimate of the magnetic field strength at the cluster wind termination shock is in line with theirs, we note that their analysis still holds. In summary, our X-ray analysis does not offer strong evidence for or against a leptonic emission scenario at the cluster termination shock or for or against a hadronic emission scenario inside Wd~1 itself. Importantly, the real scenario might also be a hybrid one with both leptonic and hadronic contributions.

\subsection{High-energy upper limit}

    As can be seen from Fig.~\ref{FigSED}, our best-fit SED violates the \textit{Fermi}-LAT HE $\gamma$-ray upper flux limit by a factor of about $6$. This is mostly driven by the contribution of the Wd~1 photon field to the IC scattering. We find that varying the temperature of this photon field within the parameter range from $10000$ to \SI{50000}{\kelvin} estimated by \citet{Haerer2023} does not have a strong effect on the discrepancy. Indeed, the contribution from the diffuse Galactic infrared radiation is discrepant, too, but at a smaller level. Most importantly, the \textit{Fermi}-LAT upper limit even lies well below the fluxes that might be expected from a smooth lower-energies continuation of the H.E.S.S. data.

    We see two main possibilities how the conflict between the data and the model might be alleviated: First of all, we note that the upper limit from \citet{Ohm2013} is model dependent. In particular, it assumes a power law index of $2$ of the $\gamma$-ray emission. This is clearly not the case for our own fit due to the steep contribution of the Wd~1 photon field. If the data points that constrain the \citet{Ohm2013} upper limit lie mainly at energies above \SI{30}{\giga\electronvolt}, a softer spectrum with a larger power law index would lead to a higher HE $\gamma$-ray upper limit. If this was really the case, agreement might be almost restored. This can be seen from the relatively small discrepancy between the HE $\gamma$-ray upper limit and the infrared contribution, which has a power slope of ${\sim}2$ around \SI{30}{\giga\electronvolt}.

    Second, our results assume a purely leptonic emission scenario. Hadronic emission from pion decay, on the other hand, would lack the steep Wd~1 photon field feature that violates the \textit{Fermi} upper limit. If there was also a hadronic contribution to the $\gamma$-ray emission from HESS~J1646$-$458, the disagreement with the data might be reduced.

    Nevertheless, this second option alone would not address the issue of how to connect the H.E.S.S. data and the \textit{Fermi}-LAT data. Indeed, one can already see the root of the problem by comparing the lowest energy H.E.S.S. data point in Fig.~7 of \citet{Aharonian2022}, which has a flux of $\SI{1.4e-11}{\tera\electronvolt\per\square\centi\meter\per\second} = \SI{2.3e-11}{\erg\per\square\centi\meter\per\second}$, with the upper limit from Fig.~2 in \citet{Ohm2013}, which takes a value of \SI{1.4e-11}{\erg\per\square\centi\meter\per\second}. Rescaling the \textit{Fermi} area of \SI{13685}{\square\arcmin} to the H.E.S.S. area of \SI{11664}{\square\arcmin} only worsens the discrepancy. Due to this data-level discrepancy, we consider the model dependence of the HE $\gamma$-ray upper limit to be the more plausible source of its violation by our SED fit.

\section{Conclusions} \label{SecConclusions}

    In this study, we have presented the first \textit{eROSITA} analysis of the star cluster Wd~1 and the associated PeVatron candidate HESS~J1646$-$458, employing data from the first four \textit{eROSITA} all-sky surveys. Our analysis has two related motivations: First, based on a spectral analysis of the diffuse X-ray emission of Wd~1, we wanted to better understand the nature of this emission, in particular if it is thermal or nonthermal. Second, we searched the source area of the PeVatron candidate HESS~J1646$-$458 for signs of X-ray synchrotron radiation. Specifically, stricter limits on or even the detection of such nonthermal radiation would help to break the degeneracy between hadronic and leptonic source scenarios for the PeVatron candidate's $\gamma$-ray emission. Consequently, such results could help to establish or refute HESS~J1646$-$458 as a source of hadronic VHE Galactic CRs.

    Diffuse emission from Wd~1 is clearly detected in the eRASS:4 data, as can be seen in Fig.~\ref{FigWd1Image}. On the other hand, the only detected point source associated with the star cluster is the magnetar CXOU~164710.2$-$455216 to its southeast. The main obstacle to our spectral analyses was the nearby bright LMXB GX~340$+$0, which has an extended halo. This halo contaminates the nearby regions, in particular Wd~1 and the entire HESS~J1646$-$458 source region. We took the contamination into account by adding a phenomenological \texttt{tbabs}$\cdot$\texttt{diskbb} component to our spectral models and by always choosing background regions with the same distance to GX~340$+$0 as the corresponding source regions.

    For the analysis of the diffuse emission of Wd~1, after fitting and fixing the background spectrum, we fitted a source model with two thermal \texttt{apec} components and another one with an \texttt{apec} and a nonthermal \texttt{powerlaw} component, as can be seen in Fig.~\ref{Fig2apec} and \ref{FigAPECPL}. We call these models \texttt{2apec} and \texttt{apec+pl}, respectively. Our main results are as follows:
    \begin{enumerate}
        \item The \texttt{2apec} and the \texttt{apec+pl} model provide similarly good fits to the data. The \texttt{2apec} model has a reduced Pearson-$\chi^2$ of $\chi^2_\mathrm{P}/\nu = 0.967$ and a goodness of 32\%, as evaluated by \texttt{Xspec}'s \texttt{goodness} command. On the other hand, the \texttt{apec+pl} model has $\chi^2_\mathrm{P} = 0.973$ and a goodness of 29\%. We conclude that the models cannot be distinguished based on their fit quality.
        \item In particular, the eRASS:4 data are insufficient to refute or confirm the presence of an Fe~XXV emission line around \SI{6.7}{\kilo\electronvolt} expected from a hot plasma, since there are no noteworthy residuals around this position in either of our two fits.
        \item Using the \texttt{2apec} model, we calculated a source luminosity of $L_\mathrm{2-\SI{8}{\kilo\electronvolt}} = 5.2_{-2.6}^{+2.6}\cdot10^{33}\,\SI{}{\erg\per\second}$ based on an assumed distance to Wd~1 of \SI{3.9}{\kilo\parsec}. We estimated that this luminosity can be explained to around 40\% by a population of unresolved PMS stars, based on a rescaling of the estimate of \citet{Kavanagh2011}, or to around 70\% by thermalized stellar wind shocks, based on the analytic wind model of \citet{Canto2000}. However, keeping in mind the large uncertainties on the data and on the calculations, these numbers can only serve as rough estimates. Furthermore, we deem a supernova explanation of the diffuse X-ray emission from Wd~1 unlikely, since apart from the magnetar CXOU~164710.2$-$455216, no direct evidence for supernova activity is detected in the cluster.
    \end{enumerate}

    The main goal of our study was the search for X-ray synchrotron radiation from the PeVatron candidate HESS~J1646$-$458, which extends up to a radius of ${\sim}\SI{30}{\arcmin}$ around Wd~1. We spectrally analyzed four circular regions of increasing radii from \SI{3}{\arcmin} out to \SI{40}{\arcmin} around the star cluster. In particular, we compared the spectra of these regions to the spectra of four corresponding background regions with the same distance to the contaminating source GX 340$+$0. The main results of this analysis are as follows:
    \begin{enumerate}
        \item We did not detect a significant source component in the source region spectra. Instead, these spectra are compatible with pure background emission. This assessment is based on the negative mean values of Gaussian fits to the source-minus-background residuals as given in Table~\ref{TabGauss}, and on the t-test results given in the same table. In particular, the t-tests force us to reject the hypothesis of at least as much flux from the source regions as from the background regions with significances $\gtrsim 2\sigma$, depending on the region pair.
        \item Next, we fitted background spectra to the source regions. These fits can be seen in Fig.~\ref{FigBackgroundFits}. Based on them, we determined upper confidence bounds on the synchrotron flux in the four regions. For this, we assumed an absorbing column density of $N_\mathrm{H} = \SI{2.2e22}{\per\cubic\centi\meter}$, compatible with our fits to the Wd~1 spectrum, and three different power law indices, namely $1.5$, $2.0$, and $2.5$. The most conservative upper flux bound at \SI{1}{\kilo\electronvolt} for the sum of the four regions is $\eta_\mathrm{X} \leq \SI{1.9e-3}{\per\kilo\electronvolt\per\square\centi\meter\per\second}$.
        \item Based on this upper bound and the VHE $\gamma$-ray data from \citet{Aharonian2022}, we performed an SED fit to HESS~J1646$-$458, assuming a leptonic emission scenario with IC scattering and synchrotron radiation. For a minimum energy of accelerated electrons $E_\mathrm{min} = \SI{500}{\mega\electronvolt}$, we find an upper $1\sigma$ confidence bound on the magnetic field strength in the source region of \SI{6.6}{\micro\gauss}.
        \item For lower values of $E_\mathrm{min}$, the upper bound becomes formally looser, though we argue that this is a computational effect of the fitting software. Since lower values of $E_\mathrm{min}$ lead to the IC radiation contributing to the X-ray upper bound, they should in reality leave less space for synchrotron radiation and therefore tighten the upper bound on $B$.
        \item Our upper confidence bound of $B \leq \SI{6.6}{\micro\gauss}$ is compatible with the theoretical upper bound of \SI{4.5}{\micro\gauss} estimated by \citet{Haerer2023}. These authors discuss in detail the leptonic emission scenario at the cluster wind termination shock. Therefore, we conclude that our results are compatible with such a leptonic emission scenario. Similarly, our results are also compatible with a hadronic emission scenario, since synchrotron radiation from hadrons is strongly suppressed. We conclude that the question for the true $\gamma$-ray emission scenario of HESS~J1646$-$458 is still open.
    \end{enumerate}

%Acknowledgements
\begin{acknowledgements}
        This work is based on data from \textit{eROSITA}, the soft X-ray instrument aboard SRG, a joint Russian-German science mission supported by the Russian Space Agency (Ros\-kos\-mos), in the interests of the Russian Academy of Sciences represented by its Space Research Institute (IKI), and the Deut\-sches Zen\-trum für Luft- und Raum\-fahrt (DLR). The SRG spacecraft was built by La\-voch\-kin Association (NPOL) and its subcontractors, and is operated by NPOL with support from the Max Planck Institute for Extraterrestrial Physics (MPE). The development and construction of the \textit{eROSITA} X-ray instrument was led by MPE, with contributions from the Dr. Karl Re\-meis Observatory Bam\-berg \& ECAP (FAU Er\-lan\-gen-N\"urn\-berg), the University of Ham\-burg Observatory, the Leib\-niz Institute for Astrophysics Pots\-dam (AIP), and the Institute for Astronomy and Astrophysics of the University of Tü\-bin\-gen, with the support of DLR and the Max Planck Society. The Ar\-ge\-lan\-der Institute for Astronomy of the University of Bonn and the Lud\-wig Ma\-xi\-mi\-li\-ans U\-ni\-ver\-si\-tät Munich also participated in the science preparation for \textit{eROSITA}. The \textit{eROSITA} data shown here were processed using the eSASS software system developed by the German \textit{eROSITA} consortium. AM is supported by the Deutsche Forschungsgemeinschaft (DFG) project number 452934793. GPo acknowledges financial support from the European Research Council (ERC) under the European Union’s Horizon 2020 research and innovation program HotMilk (grant agreement No. 865637), support from Bando per il Finanziamento della Ricerca Fondamentale 2022 dell’Istituto Nazionale di Astrofisica (INAF): GO Large program and from the Framework per l’Attrazione e il Rafforzamento delle Eccellenze (FARE) per la ricerca in Italia (R20L5S39T9). MDF, GR, and SL acknowledge Australian Research Council funding through grant DP200100784. KH would like to thank Jonathan R. Knies for advice on X-ray data analysis and Michael Freyberg for helpful comments on \textit{eROSITA} dust scattering and stray light.
\end{acknowledgements}

%%% References using the A&A style %%%
\bibliographystyle{aa}
\bibliography{references}

\begin{thebibliography}{69}
\expandafter\ifx\csname natexlab\endcsname\relax\def\natexlab#1{#1}\fi

\bibitem[{{Abdollahi} {et~al.}(2022){Abdollahi}, {Acero}, {Baldini}, {Ballet},
  {Bastieri}, {Bellazzini}, {Berenji}, {Berretta}, {Bissaldi}, {Blandford},
  {Bloom}, {Bonino}, {Brill}, {Britto}, {Bruel}, {Burnett}, {Buson}, {Cameron},
  {Caputo}, {Caraveo}, {Castro}, {Chaty}, {Cheung}, {Chiaro}, {Cibrario},
  {Ciprini}, {Coronado-Bl{\'a}zquez}, {Crnogorcevic}, {Cutini}, {D'Ammando},
  {De Gaetano}, {Digel}, {Di Lalla}, {Dirirsa}, {Di Venere}, {Dom{\'\i}nguez},
  {Fallah Ramazani}, {Fegan}, {Ferrara}, {Fiori}, {Fleischhack}, {Franckowiak},
  {Fukazawa}, {Funk}, {Fusco}, {Galanti}, {Gammaldi}, {Gargano}, {Garrappa},
  {Gasparrini}, {Giacchino}, {Giglietto}, {Giordano}, {Giroletti}, {Glanzman},
  {Green}, {Grenier}, {Grondin}, {Guillemot}, {Guiriec}, {Gustafsson},
  {Harding}, {Hays}, {Hewitt}, {Horan}, {Hou}, {J{\'o}hannesson}, {Karwin},
  {Kayanoki}, {Kerr}, {Kuss}, {Landriu}, {Larsson}, {Latronico},
  {Lemoine-Goumard}, {Li}, {Liodakis}, {Longo}, {Loparco}, {Lott}, {Lubrano},
  {Maldera}, {Malyshev}, {Manfreda}, {Mart{\'\i}-Devesa}, {Mazziotta}, {Mereu},
  {Meyer}, {Michelson}, {Mirabal}, {Mitthumsiri}, {Mizuno}, {Moiseev},
  {Monzani}, {Morselli}, {Moskalenko}, {Negro}, {Nuss}, {Omodei}, {Orienti},
  {Orlando}, {Paneque}, {Pei}, {Perkins}, {Persic}, {Pesce-Rollins},
  {Petrosian}, {Pillera}, {Poon}, {Porter}, {Principe}, {Rain{\`o}}, {Rando},
  {Rani}, {Razzano}, {Razzaque}, {Reimer}, {Reimer}, {Reposeur},
  {S{\'a}nchez-Conde}, {Saz Parkinson}, {Scotton}, {Serini}, {Sgr{\`o}},
  {Siskind}, {Smith}, {Spandre}, {Spinelli}, {Sueoka}, {Suson}, {Tajima},
  {Tak}, {Thayer}, {Thompson}, {Torres}, {Troja}, {Valverde}, {Wood}, \&
  {Zaharijas}}]{Abdollahi2022}
{Abdollahi}, S., {Acero}, F., {Baldini}, L., {et~al.} 2022, \apjs, 260, 53

\bibitem[{{Abeysekara} {et~al.}(2020){Abeysekara}, {Albert}, {Alfaro}, {Angeles
  Camacho}, {Arteaga-Vel{\'a}zquez}, {Arunbabu}, {Avila Rojas}, {Ayala
  Solares}, {Baghmanyan}, {Belmont-Moreno}, {BenZvi}, {Brisbois},
  {Caballero-Mora}, {Capistr{\'a}n}, {Carrami{\~n}ana}, {Casanova}, {Cotti},
  {Cotzomi}, {Couti{\~n}o de Le{\'o}n}, {De la Fuente}, {de Le{\'o}n},
  {Dichiara}, {Dingus}, {DuVernois}, {D{\'\i}az-V{\'e}lez}, {Ellsworth},
  {Engel}, {Espinoza}, {Fleischhack}, {Fraija}, {Galv{\'a}n-G{\'a}mez},
  {Garcia}, {Garc{\'\i}a-Gonz{\'a}lez}, {Garfias}, {Gonz{\'a}lez}, {Goodman},
  {Harding}, {Hernandez}, {Hinton}, {Hona}, {Huang}, {Hueyotl-Zahuantitla},
  {H{\"u}ntemeyer}, {Iriarte}, {Jardin-Blicq}, {Joshi}, {Kaufmann}, {Kieda},
  {Lara}, {Lee}, {Le{\'o}n Vargas}, {Linnemann}, {Longinotti}, {Luis-Raya},
  {Lundeen}, {L{\'o}pez-Coto}, {Malone}, {Marinelli}, {Martinez},
  {Martinez-Castellanos}, {Mart{\'\i}nez-Castro}, {Mart{\'\i}nez-Huerta},
  {Matthews}, {Miranda-Romagnoli}, {Morales-Soto}, {Moreno}, {Mostaf{\'a}},
  {Nayerhoda}, {Nellen}, {Newbold}, {Nisa}, {Noriega-Papaqui}, {Peisker},
  {P{\'e}rez-P{\'e}rez}, {Pretz}, {Ren}, {Rho}, {Rivi{\`e}re},
  {Rosa-Gonz{\'a}lez}, {Rosenberg}, {Ruiz-Velasco}, {Salesa Greus}, {Sandoval},
  {Schneider}, {Schoorlemmer}, {Sinnis}, {Smith}, {Springer}, {Surajbali},
  {Tabachnick}, {Tanner}, {Tibolla}, {Tollefson}, {Torres}, {Torres-Escobedo},
  {Villase{\~n}or}, {Weisgarber}, {Wood}, {Yapici}, {Zhang}, {Zhou}, \& {HAWC
  Collaboration}}]{Abeysekara2020}
{Abeysekara}, A.~U., {Albert}, A., {Alfaro}, R., {et~al.} 2020, \prl, 124,
  021102

\bibitem[{{Abramowski} {et~al.}(2012){Abramowski}, {Acero}, {Aharonian},
  {Akhperjanian}, {Anton}, {Balzer}, {Barnacka}, {Barres de Almeida},
  {Becherini}, {Becker}, {Behera}, {Bernl{\"o}hr}, {Birsin}, {Biteau},
  {Bochow}, {Boisson}, {Bolmont}, {Bordas}, {Brucker}, {Brun}, {Brun}, {Bulik},
  {B{\"u}sching}, {Carrigan}, {Casanova}, {Cerruti}, {Chadwick}, {Charbonnier},
  {Chaves}, {Cheesebrough}, {Chounet}, {Clapson}, {Coignet}, {Cologna},
  {Conrad}, {Dalton}, {Daniel}, {Davids}, {Degrange}, {Deil}, {Dickinson},
  {Djannati-Ata{\"\i}}, {Domainko}, {O'C. Drury}, {Dubois}, {Dubus}, {Dutson},
  {Dyks}, {Dyrda}, {Egberts}, {Eger}, {Espigat}, {Fallon}, {Farnier}, {Fegan},
  {Feinstein}, {Fernandes}, {Fiasson}, {Fontaine}, {F{\"o}rster},
  {F{\"u}{\ss}ling}, {Gallant}, {Gast}, {G{\'e}rard}, {Gerbig}, {Giebels},
  {Glicenstein}, {Gl{\"u}ck}, {Goret}, {G{\"o}ring}, {H{\"a}ffner}, {Hague},
  {Hampf}, {Hauser}, {Heinz}, {Heinzelmann}, {Henri}, {Hermann}, {Hinton},
  {Hoffmann}, {Hofmann}, {Hofverberg}, {Holler}, {Horns}, {Jacholkowska}, {de
  Jager}, {Jahn}, {Jamrozy}, {Jung}, {Kastendieck}, {Katarzy{\'n}ski}, {Katz},
  {Kaufmann}, {Keogh}, {Khangulyan}, {Kh{\'e}lifi}, {Klochkov}, {Klu{\.z}niak},
  {Kneiske}, {Komin}, {Kosack}, {Kossakowski}, {Laffon}, {Lamanna}, {Lennarz},
  {Lohse}, {Lopatin}, {Lu}, {Marandon}, {Marcowith}, {Masbou}, {Maurin},
  {Maxted}, {Mayer}, {McComb}, {Medina}, {M{\'e}hault}, {Moderski}, {Moulin},
  {Naumann}, {Naumann-Godo}, {de Naurois}, {Nedbal}, {Nekrassov}, {Nguyen},
  {Nicholas}, {Niemiec}, {Nolan}, {Ohm}, {de O{\~n}a Wilhelmi}, {Opitz},
  {Ostrowski}, {Oya}, {Panter}, {Paz Arribas}, {Pedaletti}, {Pelletier},
  {Petrucci}, {Pita}, {P{\"u}hlhofer}, {Punch}, {Quirrenbach}, {Raue},
  {Rayner}, {Reimer}, {Reimer}, {Renaud}, {de Los Reyes}, {Rieger}, {Ripken},
  {Rob}, {Rosier-Lees}, {Rowell}, {Rudak}, {Rulten}, {Ruppel}, {Sahakian},
  {Sanchez}, {Santangelo}, {Schlickeiser}, {Sch{\"o}ck}, {Schulz}, {Schwanke},
  {Schwarzburg}, {Schwemmer}, {Sheidaei}, {Sikora}, {Skilton}, {Sol},
  {Spengler}, {Stawarz}, {Steenkamp}, {Stegmann}, {Stinzing}, {Stycz},
  {Sushch}, {Szostek}, {Tavernet}, {Terrier}, {Tluczykont}, {Valerius}, {van
  Eldik}, {Vasileiadis}, {Venter}, {Vialle}, {Viana}, {Vincent}, {V{\"o}lk},
  {Volpe}, {Vorobiov}, {Vorster}, {Wagner}, {Ward}, {White}, {Wierzcholska},
  {Zacharias}, {Zajczyk}, {Zdziarski}, {Zech}, \& {Zechlin}}]{Abramowski2012}
{Abramowski}, A., {Acero}, F., {Aharonian}, F., {et~al.} 2012, \aap, 537, A114

\bibitem[{{Ackermann} {et~al.}(2011){Ackermann}, {Ajello}, {Allafort},
  {Baldini}, {Ballet}, {Barbiellini}, {Bastieri}, {Belfiore}, {Bellazzini},
  {Berenji}, {Blandford}, {Bloom}, {Bonamente}, {Borgland}, {Bottacini},
  {Brigida}, {Bruel}, {Buehler}, {Buson}, {Caliandro}, {Cameron}, {Caraveo},
  {Casandjian}, {Cecchi}, {Chekhtman}, {Cheung}, {Chiang}, {Ciprini}, {Claus},
  {Cohen-Tanugi}, {de Angelis}, {de Palma}, {Dermer}, {do Couto e Silva},
  {Drell}, {Dumora}, {Favuzzi}, {Fegan}, {Focke}, {Fortin}, {Fukazawa},
  {Fusco}, {Gargano}, {Germani}, {Giglietto}, {Giordano}, {Giroletti},
  {Glanzman}, {Godfrey}, {Grenier}, {Guillemot}, {Guiriec}, {Hadasch},
  {Hanabata}, {Harding}, {Hayashida}, {Hayashi}, {Hays}, {J{\'o}hannesson},
  {Johnson}, {Kamae}, {Katagiri}, {Kataoka}, {Kerr}, {Kn{\"o}dlseder}, {Kuss},
  {Lande}, {Latronico}, {Lee}, {Longo}, {Loparco}, {Lott}, {Lovellette},
  {Lubrano}, {Martin}, {Mazziotta}, {McEnery}, {Mehault}, {Michelson},
  {Mitthumsiri}, {Mizuno}, {Monte}, {Monzani}, {Morselli}, {Moskalenko},
  {Murgia}, {Naumann-Godo}, {Nolan}, {Norris}, {Nuss}, {Ohsugi}, {Okumura},
  {Orlando}, {Ormes}, {Ozaki}, {Paneque}, {Parent}, {Pesce-Rollins},
  {Pierbattista}, {Piron}, {Pohl}, {Prokhorov}, {Rain{\`o}}, {Rando},
  {Razzano}, {Reposeur}, {Ritz}, {Parkinson}, {Sgr{\`o}}, {Siskind}, {Smith},
  {Spinelli}, {Strong}, {Takahashi}, {Tanaka}, {Thayer}, {Thayer}, {Thompson},
  {Tibaldo}, {Torres}, {Tosti}, {Tramacere}, {Troja}, {Uchiyama},
  {Vandenbroucke}, {Vasileiou}, {Vianello}, {Vitale}, {Waite}, {Wang}, {Winer},
  {Wood}, {Yang}, {Zimmer}, \& {Bontemps}}]{Ackermann2011}
{Ackermann}, M., {Ajello}, M., {Allafort}, A., {et~al.} 2011, Science, 334,
  1103

\bibitem[{{Aguilar} {et~al.}(2015){Aguilar}, {Aisa}, {Alpat}, {Alvino},
  {Ambrosi}, {Andeen}, {Arruda}, {Attig}, {Azzarello}, {Bachlechner}, {Barao},
  {Barrau}, {Barrin}, {Bartoloni}, {Basara}, {Battarbee}, {Battiston}, {Bazo},
  {Becker}, {Behlmann}, {Beischer}, {Berdugo}, {Bertucci}, {Bindi},
  {Bizzaglia}, {Bizzarri}, {Boella}, {de Boer}, {Bollweg}, {Bonnivard},
  {Borgia}, {Borsini}, {Boschini}, {Bourquin}, {Burger}, {Cadoux}, {Cai},
  {Capell}, {Caroff}, {Casaus}, {Castellini}, {Cernuda}, {Cerreta}, {Cervelli},
  {Chae}, {Chang}, {Chen}, {Chen}, {Chen}, {Chen}, {Cheng}, {Chou},
  {Choumilov}, {Choutko}, {Chung}, {Clark}, {Clavero}, {Coignet}, {Consolandi},
  {Contin}, {Corti}, {Gil}, {Coste}, {Creus}, {Crispoltoni}, {Cui}, {Dai},
  {Delgado}, {Della Torre}, {Demirk{\"o}z}, {Derome}, {Di Falco}, {Di Masso},
  {Dimiccoli}, {D{\'\i}az}, {von Doetinchem}, {Donnini}, {Duranti}, {D'Urso},
  {Egorov}, {Eline}, {Eppling}, {Eronen}, {Fan}, {Farnesini}, {Feng},
  {Fiandrini}, {Fiasson}, {Finch}, {Fisher}, {Formato}, {Galaktionov},
  {Gallucci}, {Garc{\'\i}a}, {Garc{\'\i}a-L{\'o}pez}, {Gargiulo}, {Gast},
  {Gebauer}, {Gervasi}, {Ghelfi}, {Giovacchini}, {Goglov}, {Gong}, {Goy},
  {Grabski}, {Grandi}, {Graziani}, {Guandalini}, {Guerri}, {Guo}, {Haas},
  {Habiby}, {Haino}, {Han}, {He}, {Heil}, {Hoffman}, {Hsieh}, {Huang}, {Huh},
  {Incagli}, {Ionica}, {Jang}, {Jinchi}, {Kanishev}, {Kim}, {Kim}, {Kirn},
  {Korkmaz}, {Kossakowski}, {Kounina}, {Kounine}, {Koutsenko}, {Krafczyk}, {La
  Vacca}, {Laudi}, {Laurenti}, {Lazzizzera}, {Lebedev}, {Lee}, {Lee}, {Leluc},
  {Li}, {Li}, {Li}, {Li}, {Li}, {Li}, {Li}, {Li}, {Li}, {Li}, {Lim}, {Lin},
  {Lipari}, {Lippert}, {Liu}, {Liu}, {Liu}, {Lolli}, {Lomtadze}, {Lu}, {Lu},
  {Lu}, {Luebelsmeyer}, {Luo}, {Luo}, {Lv}, {Majka}, {Ma{\~n}{\'a}},
  {Mar{\'\i}n}, {Martin}, {Mart{\'\i}nez}, {Masi}, {Maurin}, {Menchaca-Rocha},
  {Meng}, {Mo}, {Morescalchi}, {Mott}, {M{\"u}ller}, {Nelson}, {Ni}, {Nikonov},
  {Nozzoli}, {Nunes}, {Obermeier}, {Oliva}, {Orcinha}, {Palmonari},
  {Palomares}, {Paniccia}, {Papi}, {Pauluzzi}, {Pedreschi}, {Pensotti},
  {Pereira}, {Picot-Clemente}, {Pilo}, {Piluso}, {Pizzolotto}, {Plyaskin},
  {Pohl}, {Poireau}, {Putze}, {Quadrani}, {Qi}, {Qin}, {Qu}, {R{\"a}ih{\"a}},
  {Rancoita}, {Rapin}, {Ricol}, {Rodr{\'\i}guez}, {Rosier-Lees}, {Rozhkov},
  {Rozza}, {Sagdeev}, {Sandweiss}, {Saouter}, {Schael}, {Schmidt}, {von
  Dratzig}, {Schwering}, {Scolieri}, {Seo}, {Shan}, {Shan}, {Shi}, {Shi},
  {Shi}, {Siedenburg}, {Son}, {Song}, {Spada}, {Spinella}, {Sun}, {Sun},
  {Tacconi}, {Tang}, {Tang}, {Tang}, {Tao}, {Tescaro}, {Ting}, {Ting},
  {Tomassetti}, {Torsti}, {T{\"u}rko{\v{g}}lu}, {Urban}, {Vagelli}, {Valente},
  {Vannini}, {Valtonen}, {Vaurynovich}, {Vecchi}, {Velasco}, {Vialle},
  {Vitale}, {Vitillo}, {Wang}, {Wang}, {Wang}, {Wang}, {Wang}, {Wang}, {Weng},
  {Whitman}, {Wienkenh{\"o}ver}, {Willenbrock}, {Wu}, {Wu}, {Xia}, {Xie},
  {Xie}, {Xiong}, {Xu}, {Xu}, {Yan}, {Yang}, {Yang}, {Yang}, {Ye}, {Yi}, {Yu},
  {Yu}, {Zeissler}, {Zhang}, {Zhang}, {Zhang}, {Zhang}, {Zhang}, {Zhang},
  {Zhang}, {Zheng}, {Zhuang}, {Zhukov}, {Zichichi}, {Zimmermann}, {Zuccon}, \&
  {AMS Collaboration}}]{AMS2015}
{Aguilar}, M., {Aisa}, D., {Alpat}, B., {et~al.} 2015, \prl, 115, 211101

\bibitem[{{Aguilar} {et~al.}(2019){Aguilar}, {Ali Cavasonza}, {Alpat},
  {Ambrosi}, {Arruda}, {Attig}, {Azzarello}, {Bachlechner}, {Barao}, {Barrau},
  {Barrin}, {Bartoloni}, {Basara}, {Ba{\c{s}}e{\v{g}}mez-du Pree}, {Battiston},
  {Becker}, {Behlmann}, {Beischer}, {Berdugo}, {Bertucci}, {Bindi}, {de Boer},
  {Bollweg}, {Borgia}, {Boschini}, {Bourquin}, {Bueno}, {Burger}, {Burger},
  {Cai}, {Capell}, {Caroff}, {Casaus}, {Castellini}, {Cervelli}, {Chang},
  {Chen}, {Chen}, {Chen}, {Cheng}, {Chou}, {Choutko}, {Chung}, {Clark},
  {Coignet}, {Consolandi}, {Contin}, {Corti}, {Crispoltoni}, {Cui}, {Dadzie},
  {Dai}, {Datta}, {Delgado}, {Della Torre}, {Demirk{\"o}z}, {Derome}, {Di
  Falco}, {Di Felice}, {Dimiccoli}, {D{\'\i}az}, {von Doetinchem}, {Dong},
  {Donnini}, {Duranti}, {Egorov}, {Eline}, {Eronen}, {Feng}, {Fiandrini},
  {Fisher}, {Formato}, {Galaktionov}, {Garc{\'\i}a-L{\'o}pez}, {Gargiulo},
  {Gast}, {Gebauer}, {Gervasi}, {Giovacchini}, {G{\'o}mez-Coral}, {Gong},
  {Goy}, {Grabski}, {Grandi}, {Graziani}, {Guo}, {Haino}, {Han}, {He}, {Heil},
  {Hsieh}, {Huang}, {Huang}, {Incagli}, {Jia}, {Jinchi}, {Kanishev}, {Khiali},
  {Kirn}, {Konak}, {Kounina}, {Kounine}, {Koutsenko}, {Kulemzin}, {La Vacca},
  {Laudi}, {Laurenti}, {Lazzizzera}, {Lebedev}, {Lee}, {Lee}, {Leluc}, {Li},
  {Li}, {Li}, {Li}, {Light}, {Lin}, {Lippert}, {Liu}, {Liu}, {Liu}, {Lu}, {Lu},
  {Luebelsmeyer}, {Luo}, {Luo}, {Luo}, {Lyu}, {Machate}, {Ma{\~n}{\'a}},
  {Mar{\'\i}n}, {Martin}, {Mart{\'\i}nez}, {Masi}, {Maurin}, {Menchaca-Rocha},
  {Meng}, {Mo}, {Molero}, {Mott}, {Mussolin}, {Nelson}, {Ni}, {Nikonov},
  {Nozzoli}, {Oliva}, {Orcinha}, {Palermo}, {Palmonari}, {Paniccia}, {Pashnin},
  {Pauluzzi}, {Pensotti}, {Perrina}, {Phan}, {Picot-Clemente}, {Plyaskin},
  {Pohl}, {Poireau}, {Popkow}, {Quadrani}, {Qi}, {Qin}, {Qu}, {Rancoita},
  {Rapin}, {Conde}, {Rosier-Lees}, {Rozhkov}, {Rozza}, {Sagdeev}, {Solano},
  {Schael}, {Schmidt}, {von Dratzig}, {Schwering}, {Seo}, {Shan}, {Shi},
  {Siedenburg}, {Song}, {Sun}, {Tacconi}, {Tang}, {Tang}, {Tian}, {Ting},
  {Ting}, {Tomassetti}, {Torsti}, {Urban}, {Vagelli}, {Valente}, {Valtonen},
  {Acosta}, {Vecchi}, {Velasco}, {Vialle}, {Viz{\'a}n}, {Wang}, {Wang}, {Wang},
  {Wang}, {Wang}, {Wang}, {Wei}, {Weng}, {Wu}, {Xiong}, {Xu}, {Yan}, {Yang},
  {Yi}, {Yu}, {Yu}, {Zannoni}, {Zeissler}, {Zhang}, {Zhang}, {Zhang}, {Zhang},
  {Zhao}, {Zheng}, {Zhuang}, {Zhukov}, {Zichichi}, {Zimmermann}, {Zuccon}, \&
  {AMS Collaboration}}]{Aguilar2019}
{Aguilar}, M., {Ali Cavasonza}, L., {Alpat}, B., {et~al.} 2019, \prl, 122,
  101101

\bibitem[{{Aharonian} {et~al.}(2022){Aharonian}, {Ashkar}, {Backes}, {Barbosa
  Martins}, {Becherini}, {Berge}, {Bi}, {B{\"o}ttcher}, {de Bony de Lavergne},
  {Bradascio}, {Brose}, {Brun}, {Bulik}, {Burger-Scheidlin}, {Cangemi},
  {Caroff}, {Casanova}, {Cerruti}, {Chand}, {Chandra}, {Chen}, {Chibueze},
  {Cristofari}, {Damascene Mbarubucyeye}, {Djannati-Ata{\"\i}}, {Ernenwein},
  {Feijen}, {Fichet de Clairfontaine}, {Fontaine}, {Funk}, {Gabici}, {Gallant},
  {Ghafourizadeh}, {Giavitto}, {Giunti}, {Glawion}, {Glicenstein}, {Goswami},
  {Grondin}, {H{\"a}rer}, {Haupt}, {Hinton}, {H{\"o}rbe}, {Hofmann}, {Holch},
  {Holler}, {Horns}, {Jamrozy}, {Joshi}, {Jung-Richardt}, {Kasai},
  {Katarzy{\'n}ski}, {Katz}, {Kh{\'e}lifi}, {Klu{\'z}niak}, {Komin}, {Kosack},
  {Kostunin}, {Kukec Mezek}, {Lang}, {Le Stum}, {Lemi{\`e}re},
  {Lemoine-Goumard}, {Lenain}, {Leuschner}, {Lohse}, {Luashvili}, {Lypova},
  {Mackey}, {Majumdar}, {Malyshev}, {Marandon}, {Marchegiani}, {Marcowith},
  {Mart{\'\i}-Devesa}, {Marx}, {Maurin}, {Meyer}, {Mitchell}, {Moderski},
  {Mohrmann}, {Montanari}, {Moulin}, {Muller}, {Murach}, {Nakashima}, {de
  Naurois}, {Nayerhoda}, {Niemiec}, {Ohm}, {Olivera-Nieto}, {de Ona Wilhelmi},
  {Ostrowski}, {Panny}, {Panter}, {Parsons}, {Peron}, {Prokhorov},
  {P{\"u}hlhofer}, {Punch}, {Quirrenbach}, {Rauth}, {Reichherzer}, {Reimer},
  {Reimer}, {Renaud}, {Reville}, {Rieger}, {Rowell}, {Rudak}, {Ruiz-Velasco},
  {Sahakian}, {Salzmann}, {Sanchez}, {Santangelo}, {Sasaki}, {Sch{\"u}ssler},
  {Schutte}, {Schwanke}, {Shapopi}, {Specovius}, {Spencer}, {Stawarz},
  {Steenkamp}, {Steinmassl}, {Steppa}, {Sushch}, {Suzuki}, {Takahashi},
  {Tanaka}, {Terrier}, {Thorpe-Morgan}, {Tsirou}, {Tsuji}, {Tuffs}, {Unbehaun},
  {van Eldik}, {van Soelen}, {Vecchi}, {Veh}, {Venter}, {Vink}, {Wagner},
  {White}, {Wierzcholska}, {Wong}, {Zacharias}, {Zargaryan}, {Zdziarski},
  {Zhu}, {Zouari}, {{\.Z}ywucka}, {Blackwell}, {Braiding}, {Burton}, {Cubuk},
  {Filipovi{\'c}}, {Tothill}, \& {Wong}}]{Aharonian2022}
{Aharonian}, F., {Ashkar}, H., {Backes}, M., {et~al.} 2022, \aap, 666, A124

\bibitem[{{Aharonian} {et~al.}(2010){Aharonian}, {Kelner}, \&
  {Prosekin}}]{Aharonian2010}
{Aharonian}, F.~A., {Kelner}, S.~R., \& {Prosekin}, A.~Y. 2010, \prd, 82,
  043002

\bibitem[{{Albert} {et~al.}(2021){Albert}, {Alfaro}, {Alvarez}, {{\'A}lvarez},
  {Angeles Camacho}, {Arteaga-Vel{\'a}zquez}, {Arunbabu}, {Avila Rojas}, {Ayala
  Solares}, {Baghmanyan}, {Belmont-Moreno}, {BenZvi}, {Brisbois},
  {Caballero-Mora}, {Capistr{\'a}n}, {Carrami{\~n}ana}, {Casanova}, {Cotti},
  {Cotzomi}, {Couti{\~n}o de Le{\'o}n}, {De la Fuente}, {de Le{\'o}n}, {Diaz
  Hernandez}, {Dingus}, {DuVernois}, {Durocher}, {D{\'\i}az-V{\'e}lez},
  {Ellsworth}, {Engel}, {Espinoza}, {Fan}, {Fern{\'a}ndez Alonso}, {Fraija},
  {Galv{\'a}n-G{\'a}mez}, {Garc{\'\i}a-Gonz{\'a}lez}, {Garfias}, {Giacinti},
  {Gonz{\'a}lez}, {Goodman}, {Harding}, {Hernandez}, {Hona}, {Huang},
  {Hueyotl-Zahuantitla}, {H{\"u}ntemeyer}, {Iriarte}, {Jardin-Blicq}, {Joshi},
  {Kieda}, {Lara}, {Lee}, {Lee}, {Le{\'o}n Vargas}, {Linnemann}, {Longinotti},
  {Luis-Raya}, {Lundeen}, {Malone}, {Marandon}, {Martinez},
  {Mart{\'\i}nez-Castro}, {Matthews}, {Miranda-Romagnoli}, {Morales-Soto},
  {Moreno}, {Mostaf{\'a}}, {Nayerhoda}, {Nellen}, {Newbold}, {Nisa},
  {Noriega-Papaqui}, {Olivera-Nieto}, {Omodei}, {Peisker}, {P{\'e}rez Araujo},
  {P{\'e}rez-P{\'e}rez}, {Rho}, {Roh}, {Rosa-Gonz{\'a}lez}, {Ruiz-Velasco},
  {Salazar}, {Salesa Greus}, {Sandoval}, {Schneider}, {Schoorlemmer},
  {Serna-Franco}, {Smith}, {Springer}, {Surajbali}, {Tanner}, {Tollefson},
  {Torres}, {Torres-Escobedo}, {Turner}, {Ure{\~n}a-Mena}, {Villase{\~n}or},
  {Weisgarber}, {Willox}, {Zhou}, \& {HAWC Collaboration}}]{Albert2021}
{Albert}, A., {Alfaro}, R., {Alvarez}, C., {et~al.} 2021, \apjl, 911, L27

\bibitem[{{Armillotta} {et~al.}(2022){Armillotta}, {Ostriker}, \&
  {Jiang}}]{2022Armillotta}
{Armillotta}, L., {Ostriker}, E.~C., \& {Jiang}, Y.-F. 2022, \apj, 929, 170

\bibitem[{{Barret} {et~al.}(2023){Barret}, {Albouys}, {Herder}, {Piro},
  {Cappi}, {Huovelin}, {Kelley}, {Mas-Hesse}, {Paltani}, {Rauw}, {Rozanska},
  {Svoboda}, {Wilms}, {Yamasaki}, {Audard}, {Bandler}, {Barbera}, {Barcons},
  {Bozzo}, {Ceballos}, {Charles}, {Costantini}, {Dauser}, {Decourchelle},
  {Duband}, {Duval}, {Fiore}, {Gatti}, {Goldwurm}, {Hartog}, {Jackson},
  {Jonker}, {Kilbourne}, {Korpela}, {Macculi}, {Mendez}, {Mitsuda}, {Molendi},
  {Pajot}, {Pointecouteau}, {Porter}, {Pratt}, {Pr{\^e}le}, {Ravera}, {Sato},
  {Schaye}, {Shinozaki}, {Skup}, {Soucek}, {Thibert}, {Vink}, {Webb}, {Chaoul},
  {Raulin}, {Simionescu}, {Torrejon}, {Acero}, {Branduardi-Raymont}, {Ettori},
  {Finoguenov}, {Grosso}, {Kaastra}, {Mazzotta}, {Miller}, {Miniutti},
  {Nicastro}, {Sciortino}, {Yamaguchi}, {Beaumont}, {Cucchetti}, {D'Andrea},
  {Eckart}, {Ferrando}, {Kammoun}, {Lotti}, {Mesnager}, {Natalucci}, {Peille},
  {de Plaa}, {Ardellier}, {Argan}, {Bellouard}, {Carron}, {Cavazzuti},
  {Fiorini}, {Khosropanah}, {Martin}, {Perry}, {Pinsard}, {Pradines}, {Rigano},
  {Roelfsema}, {Schwander}, {Torrioli}, {Ullom}, {Vera}, {Villegas},
  {Zuchniak}, {Brachet}, {Cicero}, {Doriese}, {Durkin}, {Fioretti}, {Geoffray},
  {Jacques}, {Kirsch}, {Smith}, {Adams}, {Gloaguen}, {Hoogeveen}, {van der
  Hulst}, {Kiviranta}, {van der Kuur}, {Ledot}, {van Leeuwen}, {van Loon},
  {Lyautey}, {Parot}, {Sakai}, {van Weers}, {Abdoelkariem}, {Adam}, {Adami},
  {Aicardi}, {Akamatsu}, {Alonso}, {Amato}, {Andr{\'e}}, {Angelinelli},
  {Anon-Cancela}, {Anvar}, {Atienza}, {Attard}, {Auricchio}, {Balado},
  {Bancel}, {Barusso}, {Bascu{\~n}an}, {Bernard}, {Berrocal}, {Blin}, {Bonino},
  {Bonnet}, {Bonny}, {Boorman}, {Boreux}, {Bounab}, {Boutelier}, {Boyce},
  {Brienza}, {Bruijn}, {Bulgarelli}, {Calarco}, {Callanan}, {Campello},
  {Camus}, {Canourgues}, {Capobianco}, {Cardiel}, {Castellani}, {Cheatom},
  {Chervenak}, {Chiarello}, {Clerc}, {Clerc}, {Cobo}, {Coeur-Joly}, {Coleiro},
  {Colonges}, {Corcione}, {Coriat}, {Coynel}, {Cuttaia}, {D'Ai}, {D'anca},
  {Dadina}, {Daniel}, {Dauner}, {DeNigris}, {Dercksen}, {DiPirro}, {Doumayrou},
  {Dubbeldam}, {Dupieux}, {Dupourqu{\'e}}, {Durand}, {Eckert}, {Eiriz},
  {Ercolani}, {Etcheverry}, {Finkbeiner}, {Fiocchi}, {Fossecave}, {Franssen},
  {Frericks}, {Gabici}, {Gant}, {Gao}, {Gastaldello}, {Genolet}, {Ghizzardi},
  {Gil}, {Giovannini}, {Godet}, {Gomez-Elvira}, {Gonzalez}, {Gonzalez},
  {Gottardi}, {Granat}, {Gros}, {Guignard}, {Hieltjes}, {Hurtado}, {Irwin},
  {Jacquey}, {Janiuk}, {Jaubert}, {Jim{\'e}nez}, {Jolly}, {Jourdan}, {Julien},
  {Kedziora}, {Korb}, {Kreykenbohm}, {K{\"o}nig}, {Langer}, {Laudet},
  {Laurent}, {Laurenza}, {Lesrel}, {Ligori}, {Lorenz}, {Luminari}, {Maffei},
  {Maisonnave}, {Marelli}, {Massonet}, {Maussang}, {Melchor}, {Le Mer},
  {Millan}, {Millerioux}, {Mineo}, {Minervini}, {Molin}, {Monestes},
  {Montinaro}, {Mot}, {Murat}, {Nagayoshi}, {Naz{\'e}}, {Nogu{\`e}s}, {Pailot},
  {Panessa}, {Parodi}, {Petit}, {Piconcelli}, {Pinto}, {Plaza}, {Plaza},
  {Poyatos}, {Prouv{\'e}}, {Ptak}, {Puccetti}, {Puccio}, {Ramon}, {Reina},
  {Rioland}, {Rodriguez}, {Roig}, {Rollet}, {Roncarelli}, {Roudil}, {Rudnicki},
  {Sanisidro}, {Sciortino}, {Silva}, {Sordet}, {Soto-Aguilar}, {Spizzi},
  {Surace}, {Fern{\'a}ndez S{\'a}nchez}, {Taralli}, {Terrasa}, {Terrier},
  {Todaro}, {Ubertini}, {Uslenghi}, {de Vaate}, {Vaccaro}, {Varisco},
  {Varni{\`e}re}, {Vibert}, {Vidriales}, {Villa}, {Vodopivec}, {Volpe}, {de
  Vries}, {Wakeham}, {Walmsley}, {Wise}, {de Wit}, \&
  {Wo{\'z}niak}}]{Barret2023}
{Barret}, D., {Albouys}, V., {Herder}, J.-W.~d., {et~al.} 2023, Experimental
  Astronomy, 55, 373

\bibitem[{{Beasor} {et~al.}(2023){Beasor}, {Smith}, \& {Andrews}}]{Beasor2023}
{Beasor}, E.~R., {Smith}, N., \& {Andrews}, J.~E. 2023, \apj, 952, 113

\bibitem[{{Borghese} {et~al.}(2019){Borghese}, {Rea}, {Turolla}, {Pons},
  {Esposito}, {Coti Zelati}, {Savchenko}, {Bozzo}, {Perna}, {Zane},
  {Mereghetti}, {Campana}, {Mignani}, {Bachetti}, {Rodr{\'\i}guez}, {Pintore},
  {Tiengo}, {G{\"o}tz}, {Israel}, \& {Stella}}]{Borghese2019}
{Borghese}, A., {Rea}, N., {Turolla}, R., {et~al.} 2019, \mnras, 484, 2931

\bibitem[{{Brandner} {et~al.}(2008){Brandner}, {Clark}, {Stolte}, {Waters},
  {Negueruela}, \& {Goodwin}}]{Brandner2008}
{Brandner}, W., {Clark}, J.~S., {Stolte}, A., {et~al.} 2008, \aap, 478, 137

\bibitem[{{Brunner} {et~al.}(2022){Brunner}, {Liu}, {Lamer}, {Georgakakis},
  {Merloni}, {Brusa}, {Bulbul}, {Dennerl}, {Friedrich}, {Liu}, {Maitra},
  {Nandra}, {Ramos-Ceja}, {Sanders}, {Stewart}, {Boller}, {Buchner}, {Clerc},
  {Comparat}, {Dwelly}, {Eckert}, {Finoguenov}, {Freyberg}, {Ghirardini},
  {Gueguen}, {Haberl}, {Kreykenbohm}, {Krumpe}, {Osterhage}, {Pacaud},
  {Predehl}, {Reiprich}, {Robrade}, {Salvato}, {Santangelo}, {Schrabback},
  {Schwope}, \& {Wilms}}]{Brunner2022}
{Brunner}, H., {Liu}, T., {Lamer}, G., {et~al.} 2022, \aap, 661, A1

\bibitem[{{Cant{\'o}} {et~al.}(2000){Cant{\'o}}, {Raga}, \&
  {Rodr{\'\i}guez}}]{Canto2000}
{Cant{\'o}}, J., {Raga}, A.~C., \& {Rodr{\'\i}guez}, L.~F. 2000, \apj, 536, 896

\bibitem[{{Cao} {et~al.}(2021){Cao}, {Aharonian}, {An}, {Axikegu}, {Bai},
  {Bao}, {Bastieri}, {Bi}, {Bi}, {Cai}, {Cai}, {Cao}, {Chang}, {Chang},
  {Chang}, {Chen}, {Chen}, {Chen}, {Chen}, {Chen}, {Chen}, {Chen}, {Chen},
  {Chen}, {Chen}, {Chen}, {Chen}, {Chen}, {Cheng}, {Cheng}, {Cui}, {Cui},
  {Cui}, {Dai}, {Dai}, {Dai}, {Danzengluobu}, {della Volpe}, {D'Ettorre
  Piazzoli}, {Dong}, {Fan}, {Fan}, {Fan}, {Fang}, {Fang}, {Feng}, {Feng},
  {Feng}, {Feng}, {Gao}, {Gao}, {Gao}, {Gao}, {Ge}, {Geng}, {Gong}, {Gou},
  {Gu}, {Guo}, {Guo}, {Guo}, {Guo}, {Han}, {He}, {He}, {He}, {He}, {He}, {He},
  {Heller}, {Hor}, {Hou}, {Hou}, {Hu}, {Hu}, {Hu}, {Hu}, {Huang}, {Huang},
  {Huang}, {Huang}, {Huang}, {Ji}, {Ji}, {Jia}, {Jiang}, {Jiang}, {Jin},
  {Kuleshov}, {Levochkin}, {Li}, {Li}, {Li}, {Li}, {Li}, {Li}, {Li}, {Li},
  {Li}, {Li}, {Li}, {Li}, {Li}, {Li}, {Li}, {Li}, {Li}, {Liang}, {Liang},
  {Lin}, {Liu}, {Liu}, {Liu}, {Liu}, {Liu}, {Liu}, {Liu}, {Liu}, {Liu}, {Liu},
  {Liu}, {Liu}, {Liu}, {Liu}, {Liu}, {Long}, {Lu}, {Lv}, {Ma}, {Ma}, {Ma},
  {Mao}, {Masood}, {Mitthumsiri}, {Montaruli}, {Nan}, {Pang},
  {Pattarakijwanich}, {Pei}, {Qi}, {Ruffolo}, {Rulev}, {S{\'a}iz}, {Shao},
  {Shchegolev}, {Sheng}, {Shi}, {Song}, {Stenkin}, {Stepanov}, {Sun}, {Sun},
  {Sun}, {Tam}, {Tang}, {Tian}, {Wang}, {Wang}, {Wang}, {Wang}, {Wang}, {Wang},
  {Wang}, {Wang}, {Wang}, {Wang}, {Wang}, {Wang}, {Wang}, {Wang}, {Wang},
  {Wang}, {Wang}, {Wang}, {Wang}, {Wang}, {Wang}, {Wei}, {Wei}, {Wei}, {Wen},
  {Wu}, {Wu}, {Wu}, {Wu}, {Wu}, {Xi}, {Xia}, {Xia}, {Xiang}, {Xiao}, {Xiao},
  {Xin}, {Xin}, {Xing}, {Xu}, {Xu}, {Xue}, {Yan}, {Yang}, {Yang}, {Yang},
  {Yang}, {Yang}, {Yang}, {Yang}, {Yao}, {Yao}, {Ye}, {Yin}, {Yin}, {You},
  {You}, {Yu}, {Yuan}, {Zeng}, {Zeng}, {Zeng}, {Zeng}, {Zha}, {Zhai}, {Zhang},
  {Zhang}, {Zhang}, {Zhang}, {Zhang}, {Zhang}, {Zhang}, {Zhang}, {Zhang},
  {Zhang}, {Zhang}, {Zhang}, {Zhang}, {Zhang}, {Zhang}, {Zhang}, {Zhang},
  {Zhang}, {Zhang}, {Zhao}, {Zhao}, {Zhao}, {Zhao}, {Zhao}, {Zheng}, {Zheng},
  {Zhou}, {Zhou}, {Zhou}, {Zhou}, {Zhou}, {Zhou}, {Zhu}, {Zhu}, {Zhu}, {Zhu},
  \& {Zuo}}]{Cao2021}
{Cao}, Z., {Aharonian}, F.~A., {An}, Q., {et~al.} 2021, \nat, 594, 33

\bibitem[{{Cash}(1979)}]{Cash1979}
{Cash}, W. 1979, \apj, 228, 939

\bibitem[{{Clark} {et~al.}(2008){Clark}, {Muno}, {Negueruela}, {Dougherty},
  {Crowther}, {Goodwin}, \& {de Grijs}}]{Clark2008}
{Clark}, J.~S., {Muno}, M.~P., {Negueruela}, I., {et~al.} 2008, \aap, 477, 147

\bibitem[{{Clark} {et~al.}(2005){Clark}, {Negueruela}, {Crowther}, \&
  {Goodwin}}]{Clark2005}
{Clark}, J.~S., {Negueruela}, I., {Crowther}, P.~A., \& {Goodwin}, S.~P. 2005,
  \aap, 434, 949

\bibitem[{{Clark} {et~al.}(2020){Clark}, {Ritchie}, \&
  {Negueruela}}]{Clark2020}
{Clark}, J.~S., {Ritchie}, B.~W., \& {Negueruela}, I. 2020, \aap, 635, A187

\bibitem[{{Corrales} {et~al.}(2017){Corrales}, {Mon}, {Haggard}, {Baganoff},
  {Garmire}, {Degenaar}, \& {Reynolds}}]{Corrales2017}
{Corrales}, L.~R., {Mon}, B., {Haggard}, D., {et~al.} 2017, \apj, 839, 76

\bibitem[{{Crowther} {et~al.}(2006){Crowther}, {Hadfield}, {Clark},
  {Negueruela}, \& {Vacca}}]{Crowther2006}
{Crowther}, P.~A., {Hadfield}, L.~J., {Clark}, J.~S., {Negueruela}, I., \&
  {Vacca}, W.~D. 2006, \mnras, 372, 1407

\bibitem[{{Cutri} {et~al.}(2003){Cutri}, {Skrutskie}, {van Dyk}, {Beichman},
  {Carpenter}, {Chester}, {Cambresy}, {Evans}, {Fowler}, {Gizis}, {Howard},
  {Huchra}, {Jarrett}, {Kopan}, {Kirkpatrick}, {Light}, {Marsh}, {McCallon},
  {Schneider}, {Stiening}, {Sykes}, {Weinberg}, {Wheaton}, {Wheelock}, \&
  {Zacarias}}]{Cutri2003}
{Cutri}, R.~M., {Skrutskie}, M.~F., {van Dyk}, S., {et~al.} 2003, {2MASS All
  Sky Catalog of point sources.}

\bibitem[{{Dashyan} \& {Dubois}(2020)}]{Dashyan2020}
{Dashyan}, G. \& {Dubois}, Y. 2020, \aap, 638, A123

\bibitem[{{Dickey} \& {Lockman}(1990)}]{Dickey1990}
{Dickey}, J.~M. \& {Lockman}, F.~J. 1990, \araa, 28, 215

\bibitem[{{Foreman-Mackey} {et~al.}(2013){Foreman-Mackey}, {Hogg}, {Lang}, \&
  {Goodman}}]{ForemanMackey2013}
{Foreman-Mackey}, D., {Hogg}, D.~W., {Lang}, D., \& {Goodman}, J. 2013, \pasp,
  125, 306

\bibitem[{{Freyberg} {et~al.}(2020){Freyberg}, {Perinati}, {Pacaud}, {Eraerds},
  {Churazov}, {Dennerl}, {Predehl}, {Merloni}, {Meidinger}, {Bulbul},
  {Friedrich}, {Gilfanov}, {Tenzer}, {Pommranz}, {Eckert}, {Schmitt}, {Brusa},
  \& {Santangelo}}]{Freyberg2020}
{Freyberg}, M., {Perinati}, E., {Pacaud}, F., {et~al.} 2020, in Society of
  Photo-Optical Instrumentation Engineers (SPIE) Conference Series, Vol. 11444,
  Space Telescopes and Instrumentation 2020: Ultraviolet to Gamma Ray, ed.
  J.-W.~A. {den Herder}, S.~{Nikzad}, \& K.~{Nakazawa}, 114441O

\bibitem[{{Friedman} {et~al.}(1967){Friedman}, {Byram}, \&
  {Chubb}}]{Friedman1967}
{Friedman}, H., {Byram}, E.~T., \& {Chubb}, T.~A. 1967, Science, 156, 374

\bibitem[{{Gabici} {et~al.}(2019){Gabici}, {Evoli}, {Gaggero}, {Lipari},
  {Mertsch}, {Orlando}, {Strong}, \& {Vittino}}]{Gabici2019}
{Gabici}, S., {Evoli}, C., {Gaggero}, D., {et~al.} 2019, International Journal
  of Modern Physics D, 28, 1930022

\bibitem[{{Getman} {et~al.}(2005){Getman}, {Flaccomio}, {Broos}, {Grosso},
  {Tsujimoto}, {Townsley}, {Garmire}, {Kastner}, {Li}, {Harnden}, {Wolk},
  {Murray}, {Lada}, {Muench}, {McCaughrean}, {Meeus}, {Damiani}, {Micela},
  {Sciortino}, {Bally}, {Hillenbrand}, {Herbst}, {Preibisch}, \&
  {Feigelson}}]{Getman2005}
{Getman}, K.~V., {Flaccomio}, E., {Broos}, P.~S., {et~al.} 2005, \apjs, 160,
  319

\bibitem[{{H{\"a}rer} {et~al.}(2023){H{\"a}rer}, {Reville}, {Hinton},
  {Mohrmann}, \& {Vieu}}]{Haerer2023}
{H{\"a}rer}, L.~K., {Reville}, B., {Hinton}, J., {Mohrmann}, L., \& {Vieu}, T.
  2023, \aap, 671, A4

\bibitem[{{Helder} {et~al.}(2012){Helder}, {Vink}, {Bykov}, {Ohira}, {Raymond},
  \& {Terrier}}]{2012Helder}
{Helder}, E.~A., {Vink}, J., {Bykov}, A.~M., {et~al.} 2012, \ssr, 173, 369

\bibitem[{{Hillas}(2005)}]{Hillas2005}
{Hillas}, A.~M. 2005, Journal of Physics G Nuclear Physics, 31, R95

\bibitem[{{Ipavich}(1975)}]{Ipavich1975}
{Ipavich}, F.~M. 1975, \apj, 196, 107

\bibitem[{{Kavanagh} {et~al.}(2011){Kavanagh}, {Norci}, \&
  {Meurs}}]{Kavanagh2011}
{Kavanagh}, P.~J., {Norci}, L., \& {Meurs}, E.~J.~A. 2011, \na, 16, 461

\bibitem[{{Kavanagh} {et~al.}(2019){Kavanagh}, {Vink}, {Sasaki}, {Chu},
  {Filipovi{\'c}}, {Ohm}, {Haberl}, {Manojlovic}, \& {Maggi}}]{Kavanagh2019}
{Kavanagh}, P.~J., {Vink}, J., {Sasaki}, M., {et~al.} 2019, \aap, 621, A138

\bibitem[{{Khangulyan} {et~al.}(2014){Khangulyan}, {Aharonian}, \&
  {Kelner}}]{Khangulyan2014}
{Khangulyan}, D., {Aharonian}, F.~A., \& {Kelner}, S.~R. 2014, \apj, 783, 100

\bibitem[{{Kothes} \& {Dougherty}(2007)}]{Kothes2007}
{Kothes}, R. \& {Dougherty}, S.~M. 2007, \aap, 468, 993

\bibitem[{{Lagage} \& {Cesarsky}(1983)}]{Lagage1983}
{Lagage}, P.~O. \& {Cesarsky}, C.~J. 1983, \aap, 125, 249

\bibitem[{{Lamer} {et~al.}(2021){Lamer}, {Schwope}, {Predehl}, {Traulsen},
  {Wilms}, \& {Freyberg}}]{Lamer2021}
{Lamer}, G., {Schwope}, A.~D., {Predehl}, P., {et~al.} 2021, \aap, 647, A7

\bibitem[{{Locatelli} {et~al.}(2024){Locatelli}, {Ponti}, {Zheng}, {Merloni},
  {Becker}, {Comparat}, {Dennerl}, {Freyberg}, {Sasaki}, \&
  {Yeung}}]{Locatelli2024}
{Locatelli}, N., {Ponti}, G., {Zheng}, X., {et~al.} 2024, \aap, 681, A78

\bibitem[{{Martire} {et~al.}(2022){Martire}, {Barreiro}, \&
  {Mart{\'\i}nez-Gonz{\'a}lez}}]{Martire2022}
{Martire}, F.~A., {Barreiro}, R.~B., \& {Mart{\'\i}nez-Gonz{\'a}lez}, E. 2022,
  \jcap, 2022, 003

\bibitem[{{Masui} {et~al.}(2009){Masui}, {Mitsuda}, {Yamasaki}, {Takei},
  {Kimura}, {Yoshino}, \& {McCammon}}]{Masui2009}
{Masui}, K., {Mitsuda}, K., {Yamasaki}, N.~Y., {et~al.} 2009, \pasj, 61, S115

\bibitem[{{Merloni} {et~al.}(2024){Merloni}, {Lamer}, {Liu}, {Ramos-Ceja},
  {Brunner}, {Bulbul}, {Dennerl}, {Doroshenko}, {Freyberg}, {Friedrich},
  {Gatuzz}, {Georgakakis}, {Haberl}, {Igo}, {Kreykenbohm}, {Liu}, {Maitra},
  {Malyali}, {Mayer}, {Nandra}, {Predehl}, {Robrade}, {Salvato}, {Sanders},
  {Stewart}, {Tub{\'\i}n-Arenas}, {Weber}, {Wilms}, {Arcodia}, {Artis},
  {Aschersleben}, {Avakyan}, {Aydar}, {Bahar}, {Balzer}, {Becker}, {Berger},
  {Boller}, {Bornemann}, {Br{\"u}ggen}, {Brusa}, {Buchner}, {Burwitz},
  {Camilloni}, {Clerc}, {Comparat}, {Coutinho}, {Czesla}, {Dannhauer},
  {Dauner}, {Dauser}, {Dietl}, {Dolag}, {Dwelly}, {Egg}, {Ehl}, {Freund},
  {Friedrich}, {Gaida}, {Garrel}, {Ghirardini}, {Gokus}, {Gr{\"u}nwald},
  {Grandis}, {Grotova}, {Gruen}, {Gueguen}, {H{\"a}mmerich}, {Hamaus},
  {Hasinger}, {Haubner}, {Homan}, {Ider Chitham}, {Joseph}, {Joyce},
  {K{\"o}nig}, {Kaltenbrunner}, {Khokhriakova}, {Kink}, {Kirsch}, {Kluge},
  {Knies}, {Krippendorf}, {Krumpe}, {Kurpas}, {Li}, {Liu}, {Locatelli},
  {Lorenz}, {M{\"u}ller}, {Magaudda}, {Mannes}, {McCall}, {Meidinger},
  {Michailidis}, {Migkas}, {Mu{\~n}oz-Giraldo}, {Musiimenta}, {Nguyen-Dang},
  {Ni}, {Olechowska}, {Ota}, {Pacaud}, {Pasini}, {Perinati}, {Pires},
  {Pommranz}, {Ponti}, {Poppenhaeger}, {P{\"u}hlhofer}, {Rau}, {Reh},
  {Reiprich}, {Roster}, {Saeedi}, {Santangelo}, {Sasaki}, {Schmitt},
  {Schneider}, {Schrabback}, {Schuster}, {Schwope}, {Seppi}, {Serim},
  {Shreeram}, {Sokolova-Lapa}, {Starck}, {Stelzer}, {Stierhof}, {Suleimanov},
  {Tenzer}, {Traulsen}, {Tr{\"u}mper}, {Tsuge}, {Urrutia}, {Veronica},
  {Waddell}, {Willer}, {Wolf}, {Yeung}, {Zainab}, {Zangrandi}, {Zhang},
  {Zhang}, \& {Zheng}}]{eROSITA2024}
{Merloni}, A., {Lamer}, G., {Liu}, T., {et~al.} 2024, \aap, 682, A34

\bibitem[{{Morlino} {et~al.}(2021){Morlino}, {Blasi}, {Peretti}, \&
  {Cristofari}}]{Morlino2021}
{Morlino}, G., {Blasi}, P., {Peretti}, E., \& {Cristofari}, P. 2021, \mnras,
  504, 6096

\bibitem[{{Muno} {et~al.}(2006){Muno}, {Law}, {Clark}, {Dougherty}, {de Grijs},
  {Portegies Zwart}, \& {Yusef-Zadeh}}]{Muno2006}
{Muno}, M.~P., {Law}, C., {Clark}, J.~S., {et~al.} 2006, \apj, 650, 203

\bibitem[{{Navarete} {et~al.}(2022){Navarete}, {Damineli}, {Ramirez}, {Rocha},
  \& {Almeida}}]{Navarete2022}
{Navarete}, F., {Damineli}, A., {Ramirez}, A.~E., {Rocha}, D.~F., \& {Almeida},
  L.~A. 2022, \mnras, 516, 1289

\bibitem[{{Negueruela} {et~al.}(2022){Negueruela}, {Alfaro}, {Dorda}, {Marco},
  {Ma{\'\i}z Apell{\'a}niz}, \& {Gonz{\'a}lez-Fern{\'a}ndez}}]{Negueruela2022}
{Negueruela}, I., {Alfaro}, E.~J., {Dorda}, R., {et~al.} 2022, \aap, 664, A146

\bibitem[{{Ohira} {et~al.}(2018){Ohira}, {Kisaka}, \& {Yamazaki}}]{Ohira2018}
{Ohira}, Y., {Kisaka}, S., \& {Yamazaki}, R. 2018, \mnras, 478, 926

\bibitem[{{Ohm} {et~al.}(2013){Ohm}, {Hinton}, \& {White}}]{Ohm2013}
{Ohm}, S., {Hinton}, J.~A., \& {White}, R. 2013, \mnras, 434, 2289

\bibitem[{{Padovani} {et~al.}(2020){Padovani}, {Ivlev}, {Galli}, {Offner},
  {Indriolo}, {Rodgers-Lee}, {Marcowith}, {Girichidis}, {Bykov}, \&
  {Kruijssen}}]{Padovani2020}
{Padovani}, M., {Ivlev}, A.~V., {Galli}, D., {et~al.} 2020, \ssr, 216, 29

\bibitem[{{Pedlar} {et~al.}(1999){Pedlar}, {Muxlow}, {Garrett}, {Diamond},
  {Wills}, {Wilkinson}, \& {Alef}}]{Pedlar1999}
{Pedlar}, A., {Muxlow}, T.~W.~B., {Garrett}, M.~A., {et~al.} 1999, \mnras, 307,
  761

\bibitem[{{Peschken} {et~al.}(2023){Peschken}, {Hanasz}, {Naab},
  {W{\'o}lta{\'n}ski}, \& {Gawryszczak}}]{Peschken2023}
{Peschken}, N., {Hanasz}, M., {Naab}, T., {W{\'o}lta{\'n}ski}, D., \&
  {Gawryszczak}, A. 2023, \mnras, 522, 5529

\bibitem[{{Ponti} {et~al.}(2023){Ponti}, {Zheng}, {Locatelli}, {Bianchi},
  {Zhang}, {Anastasopoulou}, {Comparat}, {Dennerl}, {Freyberg}, {Haberl},
  {Merloni}, {Reiprich}, {Salvato}, {Sanders}, {Sasaki}, {Strong}, \&
  {Yeung}}]{Ponti2023}
{Ponti}, G., {Zheng}, X., {Locatelli}, N., {et~al.} 2023, \aap, 674, A195

\bibitem[{{Predehl} {et~al.}(2021){Predehl}, {Andritschke}, {Arefiev},
  {Babyshkin}, {Batanov}, {Becker}, {B{\"o}hringer}, {Bogomolov}, {Boller},
  {Borm}, {Bornemann}, {Br{\"a}uninger}, {Br{\"u}ggen}, {Brunner}, {Brusa},
  {Bulbul}, {Buntov}, {Burwitz}, {Burkert}, {Clerc}, {Churazov}, {Coutinho},
  {Dauser}, {Dennerl}, {Doroshenko}, {Eder}, {Emberger}, {Eraerds},
  {Finoguenov}, {Freyberg}, {Friedrich}, {Friedrich}, {F{\"u}rmetz},
  {Georgakakis}, {Gilfanov}, {Granato}, {Grossberger}, {Gueguen}, {Gureev},
  {Haberl}, {H{\"a}lker}, {Hartner}, {Hasinger}, {Huber}, {Ji}, {Kienlin},
  {Kink}, {Korotkov}, {Kreykenbohm}, {Lamer}, {Lomakin}, {Lapshov}, {Liu},
  {Maitra}, {Meidinger}, {Menz}, {Merloni}, {Mernik}, {Mican}, {Mohr},
  {M{\"u}ller}, {Nandra}, {Nazarov}, {Pacaud}, {Pavlinsky}, {Perinati},
  {Pfeffermann}, {Pietschner}, {Ramos-Ceja}, {Rau}, {Reiffers}, {Reiprich},
  {Robrade}, {Salvato}, {Sanders}, {Santangelo}, {Sasaki}, {Scheuerle},
  {Schmid}, {Schmitt}, {Schwope}, {Shirshakov}, {Steinmetz}, {Stewart},
  {Str{\"u}der}, {Sunyaev}, {Tenzer}, {Tiedemann}, {Tr{\"u}mper}, {Voron},
  {Weber}, {Wilms}, \& {Yaroshenko}}]{Predehl2021}
{Predehl}, P., {Andritschke}, R., {Arefiev}, V., {et~al.} 2021, \aap, 647, A1

\bibitem[{{Ruszkowski} \& {Pfrommer}(2023)}]{Ruszkowski2023}
{Ruszkowski}, M. \& {Pfrommer}, C. 2023, \aapr, 31, 4

\bibitem[{{Smith} {et~al.}(2023){Smith}, {Abdollahi}, {Ajello}, {Bailes},
  {Baldini}, {Ballet}, {Baring}, {Bassa}, {Gonzalez}, {Bellazzini}, {Berretta},
  {Bhattacharyya}, {Bissaldi}, {Bonino}, {Bottacini}, {Bregeon}, {Bruel},
  {Burgay}, {Burnett}, {Cameron}, {Camilo}, {Caputo}, {Caraveo}, {Cavazzuti},
  {Chiaro}, {Ciprini}, {Clark}, {Cognard}, {Corongiu}, {Orestano},
  {Crnogorcevic}, {Cuoco}, {Cutini}, {D'Ammando}, {de Angelis}, {DeCesar}, {De
  Gaetano}, {de Menezes}, {Deneva}, {de Palma}, {Di Lalla}, {Dirirsa}, {Di
  Venere}, {Dom{\'\i}nguez}, {Dumora}, {Fegan}, {Ferrara}, {Fiori},
  {Fleischhack}, {Flynn}, {Franckowiak}, {Freire}, {Fukazawa}, {Fusco},
  {Galanti}, {Gammaldi}, {Gargano}, {Gasparrini}, {Giacchino}, {Giglietto},
  {Giordano}, {Giroletti}, {Green}, {Grenier}, {Guillemot}, {Guiriec},
  {Gustafsson}, {Harding}, {Hays}, {Hewitt}, {Horan}, {Hou}, {Jankowski},
  {Johnson}, {Johnson}, {Johnston}, {Kataoka}, {Keith}, {Kerr}, {Kramer},
  {Kuss}, {Latronico}, {Lee}, {Li}, {Li}, {Limyansky}, {Longo}, {Loparco},
  {Lorusso}, {Lovellette}, {Lower}, {Lubrano}, {Lyne}, {Maan}, {Maldera},
  {Manchester}, {Manfreda}, {Marelli}, {Mart{\'\i}-Devesa}, {Mazziotta},
  {McEnery}, {Mereu}, {Michelson}, {Mickaliger}, {Mitthumsiri}, {Mizuno},
  {Moiseev}, {Monzani}, {Morselli}, {Negro}, {Nemmen}, {Nieder}, {Nuss},
  {Omodei}, {Orienti}, {Orlando}, {Ormes}, {Palatiello}, {Paneque},
  {Panzarini}, {Parthasarathy}, {Persic}, {Pesce-Rollins}, {Pillera}, {Poon},
  {Porter}, {Possenti}, {Principe}, {Rain{\`o}}, {Rando}, {Ransom}, {Ray},
  {Razzano}, {Razzaque}, {Reimer}, {Reimer}, {Renault-Tinacci}, {Romani},
  {S{\'a}nchez-Conde}, {Parkinson}, {Scotton}, {Serini}, {Sgr{\`o}}, {Shannon},
  {Sharma}, {Shen}, {Siskind}, {Spandre}, {Spinelli}, {Stappers}, {Stephens},
  {Suson}, {Tabassum}, {Tajima}, {Tak}, {Theureau}, {Thompson}, {Tibolla},
  {Torres}, {Valverde}, {Venter}, {Wadiasingh}, {Wang}, {Wang}, {Wang},
  {Weltevrede}, {Wood}, {Yan}, {Zaharijas}, {Zhang}, \& {Zhu}}]{Smith2023}
{Smith}, D.~A., {Abdollahi}, S., {Ajello}, M., {et~al.} 2023, \apj, 958, 191

\bibitem[{{Sunyaev} {et~al.}(2021){Sunyaev}, {Arefiev}, {Babyshkin},
  {Bogomolov}, {Borisov}, {Buntov}, {Brunner}, {Burenin}, {Churazov},
  {Coutinho}, {Eder}, {Eismont}, {Freyberg}, {Gilfanov}, {Gureyev}, {Hasinger},
  {Khabibullin}, {Kolmykov}, {Komovkin}, {Krivonos}, {Lapshov}, {Levin},
  {Lomakin}, {Lutovinov}, {Medvedev}, {Merloni}, {Mernik}, {Mikhailov},
  {Molodtsov}, {Mzhelsky}, {M{\"u}ller}, {Nandra}, {Nazarov}, {Pavlinsky},
  {Poghodin}, {Predehl}, {Robrade}, {Sazonov}, {Scheuerle}, {Shirshakov},
  {Tkachenko}, \& {Voron}}]{Sunyaev2021}
{Sunyaev}, R., {Arefiev}, V., {Babyshkin}, V., {et~al.} 2021, \aap, 656, A132

\bibitem[{{Tarricq} {et~al.}(2021){Tarricq}, {Soubiran}, {Casamiquela},
  {Cantat-Gaudin}, {Chemin}, {Anders}, {Antoja}, {Romero-G{\'o}mez},
  {Figueras}, {Jordi}, {Bragaglia}, {Balaguer-N{\'u}{\~n}ez}, {Carrera},
  {Castro-Ginard}, {Moitinho}, {Ramos}, \& {Bossini}}]{Tarricq2021}
{Tarricq}, Y., {Soubiran}, C., {Casamiquela}, L., {et~al.} 2021, \aap, 647, A19

\bibitem[{{Vieu} {et~al.}(2020){Vieu}, {Gabici}, \& {Tatischeff}}]{Vieu2020}
{Vieu}, T., {Gabici}, S., \& {Tatischeff}, V. 2020, \mnras, 494, 3166

\bibitem[{{Vink}(2012)}]{Vink2012}
{Vink}, J. 2012, \aapr, 20, 49

\bibitem[{{Westerlund}(1961)}]{Westerlund1961}
{Westerlund}, B. 1961, \pasp, 73, 51

\bibitem[{{Westerlund}(1987)}]{Westerlund1987}
{Westerlund}, B.~E. 1987, \aaps, 70, 311

\bibitem[{{Wiener} {et~al.}(2019){Wiener}, {Zweibel}, \&
  {Ruszkowski}}]{Wiener2019}
{Wiener}, J., {Zweibel}, E.~G., \& {Ruszkowski}, M. 2019, \mnras, 489, 205

\bibitem[{{Wulf} {et~al.}(2019){Wulf}, {Eckart}, {Galeazzi}, {Jaeckel},
  {Kelley}, {Kilbourne}, {Morgan}, {Porter}, {McCammon}, \&
  {Szymkowiak}}]{Wulf2019}
{Wulf}, D., {Eckart}, M.~E., {Galeazzi}, M., {et~al.} 2019, \apj, 884, 120

\bibitem[{{Yang} {et~al.}(2018){Yang}, {de O{\~n}a Wilhelmi}, \&
  {Aharonian}}]{Yang2018}
{Yang}, R.-z., {de O{\~n}a Wilhelmi}, E., \& {Aharonian}, F. 2018, \aap, 611,
  A77

\bibitem[{{Yeung} {et~al.}(2023){Yeung}, {Freyberg}, {Ponti}, {Dennerl},
  {Sasaki}, \& {Strong}}]{Yeung2023}
{Yeung}, M.~C.~H., {Freyberg}, M.~J., {Ponti}, G., {et~al.} 2023, \aap, 676, A3

\bibitem[{{Zabalza}(2015)}]{Zabalza2015}
{Zabalza}, V. 2015, in International Cosmic Ray Conference, Vol.~34, 34th
  International Cosmic Ray Conference (ICRC2015), 922

\end{thebibliography}

%Appendix
\begin{appendix}

\section{Westerlund 1 background fit}

\begin{figure}[h!]
        \centering
        \includegraphics[width = \linewidth]{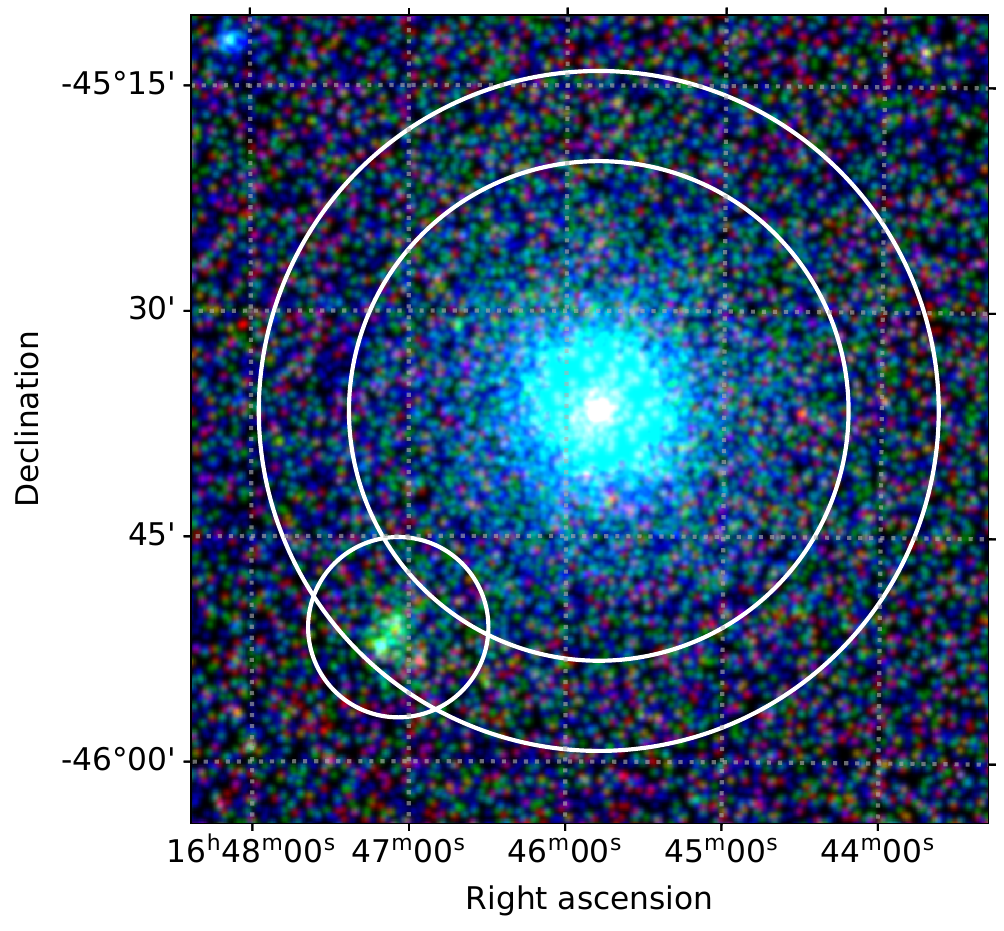}
        \caption{\textit{eROSITA} RGB image of the surroundings of Wd~1. Red, green, and blue colors correspond to $0.7$ to \SI{1.1}{\kilo\electronvolt}, $1.1$ to \SI{2.3}{\kilo\electronvolt}, and $2.3$ to \SI{10}{\kilo\electronvolt}, respectively. The image was smoothed using a Gaussian kernel with standard deviation \SI{10}{\arcsec}. The large annulus indicates the background region for the Wd~1 spectral fit, from which the white circle around Wd~1 itself was excluded.}
        \label{FigWd1Bkg}
    \end{figure}

    Figure~\ref{FigWd1Bkg} shows the background region for the spectral fit to Wd~1. The region is given by the large annulus around GX$+$340, while the smaller white circle around Wd~1 was excluded.

    \begin{table}[h!]
        \caption{Westerlund 1 background fit parameters.}
        \label{TabWd1Bkg}
        \centering
        \begin{tabular}{ll}
            \hline
            \hline
            Parameter & Value\\
            \hline
            \texttt{BACKSCAL} [\SI{}{\square\arcmin}] & 213.48\\
            \texttt{REGAREA} [\SI{}{\square\arcmin}] & 672.07\\
            $N_\mathrm{H}$ [\SI{e22}{\per\square\centi\meter}] & $2.1\pm0.1$\\
            \hline
            $\eta_\mathrm{LHB}$ [\SI{e-14}{\centi\meter\tothe{-5}\per\square\arcmin}] & $1.0_{-0.6}^{+0.4}\cdot10^{-6}$ \\        
            $N_\mathrm{H}^\mathrm{CGM}$ [\SI{e22}{\per\square\centi\meter}] & $0.4_{-0.4}^{+0.3}$\\
            $\eta_\mathrm{CGM}$ [\SI{e-14}{\centi\meter\tothe{-5}\per\square\arcmin}] & $9_{-8}^{+30}\cdot10^{-5}$\\
            $N_\mathrm{H}^\mathrm{cor}$ [\SI{e22}{\per\square\centi\meter}] & $1.2_{-0.3}^{+0.3}$\\
            $\eta_\mathrm{cor}$ [\SI{e-14}{\centi\meter\tothe{-5}\per\square\arcmin}] & $1.4_{-0.4}^{+0.6}\cdot10^{-5}$\\
            \hline
            $N_\mathrm{H}^\mathrm{LMXB}$ [\SI{e22}{\per\square\centi\meter}] & $4.3_{-0.5}^{+0.7}$\\
            $k_\mathrm{B}T_\mathrm{in}$ [\SI{}{\kilo\electronvolt}] & $10_{-6}^{100}$\\
            $\eta_\mathrm{LMXB}$ [\SI{e-2}{\square\kilo\meter\per\square\kilo\parsec}] & $4_{-4}^{+20}\cdot10^{-3}$\\
            \hline
        \end{tabular}
        \tablefoot{The \texttt{BACKSCAL} and \texttt{REGAREA} parameters were taken from the header of the background fits file. The $N_\mathrm{H}$ parameter is the arithmetic mean of eleven column densities determined along the region via \texttt{nH}, while its uncertainty is the standard deviation of this mean. All other parameters are \texttt{Xspec} best-fit parameters with 90\% confidence intervals. All normalizations except for $\eta_\mathrm{LMXB}$ are scaled with the background area in square arcminutes. Confidence intervals given without a plus or minus sign indicate that a parameter ran into a hard limit before it could sample its full 90\% confidence region.}
    \end{table}

    Table~\ref{TabWd1Bkg} gives the best-fit parameters for the background of Wd~1 in Sect. \ref{SecWd1Background}. We note the large uncertainties on $k_\mathrm{B}T_\mathrm{in}$ and $\eta_\mathrm{LMXB}$. The first one of these ran into its hard upper limit of \SI{100}{\kilo\electronvolt}. Via manual variation of these parameters, we found them to be degenerate: Similarly good fits are possible for large values of $k_\mathrm{B}T_\mathrm{in}$ and small values of $\eta_\mathrm{LMXB}$, and vice versa. However, since we only wanted to estimate the background from the LMXB's halo, we accepted this fit without further investigation.

\section{HESS J1646--458 scaling parameters}

    Table~\ref{TabJ1646Scale} gives the constant factors $w$ by which the background models of the five telescope modules were scaled during the fits to the HESS~J1646$-$458 source region in Sect. \ref{SecUpperLimits}. Furthermore, the table also lists the \texttt{BACKSCAL} parameters for the different telescope modules and source regions that were used to fix the filter wheel closed models during these fits.

    \begin{table}[h]
        \caption{X-ray background component normalizations $w$ and \texttt{BACKSCAL} parameters for the five telescope modules during the background fits to the HESS~J1646$-$458 regions.}
        \label{TabJ1646Scale}
        \centering
        \begin{tabular}{clllll}
            \hline
            \hline
            Parameter & TM 1 & TM 2 & TM 3 & TM 4 & TM 6\\
            \hline
            $w$ & 1 & 1.16 & 1.14 & 1.01 & 1.10 \\
            \texttt{BACKSCAL} region 1 & 115.2 & 114.8 & 114.5 & 112.3 & 115.6 \\
            \texttt{BACKSCAL} region 2 & 244.4 & 243.7 & 243.0 & 237.6 & 245.2 \\
            \texttt{BACKSCAL} region 3 & 312.5 & 311.0 & 315.0 & 308.5 & 311.8\\
            \texttt{BACKSCAL} region 4 & 302.8 & 300.6 & 302.4 & 297.7 & 302.0 \\
            \hline
        \end{tabular}
        \tablefoot{The telescope module normalizations were empirically determined within the \textit{eROSITA} consortium. The \texttt{BACKSCAL} parameters were taken from the headers of the four source region fits files. They are given in square arcminutes.}
    \end{table}

\newpage

\section{Minimum electron energies $E_\mathrm{min}$}

\begin{figure}[h!]
        \centering
        \includegraphics[width = \linewidth]{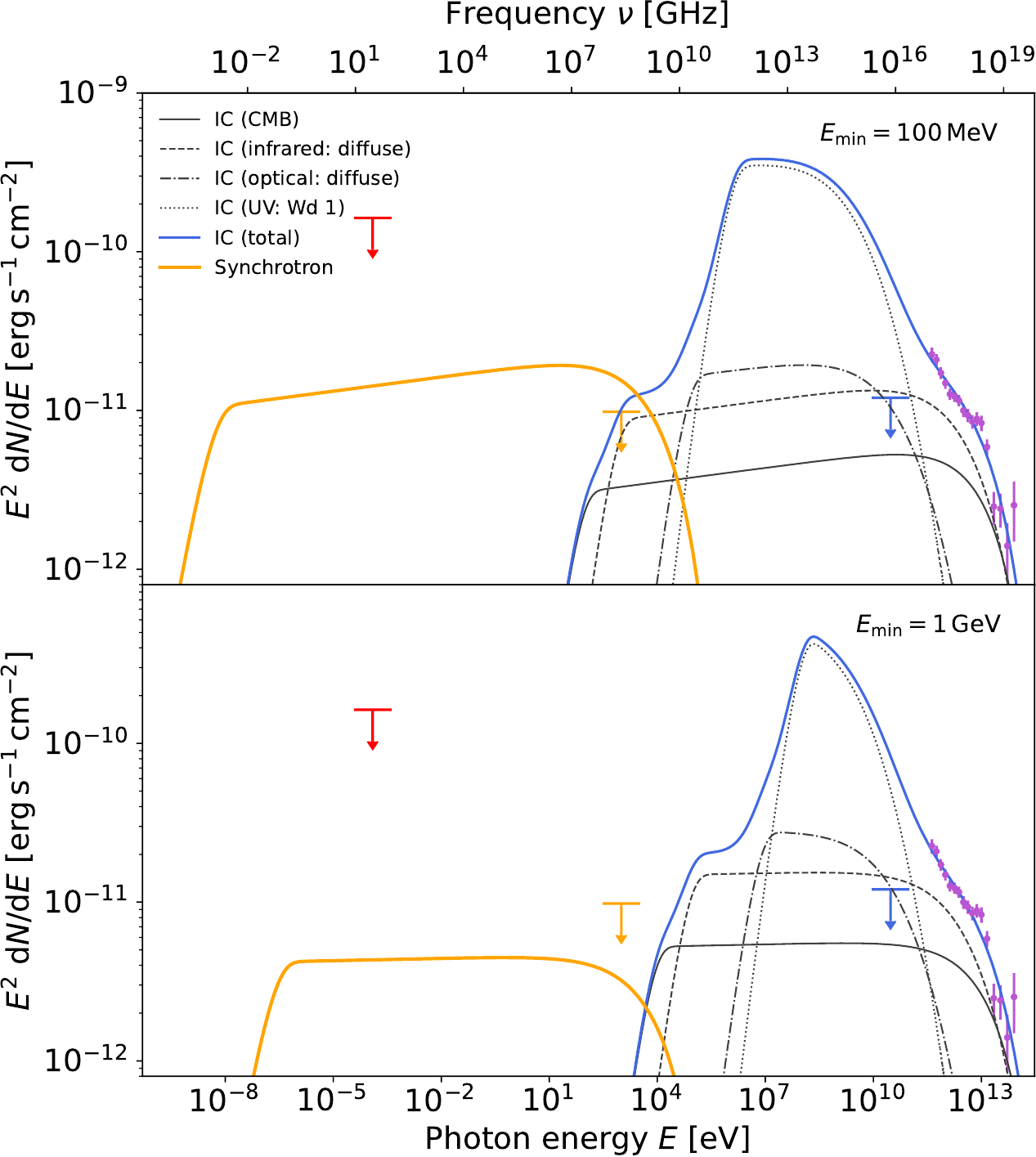}
        \caption{Spectral energy distribution fits to HESS~J1646$-$458, assuming $E_\mathrm{min} = \SI{100}{\mega\electronvolt}$ (\textit{upper panel}) and $E_\mathrm{min} = \SI{1}{\giga\electronvolt}$ (\textit{lower panel}). The data are, from left to right: the \textit{Planck} radio upper limit in red, the \textit{eROSITA} X-ray upper confidence bound in orange, the \textit{Fermi}-LAT HE $\gamma$-ray upper limit in blue, and the H.E.S.S. VHE $\gamma$-ray data from \citet{Aharonian2022} in purple. The solid orange lines are synchrotron radiation and the solid blue lines are IC scattering. The different target photon contributions to the IC scattering are given by the thin black lines. The CMB is solid, diffuse Galactic infrared radiation dashed, diffuse Galactic optical light dashed-dotted, and Wd~1's photon field dotted.}
        \label{FigSEDDouble}%
    \end{figure}

    For the SED fits in Sect. \ref{SecSED}, we tested the effect of varying the minimum energy of accelerated electrons $E_\mathrm{min}$. Apart from the fit with \SI{500}{\mega\electronvolt}, which can be seen in Fig.~\ref{FigSED}, Fig.~\ref{FigSEDDouble} shows the results for \SI{100}{\mega\electronvolt} (upper panel) and \SI{1}{\giga\electronvolt} (lower panel). Figure~\ref{FigSEDDouble} demonstrates how a lower value of $E_\mathrm{min}$ shifts the lower energy cutoff of the IC and synchrotron components to lower energies. In particular, for $E_\mathrm{min} = \SI{100}{\mega\electronvolt}$, the IC component contributes dominantly to the X-ray upper bound, which is why this is the lowest value of $E_\mathrm{min}$ discussed by us. In this case, the magnetic field strength $B$ is no longer penalized strongly for overshooting the X-ray upper bound and its main constraint becomes the radio upper limit. On the other hand, for $E_\mathrm{min} = \SI{1}{\giga\electronvolt}$, the IC component is practically irrelevant for the X-ray upper bound. This leads to results consistent with the $E_\mathrm{min} = \SI{500}{\mega\electronvolt}$ case, as can be seen in Table~\ref{TabSED}.

\end{appendix}

\end{document}